\documentclass[fleqn,usenatbib]{mnras}

\usepackage{newtxtext,newtxmath}

\usepackage[T1]{fontenc}

\DeclareRobustCommand{\VAN}[3]{#2}
\let\VANthebibliography\thebibliography
\def\thebibliography{\DeclareRobustCommand{\VAN}[3]{##3}\VANthebibliography}

\usepackage{graphicx}	
\usepackage{amsmath}	
\usepackage{float}
\usepackage{caption}
\usepackage{array}
\usepackage{booktabs}
\usepackage{mleftright}
\usepackage{bm}
\usepackage{bbm}
\usepackage{adjustbox}
\usepackage{tikz}
\usepackage{anyfontsize}
\usepackage{orcidlink}

\newcommand{\bzero}{\mathbf{0}}
\newcommand{\bone}{\mathbf{1}}
\newcommand{\bc}{\mathbf{c}}
\newcommand{\bt}{\mathbf{t}}
\newcommand{\bx}{\mathbf{x}}
\newcommand{\by}{\mathbf{y}}
\newcommand{\bC}{\mathbf{C}}
\newcommand{\bI}{\mathbf{I}}
\newcommand{\bP}{\mathbf{P}}
\newcommand{\bX}{\mathbf{X}}
\newcommand{\bY}{\mathbf{Y}}
\newcommand{\bZ}{\mathbf{Z}}
\newcommand{\bveta}{\bm{\eta}}
\newcommand{\beps}{\bm{\varepsilon}}

\newcommand{\bLambda}{\bm{\Lambda}}
\newcommand{\bSigma}{\bm{\Sigma}}
\newcommand{\bGamma}{\bm{\Gamma}}

\newcommand{\bPhi}{\bm{\Phi}}

\newcommand{\sN}{\mathcal{N}}
\newcommand{\sX}{\mathcal{X}}
\newcommand{\sY}{\mathcal{Y}}

\newcommand{\bb}[1]{{\mathbb{#1}}}
\newcommand{\R}{\bb{R}}
\newcommand{\N}{\bb{N}}
\renewcommand{\P}{\bb{P}}

\newcommand{\oneT}{_{1:T}}
\newcommand{\onet}{_{1:t}}
\renewcommand{\Vec}[1]{\mathrm{vec\mleft(#1\mright)}}

\newcommand{\tdelta}{\tilde{\delta}}
\newcommand{\tbdelta}{\tilde{\bdelta}}
\newcommand{\tbGamma}{\tilde{\bGamma}}

\newcommand{\tgamma}{\tilde{\gamma}}
\newcommand{\tL}{\tilde{L}}

\newcommand{\E}[1]{\bb{E} \left[ #1 \right]}
\newcommand{\one}[1]{\mathbbm{1}_{#1}}
\newcommand{\Var}[1]{\mathrm{Var}\left(#1\right)}
\newcommand{\argmax}[1]{\underset{#1}{\mathrm{argmax}}\,}

\newcommand{\dif}{\mathop{}\!\mathrm{d}}

\newcommand{\evlac}{{EV\,Lac\ }}
\newcommand{\evlacnospace}{{EV\,Lac}}
\newcommand{\chandra}{{\sl Chandra\ }}

\title[Separating States in Astronomical Sources Using Hidden Markov Models] {Separating States in Astronomical Sources Using Hidden Markov Models: With a Case Study of Flaring and Quiescence on \evlacnospace}

\author[R. Zimmerman et al.]{Robert Zimmerman,$^{1}$\orcidlink{0000-0002-0062-0926}
David A. van Dyk,$^{2}$\thanks{E-mail: \href{d.van-dyk@imperial.ac.uk}{d.van-dyk@imperial.ac.uk}}\orcidlink{0000-0002-0816-331X}
Vinay L. Kashyap,$^{3}$\orcidlink{0000-0002-3869-7996}
and Aneta Siemiginowska$^{3}$\orcidlink{0000-0002-0905-7375}
\\
$^{1}$Department of Statistical Sciences, University of Toronto, 700 University Avenue, Toronto, ON M5G 1Z5\\
$^{2}$Statistics Section, Department of Mathematics, Imperial College London, 180 Queen’s Gate, London, UK SW7 2AZ\\
$^{3}$Center for Astrophysics $|$ Harvard \& Smithsonian, 60 Garden Street, Cambridge, MA 02138\\
}

\date{Accepted XXX. Received YYY; in original form ZZZ}

\pubyear{2024}

\begin{document}
\label{firstpage}
\pagerange{\pageref{firstpage}--\pageref{lastpage}}
\maketitle

\begin{abstract}
We present a new method to distinguish between different states (e.g., high and low, quiescent and flaring) in astronomical sources with count data. The method models the underlying physical process as latent variables following a continuous-space Markov chain that determines the expected Poisson counts in observed light curves in multiple passbands. For the underlying state process, we consider several autoregressive processes, yielding continuous-space hidden Markov models of varying complexity. Under these models, we can infer the state that the object is in at any given time. The continuous state predictions from these models are then dichotomized with the help of a finite mixture model to produce state classifications. We apply these techniques to X-ray data from the active dMe flare star \evlacnospace, splitting the data into quiescent and flaring states. We find that a first-order vector autoregressive process efficiently separates flaring from quiescence: flaring occurs over 30--40\% of the observation durations, a well-defined persistent quiescent state can be identified, and the flaring state is characterized by higher plasma temperatures and emission measures.
\end{abstract}




\section{Introduction}\label{sec:intro}

The ubiquitous variability of astronomical sources spans large dynamic ranges in both intensity and time scale. The intensities typically vary differently in different passbands (i.e., they exhibit spectral variations as well). The causes of such variability are diverse, ranging from nuclear flashes occurring in low-mass X-ray binaries over durations of seconds, to magnetic reconnection flares on stars and accretion driven dipping in compact binaries lasting from a fraction of a ks to tens of ks, to gravitational lensing lasting for days, to abrupt changes in accretion levels onto compact objects which then persist for long durations ranging from weeks to months, to cyclic activity on stars that spans a decade, etc. The underlying physical processes that lead to such strong variations are not fully understood. In order to  model and predict these variations, we first need to identify robustly the times when the states of the sources appear to change.

We posit here that when we observe large intermittent variability, there is some identifiable characteristic in the source system -- modelled as a \emph{hidden state} -- which serves as a predictor to distinguish between different levels of activity. As an example, consider the flaring activity on stars, where we observe short duration bursts whose profiles show a rapid rise in intensity exceeding the typical intensity by several factors, followed  by a cooling-dominated exponential decay. This profile manifests as a stochastic sequence of alternating \emph{active periods} with frequent and energetic emissions at short time scales of a few ks, and \emph{quiescent periods} with periodic or smaller fluctuations.  We aim to build a model that describes the timing of such flaring, and includes a rudimentary quantification of the underlying variability. From a statistical point of view, including a latent process enables us to model observed correlations in the light curve and thus to predict and estimate the long-run proportion of time spent in flaring and quiescent states. 

Previous work on detecting or isolating such variability has focused mainly on local statistical significance testing, applying a set of somewhat ad-hoc rules, using automatic/black-box learning methods (e.g., neural networks) to identify flares in observed light curves, or modelling the intensities as a mixture distribution. In a study of $\gamma$-ray flares in blazars, for example, \citet{nalewajko2013brightest} used a simple rule that first identifies the peak flux and then defines the flare duration as the time interval with flux greater than 50\% of that observed in the peak.  \citet{robinson1995search} took a more statistical approach in their search for microflares in dMe flare stars: they computed the statistical significance of peaks in the binned data where the null distribution is determined by repeating their procedure on light curves where the bins have been randomly permuted. \citet{aschwanden2012automated} proposed an ``automated flare detection algorithm'' which is a set of criteria that are applied to a smoothed light curve; a background/quiescent level is determined using the time period before a local minimum in the light curve and the flare is associated with the interval starting at this minimum and continuing through the first subsequent local minimum that is below a background-dependent threshold. \cite {2021AGUFMSH25E2139P} adopted a similar procedure to detect flares in GOES X-ray light curves.\footnote{See also Appendix~A of the \href{https://data.ngdc.noaa.gov/platforms/solar-space-observing-satellites/goes/goes16/l2/docs/GOES-R_XRS_L2_Data_Users_Guide.pdf}{User’s Guide for GOES-R XRS L2 Products} (Machol, Codrescu, \& Peck 2023)}  A large sample of M-dwarf flares was obtained by \citet{davenport2014kepler} using an iterative smoothing procedure to remove star spot and then identifying flares as intervals that exhibit a positive flux excursion of more than $2.5\sigma$. More recently, supervised learning methods such as convolutional neural networks \citep[e.g.,][]{feinstein2020flare} have been used, while other researchers have continued to rely on visual inspection \citep[e.g.,][]{kashapova2021morphology}. Nearly all efforts to date have focused on univariate single-band light curves. A notable exception appears in \citet{fleming2022new}, who combined near UV and far UV light curves in a search for flares in M-dwarfs. They deployed a set of rules whereby a (peak) flare is identified by either two consecutive NUV data points above $3\sigma$ or two simultaneous data points above $3\sigma$, one in each band. 

While these methods include techniques that make use of statistical significance and standard deviations, they do not take advantage of principled statistical methods to model or fit features in the observed light curves. More principled statistical methods for identifying ``bursts’’ in astrophysical light curves were pioneered by Scargle’s work on Bayesian Blocks \citep{scargle1998studies,2013ApJ...764..167S}. The method assumes a piecewise constant intensity function for a Poisson process in time, and implements a fully Bayesian strategy for estimating the number of breakpoints. The time intervals with constant intensity are called blocks and their number is determined by maximizing the Bayes factor or posterior odds. The breakpoints are determined sequentially via their posterior distribution as blocks are added to the model. The Bayesian Blocks method has proved to be an invaluable tool for identifying ``bursts’’ in light curves and has recently been used to separate the quiescent and active states of $\gamma$-ray flaring blazars \citep{yoshida2023flare}. However, because the adopted model is piecewise constant, the fit results become difficult to interpret when dealing with smoothly increasing or decreasing intensities.

Large variability in astronomical sources is inevitably accompanied by spectral changes. In the case of stellar X-ray variability, \cite{wong2016detecting} proposed using a marked Poisson process for photon arrivals, treating photon wavelength as a ``mark’’. As with Bayesian Blocks, their method, called Automark, assumes a piecewise constant intensity function for the Poisson process that governs photon arrivals. Spectra are assumed to be constant \emph{between} the breakpoints, but \emph{within} each block are modelled in a flexible non-parametric manner that accounts for  spectral lines. The number of breakpoints is determined via the minimum description length principle. The method was extended to include spatial information/images by \citet{xu2021change}.

Neither Bayesian Blocks nor Automark provides a mechanism to model the underlying processes that generate the flares. With solar data the observation of individual flares enables a set of different but  also principled statistical approaches. Focusing exclusively on timing data for solar flares, for example, a number of authors have used characteristics of the distribution of waiting times between solar flares to better understand the process generating the flares. In this way, researchers have concluded that the waiting-time distribution is consistent with a time-varying Poisson process \citep[e.g.,][]{Wheatland_2000,moon:etal:01, whea:litv:02, Aschwanden_2019} or have used it to study the memory in this underlying process \citep[e.g., ][]{lepreti2001solar,Lei:etal:2020, Rivera_2022}. Unfortunately, these techniques do not apply to stars other than the Sun because individual flares are not observable. 

In this paper we consider the specific case of X-ray flares in stellar coronae, where we seek to model not the individual flares but rather the underlying flaring states, allowing us to estimate the flaring fraction and to study the spectra in different states. To this end, we employ a discrete-time \emph{hidden Markov model} (HMM) \citep{zucchini2017hidden}. This involves formulating a latent discrete-time Markov chain to represent the flaring process and is done in discrete time to match the discrete-time nature of the observed data. One novelty of our approach is that it leverages multi-band light curves to identify flaring and quiescent intervals. The flaring process evolves as a Markov chain over time and in each time interval the chain's value determines the distribution of the observed counts, and thus influences the evolution of the observed data over time. We consider both the case where the latent flare process can enter one of a finite number of states (e.g., a quiescent state and an active state) and the case of a continuum of states through which the process evolves. Mathematically, these two possibilities correspond to discrete and continuous state spaces of the latent Markov chain.

We use two \evlac light curves as a case study for our methods and find empirically that the continuous state space HMM provides a better representation of the light curves than does the discrete-state space HMM. However, the continuous-space HMM poses a computational challenge because its likelihood is intractable.  Thus, we introduce an approximation that is based on a truncated and discretized state space and that can be made arbitrarily precise. We propose three specific formulations of the continuous-space HMM for flaring stars and a method for choosing among these formulation. We then fit the preferred model and use it to estimate the underlying continuous  state variable  that indexes the transition between the quiescent and active states. Below, we denote this (possibly multivariate) indexing variable as $\mathbf{X}_t$.

The continuous-space HMM does not clearly differentiate between the quiescent and active state of the source, instead allowing for variability within the states and a smooth transition between them. Nonetheless, we aim to estimate the flaring fraction and to study the spectra within each state.  As such, we introduce a two-state analysis, where Stage~1 fits a continuous-space HMM and estimates the continuous state indexing variable $\mathbf{X}_t$ and Stage~2 fits a finite mixture model to $\mathbf{X}_t$ in order to estimate the actual intervals of quiescence and activity. The Markov process underlying the HMM allows us to model the temporal autocorrelations evident in the light curves and thus to capture them in the fitted $\mathbf{X}_t$.  In the second stage, we ignore these autocorrelations and focus instead on the marginal fitted values of $\mathbf{X}_t$ and use them to quantify the source's transitions between quiescence and activity. In this way, we can identify the long-run proportion of time spent in quiescence and flaring activity. The state predictions also allow us to estimate time intervals of quiescence and flaring, from which we obtain a comparative spectral analysis of both quiescence and flaring.

To the best of our knowledge, HMMs were first used to model time series of flare data by \cite{stanislavsky2020prediction}, who used a two-state autoregressive HMM to model continuous-valued daily \emph{solar} X-ray flux emission data in an effort to study the hidden process underlying solar flares. They focused primarily on next-day prediction of solar flare activity. More recently, \cite{esquivel2024detecting} used a similar approach with three states to model the flaring activity of an M dwarf star, in which the light curve was observed in one optical band with the TESS (Transiting Exoplanet Survey Satellite) Observatory. HMMs have also been used in other applications in astrophysics, such as distinguishing between noise-dominated and source-dominated states on strongly variable sources such as Sgr~A$^*$ \citep{meyer2014formal}.

Our approach here is more general. We use X-ray event lists containing information on photon arrival times and photon energy to construct light curves in multiple bands with low count rates in the Poisson regime, allowing us to explore short time scale events as well as spectral variations. While our method allows for prediction, our primary aim is to better understand the underlying physical process driving stellar flares.

The remainder of this paper is organized into six sections. We begin by introducing two \evlac light curves in Section~\ref{sec:motivate} to motivate our modelling choice. Section~\ref{sec:HMMs} consists of a general introduction and review of HMMs, emphasizing the notation and properties needed in the current setting. We present our Stage~1 analysis with its three HMMs in Section~\ref{sec:model}, emphasizing techniques for quantifying uncertainty and model selection.We turn to the Stage~2 analysis in Section~\ref{sec:classify} with a new proposed model-based method for classifying light curves into flaring and quiescent intervals. We illustrate the application of these models and methods with an analysis of the \evlac light curves in Section~\ref{sec:analysis}. Finally, we conclude with a discussion and suggestions for future work in Section~\ref{sec:discussion}. Several appendices review details of the algorithms used for maximum likelihood fitting of discrete-space HMMs, present technical aspects of the discrete approximation that we use for efficient fitting of continuous-space HMMs, and give additional details of our analysis of \evlacnospace.

\section{Data} \label{sec:motivate}

To motivate the development of HMMs as a modelling tool for non-periodic stochastic variability, we focus on stellar flares in particular, as those datasets often provide a clean look at a quiescent level punctuated by large, short-duration flares. Being able to separate quiescent from flaring states is crucial to understand mechanisms of stellar coronal heating, as well as the local interplanetary environment. The latter in particular affects the habitability of exoplanets, which has been flagged as an important focus of investigations in the Astro 2020 Decadal Survey \citep{2021pdaa.book.....N}.

\subsection{EV Lac}

The nearby ($5$~pc) active dMe binary \evlacnospace\ is a good candidate to test our HMM modeling. It has displayed consistent flaring across decades (at $\gtrsim{0.2-0.4}$~hr$^{-1}$ during every X-ray observation; see \cite{huenemoerder2010x} and references therein), and there are high-spectral and high-temporal resolution, long-duration datasets obtained
using the high-energy transmission gratings \citep[HETG;][]{2005PASP..117.1144C} on the \chandra\ X-ray Observatory \citep{2002PASP..114....1W}.\footnote{The datasets, obtained on 19-Sep-2001 (ObsID~01885; 100.02~ks) and 13-Mar-2009 (ObsID~10679; 95.56~ks) are available via the \chandra Data Collection (CDC) 235 at \url{https://doi.org/10.25574/cdc.235}.} This data was previously analyzed by \cite{huenemoerder2010x}, who detected 25 large individual flares across the datasets, and observed clear changes in spectral characteristics during flares, with generally higher plasma temperatures ($\gg10^6$~K) at larger emission measures; they explicitly demonstrate the value of stacking the data from flares (whether short or long) and the quiescent durations.

\begin{figure}
    \centering
        \includegraphics[width=\columnwidth]{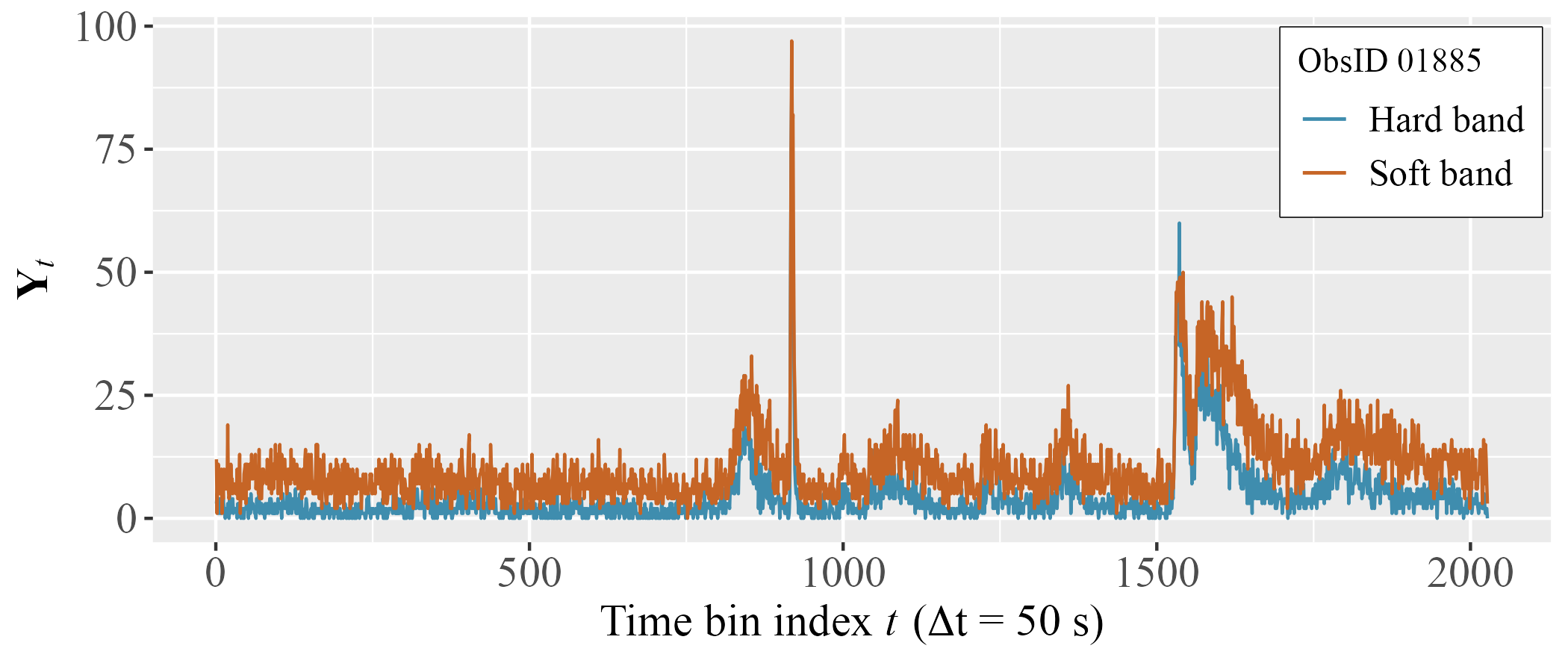}
        \includegraphics[width=\columnwidth]{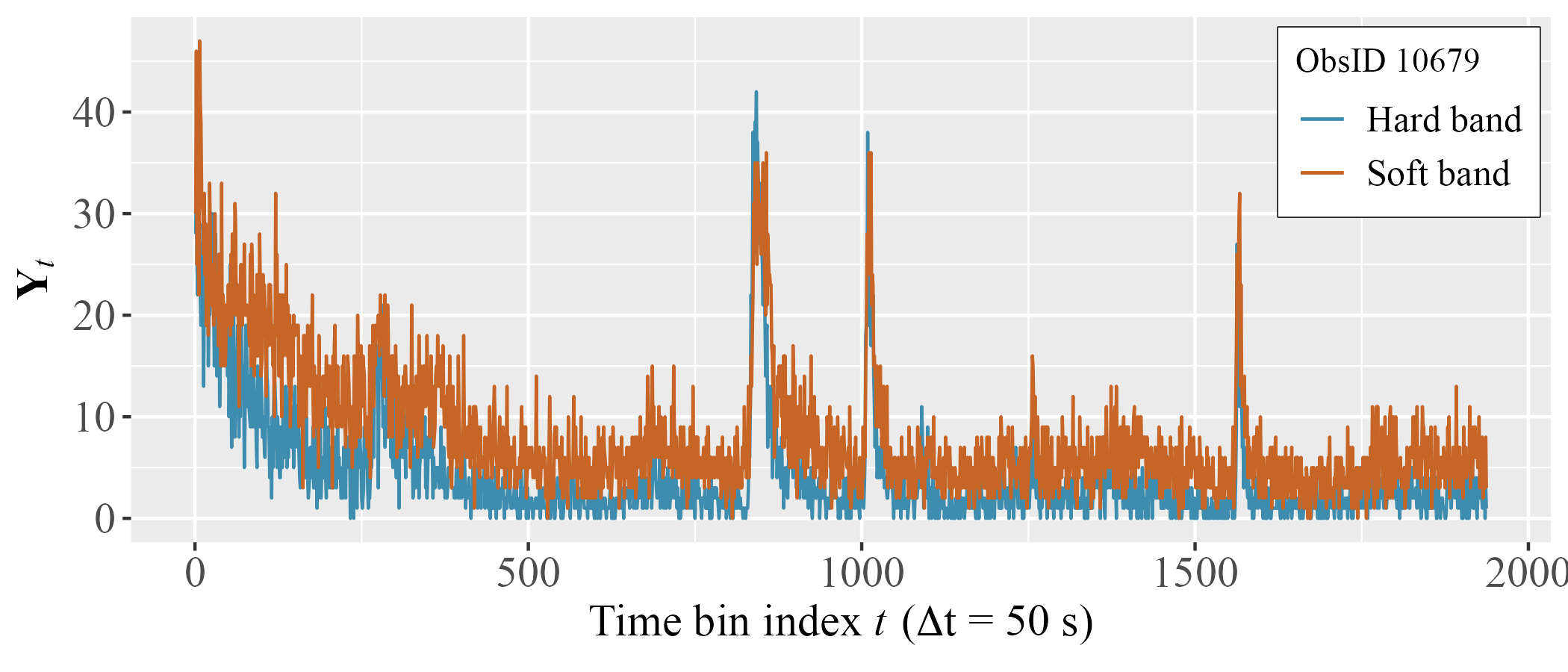}
    \caption{Bivariate time series plots of \evlac count data based on event lists where the split is based on counts in soft (0.3-1.5~keV) and hard (1.5-8~keV) passbands. Time is discretized into $50$~sec intervals; for ObsID 01855 (above), $t=0$ corresponds to 2001-09-19 19:36:23, and for ObsID 10679 (below), $t=0$ corresponds to 2009-03-13 06:47:57. The intermittent nature of \evlacnospace's flaring behavior is evident.}
    \label{fig:EVlacTS1}
\end{figure}

Here, we use the combined dispersed events from both the high-energy (HEG) and medium-energy (MEG) grating components of the first-order photons, extracted from the level-2 event list using the default extraction regions in CIAO v4.16 \citep{2006SPIE.6270E..1VF}. This allows us to avoid pileup effects \citep{2001ApJ...562..575D} on the zeroth-order data, especially during strong flares. We show the light curves for both epochs in Figure~\ref{fig:EVlacTS1}, with the data split into two passbands, a softer band covering $0.3$--$1.5$~keV and a harder band covering $1.5$--$8.0$~keV. The choice of $1.5$~keV as the split threshold is driven by the effective area peaking at that value.\footnote{We have also explored the sensitivity of our analysis to the choice of passband splitting energy value. We carried out the analysis using other astrophysically meaningful splits such $0.9$~keV -- which separates a thermal spectrum from being dominated by low-temperature and high-temperature plasma -- and found no qualitative effect on the results.} There are approximately $23{,}600$ and $17{,}900$ counts in the softer band, and approximately $9{,}800$ and $9{,}500$ counts in the harder band for ObsIDs~01885 and 10679, respectively. The counts are collected into light curves (Figure~\ref{fig:EVlacTS1}) binned at 50~s (see Appendix~\ref{app:othertimebins} for a sensitivity analysis for the choice of bins). Because these light curves are constructed from dispersed photons, pileup is entirely ignorable. The data are not affected by dead time effects, and background contamination is small and unvarying, and therefore also ignorable. The ACIS-S contamination build up at low energies over the mission \citep{2022SPIE12181E..6XP} reduces the counts in the soft band. 

We discuss the application of our model to this dataset and the relevant results in Section~\ref{sec:analysis}.

\begin{table*}
    \centering
    \begin{tabular}{cl}
        \hline \hline
        $w$ & Time bin width for grouping observations into discrete counts; usually $w = 50$ s \\
        $t$ & Index of time bin \\
        $Y_{t,1}$ & Observed soft band count at time index $t$ \\
        $Y_{t,2}$ & Observed hard band count at time index $t$ \\
        $\bY_t$ & Observed bivariate vector of counts (soft and hard band) at time index $t$ (i.e., $\bY_t = (Y_{t,1}, Y_{t,2})$) \\
        $\bY_{s:s'}$ & Collection of observed $\bY_t$ from $t=s$ to $t=s'$ \\
        $X_t$ & State of underlying Markov chain at time index $t$\\
        $\bX_{s:s'}$ & Collection of underlying states from $t=s$ to $t=s'$ \\
        $\sX$ & Underlying state-space which each $X_t$ takes values within \\ 
        $\bdelta$ & Initial distribution for a discrete-space Markov chain, represented as a vector \\
        $\tilde{\bdelta}$ & Discrete approximation to an initial distribution to a continuous-space Markov chain \\
        $\gamma_{i,j}$ & Transition probability from state $i$ to state $j$ for a discrete-space Markov chain \\
        $\bGamma$ & Transition matrix for a discrete-space Markov chain \\
        $\tilde{\bGamma}$ & Discrete approximation to a transition density of a continuous-space Markov chain \\
        $\lambda_{k,1}$ & Parameter for $k$'th state-dependent distribution of soft band count \\
        $\lambda_{k,2}$ & Parameter for $k$'th state-dependent distribution of hard band count \\
        $\blambda_k$ & Parameter vector for $k$'th state-dependent distribution (i.e., $\blambda_k = (\lambda_{k,1}, \lambda_{k,2})$) \\
        $h_k(\cdot \mid \blambda_k)$ & State-dependent density or mass function of $\bY_t$ (i.e., conditional on $X_t = k$) \\
        $\bveta$ & Vector of all unknown parameters in a given model \\
        $L(\bveta \mid \by\oneT)$ & Likelihood function (as a function of $\bveta$) for a given model \\
        $\delta(\cdot)$ & Initial distribution for a continuous-space Markov chain, represented as a density function \\
        $\gamma(\cdot, \cdot)$ & Transition density for a continuous-space Markov chain \\
        $\tilde{L}(\bveta \mid \by\oneT)$ & Likelihood function (as a function of $\bveta$) for discrete approximation \\
        $\pi$ & Stationary distribution for a given Markov chain \\
        $\beta_h$ & Mean emission rate for band $h$ when $X_{t,h} = 0$, scaled by $1/w$ (for $h=1,2$)\\
        $\phi_h$ & Autocorrelation parameter for $X_{t,h}$ (for $h=1,2$) \\
        $\bPhi$ & Autocorrelation matrix with $(\phi_1$, $\phi_2)$ along the diagonal and off-diagonal entries equal to $0$\\
        $\varepsilon_{t,1}$ & Soft band error/innovation at time index $t$ given by $X_{t,1} - \phi_1 X_{t-1,1}$ \\
        $\varepsilon_{t,2}$ & Hard band error/innovation at time index $t$ given by $X_{t,2} - \phi_1 X_{t-1,2}$ \\
        $\beps_t$ & Bivariate error/innovation term at time index $t$ (i.e., $\beps_t = (\varepsilon_{t,1}, \varepsilon_{t,2})$) \\
        $\sigma^2_h$ & Variance of $\varepsilon_{t,h}$ (for $h=1,2$) \\
        $\rho$ & Correlation between $\varepsilon_{t,1}$ and $\varepsilon_{t,2}$ \\
        $\bzero$ & Vector of zeros of length 2 (i.e., $\bzero = (0,0)$) \\
        $\bSigma$ & Covariance matrix with $(\sigma_1^2, \sigma_2^2)$ along the diagonal and off-diagonal entries equal to $\sigma_1 \sigma_2 \rho$ \\
        $\sN(\mu, \sigma^2)$ & Univariate normal distribution with mean $\mu$ and variance $\sigma^2$ \\
        $\sN_2(\bzero, \bSigma)$ & Bivariate normal distribution with mean vector $\bzero$ and covariance matrix $\bSigma$ \\
        $\hat{Y}_{t,h}$ & Predicted mean (Poisson rate) of distribution of $Y_{t,h}$ (for $h=1,2$) \\
        $\hat{X}_{t,h}$ & Prediction of $X_{t,h}$ conditional on $\bY\oneT = \by\oneT$ (for $h=1,2$) \\
        $\R$ & Set of real numbers \\
        $\N_{\geq 1}$ & Set of positive integers \\
        $\mathbb{P}_{\bveta}(A)$ & Probability of an event $A$ given distributional parameter values $\bveta$ \\
        $\mathbb{E}_{\bveta}\mleft[X\mright]$ & Expectation of a random variable $X$ given distributional parameter values $\bveta$ \\
        $A_i$ & Sub-rectangle $i$ used to partition continuous state-space in discrete HMM approximation \\
        $\bc_i^*$ & Representative point within $A_i$ used to define states in discrete HMM approximation \\
        $\bone$ & Column vector of ones (i.e., $\bone = (1, 1, \ldots, 1)^\top$) \\
        $\alpha$ & Mixing parameter for first component of a 2-component finite mixture model \\
        $\alpha_j$ & Mixing parameter for $j$'th component of a $K$-component finite mixture model \\
        $\bpi$ & Vector of parameters $\bpi = (\pi_1, \ldots, \pi_K)$ in density used for semi-supervised classification \\
        \hline \hline
    \end{tabular}
    \caption{Table of notation used throughout the paper}
    \label{tab:notation}
\end{table*}

\section{Hidden Markov Models}
\label{sec:HMMs}

We begin with a brief review of discrete-time HMMs, in order to present the relevant theory and notation required to understand the models and methods developed in this paper. A readable, but more comprehensive, introduction to HMMs can be found in \cite{zucchini2017hidden} while \cite{CappeHMM} provides a more advanced treatment.

\subsection{Discrete-Time Hidden Markov Models}
\label{sec:discrete-time-HMM}

Heuristically, we employ discrete-time HMMs when we believe that there is an unobserved underlying process governing the distribution of an observed time series of data at each discrete observation time. For example, we might postulate that a stellar corona is in either a quiescent state or active state at any given time, and that the distribution of observed counts differs between these two states. The underlying state (quiescent or active) is unobserved but governs the distribution of the observed photon counts. Mathematically, the underlying process is modelled as a Markov chain: informally, a sequence of random variables, $X_1, X_2,\ldots$, for which the distribution of any $X_t$ depends on the history of the chain only through the value of $X_{t-1}$. The variables $X_1, X_2,\ldots$, determine the overall state of the process (e.g., whether the stellar corona is in a quiescent or active state); thus we refer to the $X_t$ as \emph{state variables} (or simply \emph{states}).
 
Inferences about the Markov chain, such as the determination of its values at any time (a process known as \emph{state decoding}) are performed using only the observed data. Domains in which HMMs commonly appear include meteorology (in which the daily occurrence of rainfall is generated by underlying ``wet'' and ``dry'' states of nature \citep{zucchini2017hidden}), animal movement ecology (in which an animal's behavioural states are inferred from telemetry data capturing its physical movements \citep{langrock2012flexible}) and finance (in which stock returns are influenced by the underlying state of the economy). In astronomy, \cite{stanislavsky2020prediction} modelled solar X-ray flux as being generated by underlying ``flaring'' and ``non-flaring'' states of the sun, as discussed in Section~\ref{sec:intro}.

More formally, the basic discrete-time HMM has two key components. The first component is an unobserved Markov chain, $X\oneT = (X_1, \ldots, X_T)$, where each $X_t$ takes values in a common \emph{state space} $\sX$ and the chain is subject to the Markov property,  
\begin{equation}
    {\bb{P}}(X_t \in A \mid X_{t-1} = x_{t-1},\ldots, X_t = x_1) = {\bb{P}}(X_t \in A \mid X_{t-1} = x_{t-1}), 
    \label{eq:markov}
\end{equation}
for all $A \subseteq \sX$ (for notational convenience, we start by assuming the $X_t$ are univariate). The second component is a sequence of observed data, $\bY\oneT = (\bY_1, \bY_2, \ldots, \bY_T)$, where each $\bY_t$ takes values in a common \emph{observation space} $\sY$. For \evlacnospace, we consider soft and hard passband counts within each time bin; thus each $\bY_t$ is bivariate (i.e., a 2-component vector),  $\sY = \R^2$, and we set $\bY_t$ in bold throughout the paper. The two components are subject to the following conditional independence rules: 
\begin{enumerate}
    \item $\bY_t$ and $\bY_s$ are conditionally independent given the underlying Markov chain $X\oneT$, for any $s \neq t$, and 
    \item the distribution of  $\bY_t$ depends on  $X\oneT$ only through the state, $X_t$, at time index $t$.   
\end{enumerate}
It follows that $\bY_t$ and $\bY_s$ are conditionally independent given $(X_t,X_s)$ for any $t\neq s$. This means that, conditional on the state of the Markov chain at time index $t$, the observation $\bY_t$ is independent of all other observations; see Figure~\ref{fig:basicHMM}. Note that (ii) implies that the distribution of each $\bY_t$ is fully characterized by the underlying state $X_t$; often, the distributions of the individual observations,  $\bY_t$,  all belong to the same parametric family (such as a normal distribution), and the state $X_t$ manifests itself in the particular parameters of the distribution of $\bY_t$ (such as the mean and variance, in the case of state-dependent normal distributions). In most cases, the state space $\sX$ is either finite or a continuum; we describe these cases separately.

\definecolor{lightblue}{RGB}{225, 232, 239}
\begin{figure*}
  \centering
  \scalebox{0.7}{
    
\begin{tikzpicture}[shorten >=1pt,node distance=3cm,auto]
  \tikzstyle{state}=[shape=circle,thick,draw, minimum size = 1.5cm, fill=lightblue]
  \tikzstyle{obs}=[shape=circle,thick,draw, minimum size = 1.5cm]

  \node (Cold) {$\cdots$};
  \node[state, right of=Cold] (Ctm2) {\Large $X_{t-2}$};
  \node[state, right of=Ctm2] (Ctm1) {\Large $X_{t-1}$};
  \node[state, right of=Ctm1] (Ct) {\Large $X_{t}$};
  \node[state, right of=Ct] (Ctp1) {\Large $X_{t+1}$};
  \node[state, right of=Ctp1] (Ctp2) {\Large $X_{t+2}$};
  \node[right of=Ctp2] (Cnew) {$\cdots$};

  \node[obs, below of=Ctm2] (Xtm2) {\Large $\bY_{t-2}$};
  \node[obs, below of=Ctm1] (Xtm1) {\Large $\bY_{t-1}$};
  \node[obs, below of=Ct] (Xt) {\Large $\bY_{t}$};
  \node[obs, below of=Ctp1] (Xtp1) {\Large $\bY_{t+1}$};
  \node[obs, below of=Ctp2] (Xtp2) {\Large $\bY_{t+2}$};

  \path[->, draw, thick, -latex] (Cold) -- (Ctm2);
  \path[->, draw, thick, -latex] (Ctm2) -- (Ctm1);
  \path[->, draw, thick, -latex] (Ctm1) -- (Ct);
  \path[->, draw, thick, -latex] (Ct) -- (Ctp1);
  \path[->, draw, thick, -latex] (Ctp1) -- (Ctp2);
  \path[->, draw, thick, -latex] (Ctp2) -- (Cnew);

  \path[->, draw, thick, -latex] (Ctm2) -- (Xtm2);
  \path[->, draw, thick, -latex] (Ctm1) -- (Xtm1);
  \path[->, draw, thick, -latex] (Ct) -- (Xt);
  \path[->, draw, thick, -latex] (Ctp1) -- (Xtp1);
  \path[->, draw, thick, -latex] (Ctp2) -- (Xtp2);
\end{tikzpicture}  } \caption{A graphical model representing the standard discretized HMM dependence structure. In this graph, open nodes represent observed quantities and shaded nodes represent unobserved quantities. Generally, an arrow from node $X$ to node $Y$ indicates that the random variables $X$ and $Y$ are not independent, and that the joint distribution of $(X,Y)$ is analyzed via the factorization $f_{X,Y}(x,y) = f_{Y \mid X}(y \mid x) \cdot f_X(x)$ rather than $f_{X,Y}(x,y) = f_{X \mid Y}(x \mid y) \cdot f_Y(y)$. In the unobserved Markov chain $X_1, X_2, \ldots $,  each $X_t$ determines the distribution of its successor $X_{t+1}$ (represented by the forward-pointing arrows). In the observed process $\mathbf{Y}_1, \mathbf{Y}_2, \ldots$ each $X_t$ determines the distribution of each $\mathbf{Y}_t$ (represented by the downward-pointing arrows), such that, conditional on these determinations, the $\mathbf{Y}_t$ are independent (represented by the lack of arrows between the $\mathbf{Y}_t$).} \label{fig:basicHMM}
\end{figure*}

\subsection{Discrete-Space Hidden Markov Models}
\label{sec:discrete:hmm}

When the state space $\sX$ is finite, it is commonly represented as $\sX = \{1,\ldots,K\}$ for some $K \in \N$, where each value in $\sX$ plays the role of a label for an underlying state of nature (for example, when $K=2$, ``flaring'' and ``quiescent'' can simply be represented as ``1'' and ``2'', respectively). In this case, the resulting HMM is referred to as a \emph{discrete-space HMM}. The specification of a (time-homogeneous) discrete-space HMM consists of three ingredients: 
\begin{enumerate}
    \item An \emph{initial distribution} on $\sX$, represented by a vector $\bdelta = (\delta_1, \ldots, \delta_K)$
    with $\delta_i = {{\bb P}}\left(X_1 = i\right)$,
    \item A set of \emph{transition probabilities}, $\gamma_{i,j} = {{\bb P}}\left(X_{t+1} = j \mid X_t = i\right)$ for any $t \geq 1$, represented by a $K \times K$ \emph{transition matrix}, $\bGamma$, with element $(i,j)$ given by $\gamma_{i,j}$, and 
    \item A set of \emph{state-dependent distributions}, each characterized by a density or mass function $h_k(\by \mid \blambda_k)$ determining the conditional distribution of $\bY_t \mid X_t = k$ for any $t$. Here $\blambda_k$ is a state-specific distributional parameter, which may consist of several components.
\end{enumerate}

Let $\bveta$ denote the set of HMM model parameters, including the initial distribution, the transition probabilities, and the parameters of the state-dependent distributions, i.e., $\bveta=(\bdelta, \bGamma, \blambda_1, \ldots, \blambda_K)$. The likelihood function for the discrete-space HMM is given by
\begin{multline}
\label{eq:DSHMMlike}
    L(\bveta \mid \by\oneT) =  \\
    \sum_{x_1=1}^K \cdots \sum_{x_T=1}^K \left(\delta_{x_1} \cdot h_{x_1}(\by_1 \mid \blambda_{x_1}) \prod_{t=2}^T \left(\gamma_{x_{t-1},x_t} \cdot h_{x_t}(\by_t \mid \blambda_{x_t})\right)\right).
\end{multline}
The sums in \eqref{eq:DSHMMlike} ``marginalize'' the unknown state sequence $X\oneT$ out of the likelihood by summing over all possible state sequences which could have generated the observed data.

Standard algorithms are available for computing the maximum likelihood estimate of $\bveta$ under \eqref{eq:DSHMMlike}. While the number of terms summed in \eqref{eq:DSHMMlike} is exponential in $T$, an efficient algorithm known as the \emph{forward algorithm} allows the likelihood to be computed in polynomial time; see Appendix~\ref{subsub:fwdalgo} for details. Embedding this algorithm within the E-step of the well-known EM algorithm (see Appendix~\ref{app:EMalgos}) produces the \emph{Baum-Welch algorithm}, which allows for fast maximization of \eqref{eq:DSHMMlike}; see \cite{zucchini2017hidden} for details. Once the model parameters have been estimated, the \emph{forward-backward algorithm} (detailed in Appendix~\ref{subsub:fwdbckwdalgo}) can be used to compute posterior state membership probabilities of the form $\hat{p}_{t,k} = {{\bb P}}\left({X_t = k \mid \bY\oneT = \by\oneT}\right)$ for each $t=1,\ldots,T$, and from these, the posterior state membership classifications given by $\mathrm{argmax}_{k} \, \hat{p}_{t,k}$.

\subsection{Continuous-Space Hidden Markov Models}

When the state space $\sX$ is a continuum (such as $\R$ or, more generally, $\R^d$ for some $d \geq 1$), the resulting HMM is called a \emph{continuous-space HMM}. In this case, the first two ingredients in the discrete-space HMM specification are replaced by continuous analogues, while the third is essentially unchanged:
\begin{enumerate}
    \item An \emph{initial distribution} on $\sX$, represented by a probability density function $\delta(x)$ satisfying ${\bb{P}}(\bX_1 \in A) = \int_A \delta(x) \dif x$ for $A \subseteq \sX$, 
    \item A \emph{transition density function}, $\gamma:\sX^2 \to [0,\infty)$ satisfying ${\bb{P}}(\bX_{t+1} \in A \mid \bX_t = \bx) = \int_A \gamma(\bx, \bx') \dif \bx'$ for any $t \geq 1$ and $\bx' \in \sX$, and
    \item A set of \emph{state-dependent distributions}, each characterized by a density or mass function $h_\bx(\by \mid \blambda_\bx)$ determining the conditional distribution of $\bY_t \mid \bX_t = \bx$ for any $t$. Here $\blambda_\bx$ is the parameter specifying the distribution of $\bY_t$ given that $\bX_t=\bx$; this parameter may consist of several components.
\end{enumerate}

The likelihood function for the continuous-space HMM is 
\begin{multline}
\label{CSHMMlike}
L(\bveta \mid \by\oneT) =  \\
\int_{\sX} \cdots \int_{\sX}  \delta(\bx_1) \cdot h_{\bx_1}(\by_1 \mid \blambda_{\bx_1}) \prod_{t=2}^T \gamma(\bx_{t-1}, \bx_t) \cdot h_{\bx_t}(\by_t \mid \blambda_{\bx_t}) \dif \bx_{T:1},
\end{multline}
where the iterated integrals over $\sX$ have replaced the sums in \eqref{eq:DSHMMlike} and $\dif \bx_{T:1} = \dif \bx_T \cdots \dif \bx_1$.

In both discrete-space and continuous-space Markov chains, the corresponding transition probabilities or transition density may induce a \emph{stationary distribution} for the underlying Markov chain -- a distribution $\pi$ where $X_t \sim \pi$ implies that  $X_{t+1} \sim \pi$ (i.e., if one iterate of the chain is marginally distributed according the stationary distribution, all subsequent iterates are also marginally distributed according to $\pi$). Under broadly realistic assumptions, the stationary distribution is equal to the asymptotic distribution of the chain, i.e., the limiting distribution of $X_t$ as $t \to \infty$ \citep[e.g.,][]{resnick2013adventures}.

\subsection{Approximation to the Continuous-Space HMM Likelihood}\label{sub:approximation}

In contrast to the situation for the discrete-space HMM, computing the maximum likelihood estimate under a continuous-space HMM by maximizing \eqref{CSHMMlike} poses considerable challenges. With the sums over $\{1,\ldots,K\}$ replaced by integrals over $\sX$, no efficient algorithms are known that can compute \eqref{CSHMMlike}, let alone maximize it. Fortunately, however, we can  \emph{approximate} \eqref{CSHMMlike} to arbitrary high level of accuracy by replacing the continuous-space Markov chain with a suitably chosen discrete-space one; this idea originates from the work of \cite{kitagawa1987non} and was developed for state-space models by \cite{langrock2011some}. We provide a brief outline of the method and its derivation here, with additional details in Appendix~\ref{app:approximation}; see also \cite{langrock2011some} for a complete exposition in the univariate case and \cite{langrock2012some} for several illustrative examples. 

First, we must identify an \emph{essential domain} $A$, which is a bounded subset of $\sX$ such that ${\bb{P}}(\bX_t  \in A)$ is nearly one for all $t$ \citep{kitagawa1987non}. Next, $A$ must be partitioned into subsets $A_1, \ldots, A_m$ and a representative point, ${\bc}^*_i$, chosen for each $A_i$, e.g., ${\bc}^*_i$ can be set to the center of $A_i$. If all of the $A_i$ are small, then 
\begin{multline}
\label{eq:CSHMMlikeapp1}
    L(\bveta \mid \by\oneT) \approx \sum_{i_1=1}^m \cdots \sum_{i_T=1}^m \left( {\bb{P}}(\bX_1 \in A_{i_1}) \cdot h_{\bc^*_{i_1}}(\by_1 \mid \blambda_{\bc^*_{i_1}}) \cdot \vphantom{\prod_{t=2}^T}\right. \\  
    \left.\prod_{t=2}^T \left( {\bb{P}}(\bX_t \in A_{i_t} \mid \bX_{t-1} = \bc^*_{i_{t-1}}) \cdot h_{\bc^*_{i_t}}(\by_t \mid \blambda_{\bc^*_{i_t}})\right)\right),
\end{multline}
where the approximation becomes exact as $A$ approaches $\cal X$ and each of the $A_i$ decrease in size.  (See Appendix~\ref{app:approximation} for details.) Defining the vector $\tbdelta \in \R^m$ and matrix $\tbGamma \in \R^{m \times m}$ by the entries
\begin{equation}\label{eq:normalizers}
    \tdelta_j = {\bb{P}}(\bX_1 \in A_j) \quad \text{and} \quad \tgamma_{i,j} = {\bb{P}}(\bX_t \in A_j \mid \bX_{t-1} = \bc^*_i),
\end{equation}
the approximation \eqref{eq:CSHMMlikeapp1} can be written
\begin{multline}\label{eq:CSHMMlikeapp}
    \tL(\bveta \mid \by\oneT) \approx \\
    \sum_{i_1=1}^m \cdots \sum_{i_T=1}^m \left( \tdelta_{i_1} \cdot h_{\bc^*_{i_1}}(\by_1 \mid \blambda_{\bc^*_{i_1}}) \prod_{t=2}^T \left( \tgamma_{i_{t-1},i_t} \cdot h_{\bc^*_{i_t}}(\by_t \mid \blambda_{\bc^*_{i_t}})\right)\right),
\end{multline}
where $\bveta$ is a vector consisting of the unknown parameters in the state-space model, including the state-dependent parameters $\blambda_{\bc^*_{i,1}}, \ldots, \blambda_{\bc^*_{i,T}}$ and any parameters associated with the distribution of the underlying Markov chain $\bX\oneT$. If we replace the initial density $\delta$ and transition density $\gamma$ with the discretized functions in \eqref{eq:normalizers}, the approximation in \eqref{eq:CSHMMlikeapp} is precisely of the form of the discrete-space HMM likelihood given in  \eqref{eq:DSHMMlike}, and so, up to the renormalization of $\tbdelta$ and the rows of $\tbGamma$, \eqref{eq:CSHMMlikeapp} is the likelihood of an $m$-state discrete-space HMM in which the chain being in ``state'' $i$ at time index $t$ corresponds to the event that $\bX_t \in A_i$. 

With all elements in the approximation specified in this way, evaluation of \eqref{eq:CSHMMlikeapp} can proceed using the forward algorithm discussed in Section~\ref{sec:discrete:hmm}. When $\sX = \R^d$ for $d > 1$ and the size of the partition $m$ is large, mapping the unordered partition of $A$ to an ordered set of states $\{1,\ldots,m\}$ poses its own challenges. When $d = 2$, this mapping can be accomplished by a \emph{pairing function} -- that is, a bijection from $\N_{\geq 1} \times \N_{\geq 1}$ to $\N_{\geq 1}$. We slightly modify Szudzik's ``Elegant'' bijection between $\mathbb{N} \times \mathbb{N}$ and $\mathbb{N}$ \citep{szudzik2006elegant} so that the original function and its inverse have the required domain and range. The modification and its inverse are respectively given by
\[
    \text{pair}(i,j) = \begin{cases}
        j^2 - 2j + i + 1, & i \neq \max\{i,j\}\\
        i^2 + j - i, & i = \max\{i,j\}\\
    \end{cases}
\]
and
\[
    \text{unpair}(j) = \begin{cases}
        \left(j - g(j)^2, g(j) + 1\right), & j- g(j)^2 - 1 < g(j)\\
        \left( g(j) + 1, j -g(j)^2 - g(j)\right), & j- g(j)^2 - 1 \geq g(j)\\
    \end{cases},
\]
where $g(j) = \lfloor\sqrt{j-1}\rfloor$.

In practice, one can manually verify that the range of the chosen essential domain is sufficient for the data at hand by inspecting a histogram of the predicted states produced by any state decoding algorithm (see Appendix~\ref{subsub:fwdbckwdalgo}) after the model has been fit \citep{zucchini2017hidden}.

\section{Stage 1: HMM\MakeLowercase{s} for Flaring Sources}
\label{sec:model}

In this section we propose three new HMMs which are well suited to model flares in stellar coronae. These models are more generally applicable, but because we focus on datasets of flaring stellar light curves (see Section~\ref{sec:motivate}), and because other model choices are possible, we caution that it is necessary to consider carefully the particular scenario before adopting these models without suitable modifications. Indeed, we are actively engaged in applying the models to flaring sources other than stars and exploring what generalizations to the models might be appropriate for these application; see Section~\ref{sec:discussion-improve} for discussion. All of the models consider photon counts recorded in a sequence of time intervals indexed by $t$ and tabulated into soft passband counts, $Y_{t,1}$, and hard passband counts, $Y_{t,2}$, for $t=1,\ldots, T$. We start by considering the relative merits of discrete and continuous state spaces as the basis for modelling the flaring behaviour of stars.

\subsection{Discrete-Space HMMs for Flaring Stellar Coronae}
\label{sub:discretespaces}

With a discrete state space, a state-dependent bivariate Poisson distribution can be written
\begin{equation}\label{eq:2statePoisHMM}
\bY_t \mid X_t = k \sim \text{Poisson}(\lambda_{k,1}) \cdot \text{Poisson}(\lambda_{k,2})
\end{equation}
for $t=1,\ldots,T$ and $k=1,\ldots,K$, where here and below the notation $\bY_t \mid X_t = k \sim \text{Poisson}(\lambda_{k,1}) \cdot \text{Poisson}(\lambda_{k,2})$ indicates that the Poisson distributions of the passbands $Y_{t,1}$ and $Y_{t,2}$, conditional on the event $X_t = k$, are independent for all $t$. There are many possible alternatives to \eqref{eq:2statePoisHMM} for count data, including combinations of various bivariate Poisson and negative binomial distributions (see \cite{johnson1997discrete} for examples) and state-dependent copulas \citep[see][]{zimmerman2023copula}, all of which induce dependence structures between $Y_{t,1}$ and $Y_{t,2}$. In principle, a two-state HMM could be used to model a star's states as ``quiescent'' and ``flaring'', roughly in the manner of \cite{stanislavsky2020prediction}. Alternatively, a three-state HMM might split the ``flaring'' state into states of rising and falling flaring activity \citep{esquivel2024detecting}.

\begin{figure}
    \centering
    \includegraphics[width=\columnwidth]{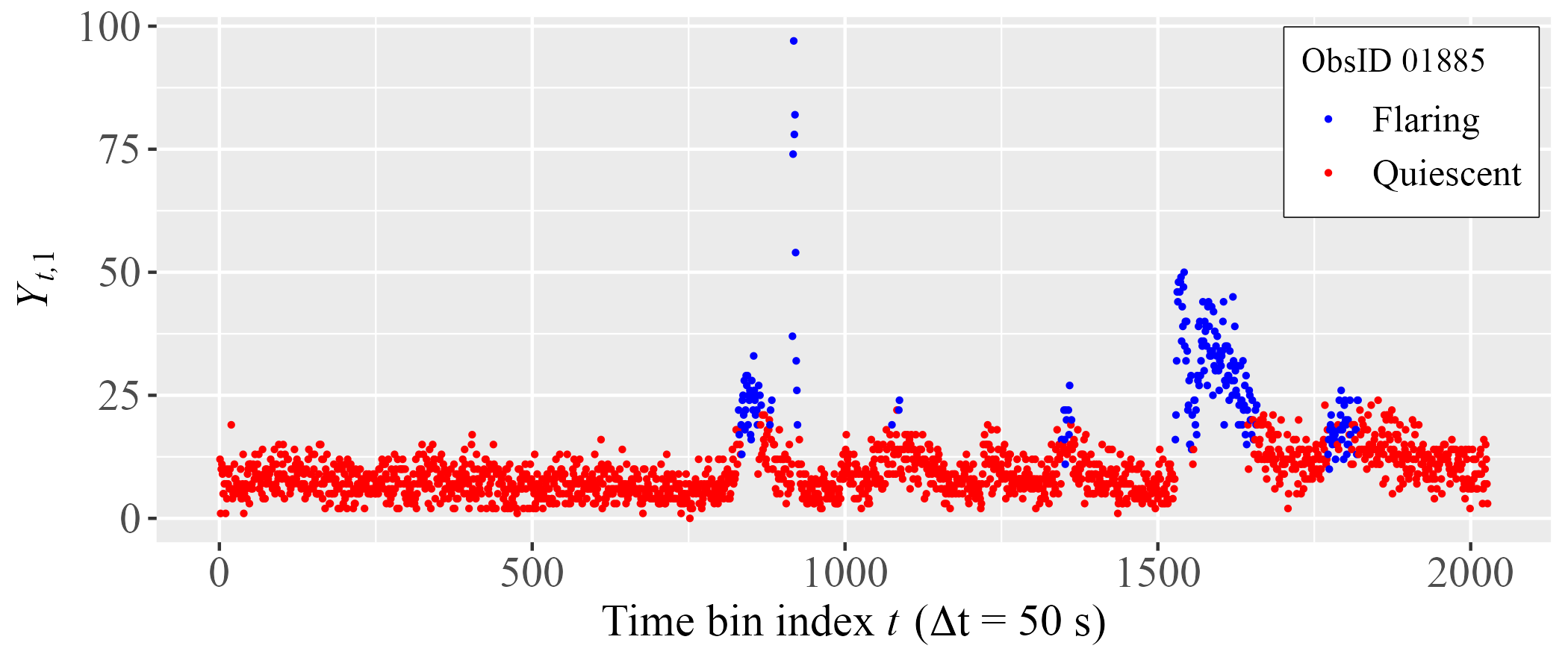}
    \includegraphics[width=\columnwidth]{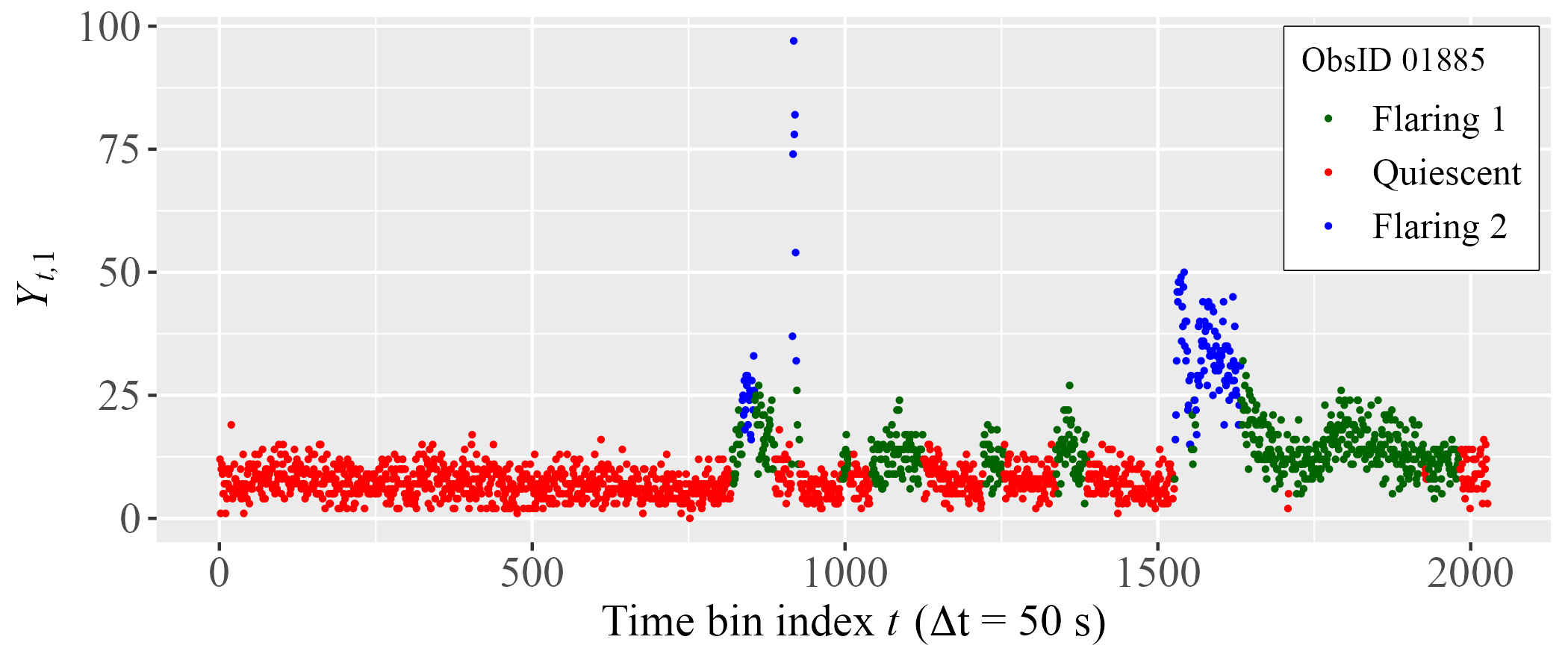}
    \caption{Soft band ObsID 01885 light curve coloured with classifications based on 2-state (above) and 3-state (below) HMMs fit directly to the observed data $\bY\oneT$}
    \label{fig:EVlac_2HMM}
\end{figure}

We fit the model specified in \eqref{eq:2statePoisHMM} to ObsID~01885 light curve for both $K=2$ and $K=3$ via maximum likelihood as described in Section~\ref{sec:discrete:hmm}. Figure~\ref{fig:EVlac_2HMM} illustrates the fitted predicted classifications for each time interval, computed as $\mathrm{argmax}_{k} \, \hat{p}_{t,k}$, again as described in Section~\ref{sec:discrete:hmm}. Inspection of Figure~\ref{fig:EVlac_2HMM} (or indeed of Figure~\ref{fig:EVlacTS1}) reveals a theoretical defect of using a discrete-space HMM to model the stellar flare process of \evlacnospace. Under the conditional independence rules of Section~\ref{sec:discrete-time-HMM}, all observations generated by the same state are independent and identically distributed. Indeed, this implies that the red observations in Figure~\ref{fig:EVlac_2HMM} must be independent and identically distributed, as are the green and blue ones. This implication is contradicted by the clear temporal trend of the red observations, as well as the sharp rise and fall of the blue ones. Thus, the conditional independence rule is not satisfied and the standard discrete-space HMM is not directly suitable for our data. 

This time series is comprised of jumps between two clearly distinguished levels, pushed by a gradual trend over time \cite[see Figures~1 and 2 of][]{stanislavsky2020prediction}.

\subsection{Continuous-Space HMMs for Flaring Stellar Coronae}
\label{sec:cont-HMM}

There is no reason to assume that the underlying physical process generating stellar flare activity is binary and is either ``on'' or ``off''. Here we consider a more realistic model that allows the expected photon count at time index $t$ to depend on a \emph{continuous} underlying process. This enables us to model gradual and/or smooth transitions between a quiescent and an active corona (e.g., with long periods of quiescence interrupted by more intense signals at random intervals). We also weaken the assumption that a single underlying univariate process $X_t$ drives both the hard and soft band photon counts. Specifically, we replace $X_t$ with a bivariate vector $\bX_t$ whose components $X_{t,1}$ and $X_{t,2}$ may be correlated with each other. We maintain the Markov assumption expressed as a bivariate version of (\ref{eq:markov}).

We specify a \emph{Poisson state-space model}\footnote{The term ``state-space model'' -- unlike ``hidden Markov model'' -- is not consistently defined in the literature. Here, we simply regard state-space models as those with observation processes (partially) driven by some hidden linear state process defined on a continuous state space. In other domains such as control theory, this term commonly refers to more specific models in which the observation process is itself a linear function of the state process.} that satisfies these requirements. First, the state-dependent distribution models flux measurement (i.e., the counts in two passbands) via a Poisson (error) distribution conditional on the underlying $\R^2$-valued state-space variable $\bX_t$:
\begin{equation}
\mathbf{Y}_t \mid \mathbf{X}_t \sim \text{Poisson}(w \cdot \beta_1 \cdot e^{X_{t,1}}) \cdot \text{Poisson}(w \cdot \beta_2 \cdot e^{X_{t,2}}).
\label{eq:PoissonSSM1}    
\end{equation}
Second, the astrophysical source variability or signal is modelled via an autoregressive process for $\bX_t$, specified as
\begin{subequations}
\begin{align}
\mathbf{X}_t &= \bPhi \mathbf{X}_{t-1} + \boldsymbol{\varepsilon}_t,\label{eq:PoissonSSM2}\\
\bPhi &= \begin{bmatrix} \phi_1 & 0 \\ 0 & \phi_2 \\ \end{bmatrix}, \label{eq:PoissonSSM3}\\
\boldsymbol{\varepsilon}_t &\stackrel{\rm iid}{\sim} \mathcal{N}_2(\mathbf{0},\boldsymbol{\Sigma}),\ \hbox{and} \label{eq:PoissonSSM4}\\
\boldsymbol{\Sigma} &= \begin{bmatrix} \sigma_1^2 & \sigma_1 \sigma_2 \rho\\ \sigma_1 \sigma_2 \rho & \sigma_2^2 \\ \end{bmatrix}. \label{eq:PoissonSSM5}
\end{align}
\end{subequations}
The $\boldsymbol{\varepsilon}_t$ term in \eqref{eq:PoissonSSM2} does not represent observational noise; rather, it represents the random innovation in the underlying source or signal variability. Observational uncertainty, on the other hand, is captured implicitly by the Poisson distribution in \eqref{eq:PoissonSSM1}, and not by any explicit additive term in the model. Fitting this Poisson state-space model allows us to go beyond simply fitting the raw light curves. Ultimately, this will allow us to identify time intervals with different statistical behaviours (e.g., quiescence and flaring); see Section~\ref{sec:classify}. Note that the notation $\mathcal{N}_2(\mathbf{0},\boldsymbol{\Sigma})$ in \eqref{eq:PoissonSSM4} represents a bivariate Gaussian distribution with mean vector equal to $\mathbf{0}$ and covariance matrix equal to $\bSigma$.

The parameters to be estimated in \eqref{eq:PoissonSSM1}--\eqref{eq:PoissonSSM5} are $\beta_1, \beta_2 >0$, the coefficient matrix $\bPhi$ with diagonal entries $\phi_1, \phi_2 \in (-1,1)$, and the covariance matrix $\bSigma$ built up of components $\sigma_1, \sigma_2 > 0$ and $\rho \in (-1,1)$. The remaining term, $w$, is the time bin used to group the original photon event list into discrete counts; including $w$ facilitates the study of dependence on bin size (see Appendix~\ref{app:othertimebins}) and also helps to avoid numerical underflow in the estimation process. 

Under the model in \eqref{eq:PoissonSSM1}--\eqref{eq:PoissonSSM5}, the expected photon counts $\E{Y_{t,1}}$ and $\E{Y_{t,2}}$ at time index $t$ in the soft and hard bands are monotone increasing functions of $X_{t,1}$ and $X_{t,2}$, respectively. The parameter $\beta_h$ is proportional to the expected Poisson photon count when $X_{t,h} = 0$. (Since $X_{t,h}$ can take on negative values, $X_{t,h} = 0$ does not necessarily correspond to a state of particularly low or high flaring activity.) The coefficient matrix $\bPhi$ determines the extent to which each $X_{t,h}$ is correlated with its predecessor, $X_{t-1,h}$. A slight generalization of \eqref{eq:PoissonSSM3} allows the off-diagonal entries of $\bPhi$ to be nonzero, thereby allowing $X_{t,1}$ to depend on $X_{t-1,2}$ and vice versa (see Section~\ref{sec:discussion}).

The state process $\bX_t$ of the model described by \eqref{eq:PoissonSSM2}--\eqref{eq:PoissonSSM5} is a \emph{first-order vector autoregressive process}, denoted as a VAR(1) process in the statistical literature. VAR models are commonly applied in areas such as mathematical finance, where they play important roles in stochastic volatility modelling \citep[e.g.,][]{primiceri2005time}.

To compute the (approximate) maximum likelihood estimate under the model in \eqref{eq:PoissonSSM1}--\eqref{eq:PoissonSSM5}, we maximize the discrete-state-space approximation to the likelihood; see  \eqref{eq:CSHMMlikeapp}. Because the state space is $\R^2$, it is convenient -- although not strictly necessary -- to choose the essential domain $A$ to be a rectangle. Similarly, we partition $A$ into a large number of sub-rectangles, $A_1, \ldots, A_m$, and set the representative point, ${\bc}^*_i$, of each to be its center.

To numerically optimize (\ref{eq:CSHMMlikeapp}), we use a parallelized version of the popular L-BFGS routine as implemented in the \texttt{optimParallel} package \citep{optimPar2019} within \texttt{R}. We prefer to use unconstrained optimization to avoid numerical issues caused by parameter inputs lying on the boundaries of their respective domains; thus instead of optimizing the parameters $\phi_1$, $\phi_2$ and $\rho$ in the approximate likelihood over $(-1,1)$, we optimize $\tanh^{-1}(\phi_1)$, $\tanh^{-1}(\phi_2)$ and $\tanh^{-1}(\rho)$ over $\R$, and then transform the optimizing values back to their natural domain via the inverse function $x \mapsto \tanh(x)$. Similarly, we optimize $\log{\beta_1}$, $\log{\beta_2}$, $\log{\sigma_1}$ and $\log{\sigma_2}$ over $\R$, and replace the results with their exponentiated values.

The (approximate) maximum likelihood estimates may be slightly biased due to small sample sizes. (maximum likelihood estimates are \emph{asymptotically} unbiased for most ``smooth'' models, but are generally not unbiased with finite samples.) Similarly, with a small sample size the negative Hessian matrix of the log-likelihood function evaluated at the maximum likelihood estimate may yield an inadequate approximation to the Fisher information matrix, which is normally used to produce confidence intervals. In order to remedy both issues, we appeal to the parametric bootstrap, which allows us to estimate simultaneously the standard errors of parameter estimates and their biases \citep{tibshirani1993introduction}. 

Specifically, after computing the maximum likelihood estimate of the parameters, $\hat{\bveta}_\text{mle}$,
using the actual data, $\bY\oneT$, we independently generate $B$ replicate datasets, $\bY_{1:T}^{(1)}, \ldots, \bY_{1:T}^{(B)}$, under the model, each with parameter fixed at $\hat{\bveta}_\text{mle}$. In the context of an HMM, this requires first simulating the underlying state sequences, $\bX_{1:T}^{(1)}, \ldots, \bX_{1;T}^{(B)}$, and then generating each $\bY_t^{(b)} \mid \bX_t^{(b)}$ according to the conditional distribution \eqref{eq:PoissonSSM1}. For each $b=1,\ldots,B$, we then re-fit the model using $\bY_{1:T}^{(b)}$ to produce a replicate estimate, $\hat{\bveta}^{(b)}_\text{bs}$. Next, we estimate the bias $\boldsymbol{b}$ and covariance matrix $\bC$ of the maximum likelihood estimator via 
\begin{equation}
    \hat{\boldsymbol{b}}_\text{bs} = \bar{\hat{\bveta}}_\text{bs} - \hat{\bveta}_\text{mle}
\end{equation}
and
\begin{equation}
    \hat{\bC}_\text{bs} = \frac{1}{B-1} \sum_{b=1}^B (\hat{\bveta}^{(b)}_\text{bs} - \bar{\hat{\bveta}}_\text{bs}) (\hat{\bveta}^{(b)}_\text{bs} - \bar{\hat{\bveta}}_\text{bs})^\top,
\end{equation}
where
\begin{equation}
   \bar{\hat{\bveta}}_\text{bs} = \frac{1}{B}\sum_{b=1}^B \hat{\bveta}^{(b)}_\text{bs}
\end{equation}
is the mean of the bootstrap replicate estimates. Finally, the bootstrap-corrected estimate is  $\hat{\bveta}_\text{corr} = \hat{\bveta}_\text{mle} - \hat{\boldsymbol{b}}_\text{bs}$, and standard errors for the components of $\hat{\bveta}_\text{corr}$ are equal to the square roots of the diagonal elements of $\hat{\bC}_\text{bs}$. Approximate 95\% confidence intervals for the components are computed with these standard errors.

We conducted a brief simulation study to confirm the veracity of the bootstrap estimates and errors. We simulated data under Model~2 (see Section~\ref{subsub:model2} below) with a pre-specified $\bveta = (\phi_1, \sigma_1, \sigma_2, \beta_1, \beta_2)$ chosen relatively close to the values given in the second column of Table~\ref{tab:M2_w50main}. Choosing $B = 100$, we independently repeated the bootstrapping procedure 100 times, producing 100 bootstrap 95\% confidence intervals centered around 100 bias-corrected maximum likelihood estimates. The coverage probabilities for the five parameters (i.e., the number of times each true parameter $\phi_1,\ldots,\beta_2$ fell inside the bootstrap confidence intervals, divided by 100) were $0.93$, $0.97$, $0.92$, $0.92$, and $0.91$, respectively, which all agree with the expected value of $0.95$ at the $95\%$ confidence level.

\subsection{Three State-Space Models for Flaring Stellar Coronae}\label{sub:nestedHMMs}

While the Poisson state-space model in \eqref{eq:PoissonSSM1}--\eqref{eq:PoissonSSM5} includes features 
well-suited to stellar flare data, it may be more general than necessary; for example, it is not immediately clear that separate underlying processes, $X_{t,1}$ and $X_{t,2}$, are necessary for the hard and soft bands. We therefore consider two special cases of the model, the first itself a special case of the second, before considering the model in \eqref{eq:PoissonSSM1}--\eqref{eq:PoissonSSM5} in its full generality as a third model. Thus, the three models we consider form a nested sequence. For each model, we first provide a stochastic representation, and then give the initial distribution (as characterized by $\tilde{\delta}_j$, for $j \in \{1,\ldots,m\}$) and transition probabilities (as characterized by $\tilde{\gamma}_{i,j}$, for $i,j \in \{1,\ldots,m\}$) of the associated discrete-space HMM approximation to the continuous-space model. This involves expressing both the $\tilde{\delta}_j$ and the $\tilde{\gamma}_{i,j}$ as functions of the parameters involved with the stochastic representation of the underlying state process.

Note that the initial density plays a relatively minor role in the likelihood, and that its impact diminishing as $T$ grows. We follow \cite{langrock2011some} and use the stationary distribution of the state process, $\bX\oneT$, for the initial distribution $\tilde{\delta}_j$. Statistically, this is tantamount to assuming that the distribution of the states that the star inhabits is in equilibrium, and is not evolving over time.\footnote{This is a reasonable choice for a steadily flaring star like \evlacnospace, which has not shown evidence of drastic changes in X-ray luminosity during observations over the past several decades \citep{huenemoerder2010x}. This choice is also supported by the steadiness of the spectra in the quiescent and flaring states that we find \emph{post facto} across epochs (see Section~\ref{subsub:EVLacstates}).} The transition probabilities $\tilde{\gamma}_{i,j}$ are derived from the stochastic representation of the model.

\subsubsection{Model 1: AR(1) Process}

To reduce the underlying state process to one dimension, we set  $X_{t,1} = X_{t,2} =: X_t$ for all $t$, in which case the latent process reduces to a univariate \emph{first-order autoregressive process}, denoted as an AR(1) process for short. The entire state-space model can be written in the simplified form
\begin{align}
\begin{split}
\mathbf{Y}_t \mid X_t &\sim \text{Poisson}(w \cdot \beta_1 \cdot e^{X_t}) \cdot \text{Poisson}(w \cdot \beta_2 \cdot e^{X_t}),\\
X_t &= \phi X_{t-1} + \varepsilon_t,
\ \hbox{and}
\label{eq:modelone}\\
\varepsilon_t &\stackrel{\rm iid}{\sim} \mathcal{N}(0,\sigma^2).
\end{split}
\end{align}
The vector of unknown parameters for Model~1, $\bveta_\text{M1} = (\phi, \sigma, \beta_1, \beta_2)$, is fit to the data.

Under Model 1, $X_t = \phi X_{t-1} + \varepsilon_t$ with $\varepsilon_t \sim \mathcal{N}(0,\sigma^2)$, and it can be shown that this process admits a stationary distribution if and only if $\phi \in (-1,1)$, whence the stationary distribution is given by the $\sN\mleft(0, \sigma^2/(1-\phi^2) \mright)$ distribution. Thus, if $A_j = [a_j, b_j]$, then the initial distribution for the discrete-space HMM approximation of Model 1 is taken to be the vector $\tilde{\bdelta}$ comprised of entries
\begin{equation}
    \tilde{\delta}_j = \mathbb{P}(X_t \in A_j) = G_{X}(b_j) - G_X(a_j),
\end{equation}
where
\begin{equation}\label{eq:M1_GofX}
    G_X(x) = \int_{-\infty}^x \sqrt{\frac{1-\phi^2}{2\pi \sigma^2}} \exp\left\{-\frac{t^2(1-\phi^2)}{2\sigma^2}\right\} \dif t
\end{equation}
is the cumulative distribution function (cdf) of the $\sN\mleft(0, \sigma^2/(1-\phi^2) \mright)$ distribution.

The transition density $\gamma(x_{t-1}, \cdot)$ for Model 1 is defined as the conditional density of $X_t \mid (X_{t-1} = x_{t-1})$. Under this model, it can be shown that $X_t \mid (X_{t-1} = x_{t-1}) \sim \sN(\phi x_{t-1}, \sigma^2)$, and so, if $c_i^*$ is the representative point chosen within the interval $A_i$, then the transitions probabilities between states $\gamma_{i,j} = \mathbb{P}(X_t \in A_j \mid X_{t-1} \in A_i)$ are approximated by
\begin{equation}
\tgamma_{i,j} = \mathbb{P}(X_t \in A_j \mid X_{t-1} = c_i^*)
= F_{X,i}(b_j) - F_{X,i}(a_j),
\end{equation}
where
\begin{equation}
    F_{X,i}(x) = \int_{-\infty}^x \frac{1}{\sqrt{2\pi \sigma^2}}\exp\left\{-\frac{(t-\phi c^*_i)^2}{2\sigma^2}\right\} \dif t
\end{equation}
is the cdf of the $\sN\mleft( \phi c^*_i, \sigma^2\mright)$ distribution. The $\tgamma_{i,j}$ are then taken as the entries of the transition matrix in the discrete-space HMM approximation of the model.

\subsubsection{Model 2: VAR(1) Process On a Line}\label{subsub:model2}

Model~1 can be viewed as a special case of the general Poisson state-space model \eqref{eq:PoissonSSM1}--\eqref{eq:PoissonSSM5}, where $X_{t,1}$ is forced to be equal to $X_{t,2}$ with probability 1 for all $t$. In Model~2 we relax this restriction and allow $X_{t,2}$ to depend positively and linearly on $X_{t,1}$; specifically, we set $X_{t,2} = {\sigma_2}X_{t,1}/\sigma_1$ with probability 1, where each $\sigma_h > 0$ is given by $\sigma_h^2 = \Var{X_{t,h} \mid X_{t-1,h}}$ for all $t$. (The assumption of stationarity implies that this variance does not depend on $t$.) Formally, this can be written as a bivariate state-space model where the $\bX_t$ follow the degenerate distribution implied by 
\begin{align}
\begin{split} \label{eq:var1online}
\mathbf{X}_t &= \bPhi \mathbf{X}_{t-1} + \beps_t,\\
\bPhi &= \begin{bmatrix} \phi & 0 \\ 0 & \phi \\ \end{bmatrix}, \ \hbox{and} \\
\beps_t &\stackrel{\rm iid}{\sim} \lim_{\rho \to 1} \, \mathcal{N}_2\left(\bzero,\begin{bmatrix} \sigma_1^2 & \sigma_1 \sigma_2 \rho\\ \sigma_1 \sigma_2 \rho & \sigma_2^2 \\ \end{bmatrix}\right).
\end{split}
\end{align}
The bivariate distribution for $\beps_t$ lacks a density with respect to Lebesgue measure on $\R^2$, but admits a density on the line $y ={\sigma_2}x/\sigma_1$. However, it is more convenient to write the state-space model entirely in terms of the univariate state process $X_t := X_{t,1}$ as
\begin{align}
\begin{split}
\mathbf{Y}_t \mid X_t &\sim \text{Poisson}(w \cdot \beta_1 \cdot e^{X_t}) \cdot \text{Poisson}(w \cdot \beta_2 \cdot e^{{\sigma_2}X_t / \sigma_1} ),\\
X_t &= \phi X_{t-1} + \varepsilon_{t}, \ \hbox{and} \label{eq:model2}\\
\varepsilon_{t} &\stackrel{\rm iid}{\sim} \mathcal{N}(0,\sigma_1^2).
\end{split}
\end{align}
The vector of unknown parameters for Model~2, $\bveta_\text{M2} = (\phi, \sigma_1, \sigma_2, \beta_1, \beta_2)$, is fit to the data.

In the bivariate formulation \eqref{eq:var1online}, $X_{t,1}$ lies within the interval $[a,b] \subset \R$ if and only if $X_{t,2}$ lies within $[{\sigma_2}a /\sigma_1, {\sigma_2}b/\sigma_1]$ with probability 1. Thus, the transition probabilities for $X_{t,2}$ are determined by those of $X_{t,1}$ alone, as is the initial distribution of $X_{t,2}$ (since we assume $X_{t,1}$ -- and therefore $X_{t,2}$ -- is stationary). It follows that the initial distribution $\tilde{\bdelta}$ and the transition probabilities $\tilde{\gamma}_{i,j}$ for Model~2 are exactly the same as those in Model~1, but with $\sigma$ replaced by $\sigma_1$; effectively, the only difference between Model~1 and Model~2 is the inclusion of $\sigma_2/\sigma_1$ in the state-dependent Poisson distribution corresponding to the hard-band photons. For the process $X_t = \phi X_{t-1} + \varepsilon_t$ to be stationary, we again require that $\phi \in (-1,1)$.

\subsubsection{Model 3: Uncorrelated VAR(1) Process}\label{subsub:model3}

Model~3 further generalizes Model~2 by removing the restriction that $X_{1,t}$ and $X_{2,t}$ depend on each other linearly. In particular, Model~3 allows $X_{1,t}$ and $X_{2,t}$ to move freely in their own ``directions'', but ensures dependence between them by way of correlated errors. Specifically, 
\begin{align}
\begin{split} \label{eq:uncorVAR1}
\mathbf{Y}_t \mid \mathbf{X_t} &\sim \text{Poisson}(w \cdot \beta_1 \cdot e^{X_{t,1}}) \cdot \text{Poisson}(w \cdot \beta_2 \cdot e^{X_{t,2}}),\\
\mathbf{X}_t &= \bPhi \mathbf{X}_{t-1} + \boldsymbol{\varepsilon}_t,\\
\bPhi &= \begin{bmatrix} \phi_1 & 0 \\ 0 & \phi_2 \\ \end{bmatrix},\\
\boldsymbol{\varepsilon}_t &\stackrel{\rm iid}{\sim} \mathcal{N}_2(\mathbf{0},\boldsymbol{\Sigma}),\ \hbox{and} \\
\boldsymbol{\Sigma} &= \begin{bmatrix} \sigma_1^2 & \sigma_1 \sigma_2 \rho\\ \sigma_1 \sigma_2 \rho & \sigma_2^2 \\ \end{bmatrix}.
\end{split}
\end{align}
The vector of unknown parameters for Model~3, $\bveta_\text{M3} = (\phi_1, \phi_2, \sigma_1, \sigma_2, \beta_1, \beta_2, \rho)$, is fit to the data.

Model~3 includes two more parameters than Model~2, namely, $\rho \in (-1,1)$ and $\phi_2 > 0$. In contrast to Model~2, here densities with respect to $\R^2$ exist for the bivariate conditional and stationary distributions of the $\bX_t$. Since $\bX_t$ can lie within any open set of $\R^2$ with positive probability, the resulting initial distribution and transition probabilities in the discrete-space HMM approximation to the model must be derived anew.

Under Model 3, the existence of a stationary distribution for the process $\mathbf{X}_t = \bPhi \mathbf{X}_{t-1} + \boldsymbol{\varepsilon}_t$ requires that $\phi_1, \phi_2 \in (-1,1)$. The corresponding distribution is well-known \citep[e.g.,][]{hamilton2020time} and is given by the $\sN_2(\bzero, \bLambda)$ distribution, where $\Vec{\bLambda} = (\bI - \bPhi \otimes \bPhi)^{-1} \Vec{\bSigma}$, $\bI$ is the $4 \times 4$ identity matrix, $\otimes$ is the Kronecker product between matrices, and $\Vec{\cdot}$ is the vectorization operator that stacks the columns of an $m \times n$ matrix into a $mn \times 1$ vector. Thus, if $A_j = [a_{j,1}, b_{j,1}] \times [a_{j,2}, b_{j,2}]$ which, rather than an interval in $\R$ as in Models 1 and 2, is now a rectangle in $\R^2$, then the initial distribution for the discrete-space HMM approximation of Model 3 is taken to be the vector $\tilde{\bdelta}$ comprised of entries
\begin{align}
\tdelta_j &= \P(\bX_t \in A_j) \nonumber\\
&= G_{\bX}(a_{j,2}, b_{j,2}) 
- G_{\bX}(a_{j,2}, b_{j,1}) 
- G_{\bX}(a_{j,1}, b_{j,2})  \nonumber\\
&  \qquad \qquad \qquad \qquad \qquad \qquad \qquad \phantom{n}
+ G_{\bX}(a_{j,1}, b_{j,1})
\end{align}
where
\begin{equation}\label{eq:M3_stationarycdf}
G_{\bX}(x_1,x_2) = \int_{-\infty}^{x_1} \int_{-\infty}^{x_2} \frac{1}{2\pi \sqrt{|\det{\bLambda}|}} \exp\mleft(\bt^\top \bLambda^{-1}\bt\mright) \dif \bt
\end{equation}
is the cdf of the $\sN_2(\bzero, \bLambda)$ distribution.

The transition density $\gamma(\bx_{t-1}, \cdot)$ for Model~3 is now defined as the conditional density of $\bX_t \mid (\bX_{t-1} = \bx_{t-1})$. Under this model, it can be shown that $\bX_t \mid (\bX_{t-1} = \bx_{t-1}) \sim \sN_2(\bPhi \bx_{t-1}, \bSigma)$ and so, if $\bc_i^*$ is the representative point chosen within the rectangle $A_i$, then the transitions between states $\gamma_{i,j} = \mathbb{P}(\bX_t \in A_j \mid \bX_{t-1} \in A_i)$ are approximated by 

\begin{align}
\tgamma_{i,j} &= \mathbb{P}(\bX_t \in A_j \mid \bX_{t-1} = \bc_i^*) \nonumber\\
&= F_{\bX,i}(a_{j,2}, b_{j,2}) 
- F_{\bX,i}(a_{j,2}, b_{j,1}) 
- F_{\bX,i}(a_{j,1}, b_{j,2}) \nonumber\\
&  \qquad \qquad \qquad \qquad \qquad \qquad \qquad \quad \phantom{l}
+ F_{\bX,i}(a_{j,1}, b_{j,1}),
\end{align}
where
\begin{multline}\label{eq:M3_transitioncdf}
F_{\bX,i}(x_1,x_2) = \\
\int_{-\infty}^{x_1} \int_{-\infty}^{x_2} \frac{1}{2\pi \sqrt{|\det{\bSigma}|}} \exp\mleft\{(\bt - \bPhi \bc_i^*)^\top \bSigma^{-1}(\bt - \bPhi \bc_i^*)\mright\} \dif \bt
\end{multline}
is the $\sN_2(\bPhi \bc^*_i, \bSigma)$ cdf. The bivariate normal cdfs \eqref{eq:M3_stationarycdf} and \eqref{eq:M3_transitioncdf} can be computed efficiently using any statistical software package.

\subsection{State-Space Model Selection}\label{sec:modelselect}

The three models discussed in  Section~\ref{sub:nestedHMMs} are nested within each other: Model~1 is a special case of Model~2 subject to the constraint $\sigma_1 = \sigma_2$, and Model~2 is a special case of Model~3 subject to the constraints $\phi_1 = \phi_2$ and $\rho = 1$. Thus, any two of these models can, at least in principle, be compared using a likelihood ratio test (LRT). Under certain conditions, if the data are generated under the ``simpler'' of the two models being compared (i.e., the model with fewer parameters), the LRT statistic is asymptotically\footnote{The distribution function of the LRT statistic converges pointwise to that of a $\chi_{(\nu)}^2$ random variable as the size of the dataset increases (e.g., as the total time duration of the light curve increases). It is in this sense that the LRT statistic is {\sl asymptotically} $\chi_{(\nu)}^2$-distributed. This assumes that the necessary theoretical conditions are met \citep[e.g.,][]{prot:etal:02} and that the data are generated under the simpler} model. distributed $\chi_{(\nu)}^2$ with degrees of freedom $\nu$ equal to the difference in the number of parameters between the two models. Under certain conditions \citep[e.g.,][]{prot:etal:02}, this result allows a $p$-value to be computed; when the LRT statistic is sufficiently large relative to its asymptotic $\chi^2_{(\nu)}$ distribution, a small $p$-value is obtained and we can conclude that the data is inconsistent with the simpler model. The LRT statistic is equal to $-2$ times the difference of the maximized log-likelihood functions of the two models under comparison. Thus, we reject the smaller model when the larger model sufficiently improves the fit to a degree as measured by the log-likelihood function.

Among the conditions required for the LRT's asymptotic $\chi^2_{(\nu)}$ distribution are that (i) the models under comparison are nested and (ii) the parameters of the smaller model are not constrained to be on the boundary of the set of possible parameter values under the larger model. These conditions are met for Models~1 and 2 and the standard LRT is thus a suitable means of comparing them. Unfortunately, the comparison of Models~2 and 3 does not satisfy the second of these conditions because one parameter in the smaller Model~2 lies on the boundary of the parameter space of the larger Model~3 (i.e., $\rho = 1$). In fact, the asymptotic distribution of the LRT statistic is not known in this case. While \cite{self1987asymptotic} provided a generalized LRT statistic that helps to account for such situations, its implementation can be computationally difficult.

When the choice between Model~2 and Model~3 is not clear from the results of the model estimation procedure (as is the case for the \evlac data; see Section~\ref{sub:SSMfit}), one can again use the parametric bootstrap, this time to approximate the finite-sample distribution of the LRT statistic by way of simulations. Assuming that  model fitting produces the MLEs $\hat{\bveta}_\text{M2}$ for Model~2 and $\hat{\bveta}_\text{M3}$ for Model~3, this bootstrap procedure generates a large number $B$ of independent replicate datasets $\bY\oneT^{(1)}, \ldots, \bY\oneT^{(B)}$ under Model~2 with parameter $\hat{\bveta}_\text{M2}$. For each $b=1, \ldots, B$, both Models 2 and 3 are fit to $\bY\oneT^{(b)}$, producing the respective MLEs $\hat{\bveta}_\text{M2}^{(b)}$ and $\hat{\bveta}_\text{M3}^{(b)}$. The bootstrapped LRT statistics $\hat{\psi}^{(b)} = -2( \ell_\text{M2}(\hat{\bveta}_\text{M2}^{(b)}) - \ell_\text{M3}(\hat{\bveta}_\text{M3}^{(b)}))$ are computed, where $\ell_{\text{M2}}$ and $\ell_{\text{M3}}$ are the log-likelihood functions for Model~2 and Model~3, respectively. The statistics $\hat{\psi}^{(1)}, \ldots, \hat{\psi}^{(B)}$ are then used to construct an approximate distribution $\hat{F}_\psi$, perhaps using a kernel density estimate (see Section~\ref{subsub:semisupervised}). This distribution is used in place of the $\chi^2_{(\nu)}$ distribution to compute a $p$-value. Specifically, Model~2 can be rejected in favor of Model~3 at the $95\%$ confidence level if the LRT statistic produced from the original data, $\hat{\psi} = -2( \ell_\text{M2}(\hat{\bveta}_\text{M2}) - \ell_\text{M3}(\hat{\bveta}_\text{M3}))$, is such that $1 - \hat{F}_\psi(\hat{\psi}) < 0.05$. 

In addition to overcoming theoretical roadblocks associated with the standard LRT approach, the bootstrap technique helps to account for potential numerical inaccuracies (e.g., stemming from the discrete-space HMM approximation of the state-space likelihood or its optimization, which is especially relevant when the dimension of state space $\sX$ is greater than $1$). Because the same numerical inaccuracies affect the LRT statistic as computed on the data and as computed on the bootstrap replicates, the bootstrap provides the null distribution of the LRT statistic \emph{as it is computed}. This allows us to define the statistic to be \emph{as computed} (including potential numerical inaccuracies) and correctly calibrate its null distribution and the p-value. Specifically, the Monte Carlo nature of the bootstrapped $p$-values takes the entire approximation procedure into account, whereas the standard LRT approach assumes the use of genuine log-likelihood functions which are perfectly optimized in the involved calculations. 

Having fit the state-space model, standard HMM algorithms (see Appendix~\ref{app:HMMalgos}) allow one to decode the observations, that is, to make predictions, $\hat{\bX}_1, \ldots, \hat{\bX}_T$, of the underlying states, $\bX_1, \ldots, \bX_T$. With a continuous-space HMM, predictions take values in the set of representative points $\{\bc_1^*, \ldots, \bc_m^*\}$ defined in the discrete approximation to the continuous-space likelihood, see Section~\ref{sub:approximation}.

\section{Stage 2: Classifying Light Curves into Flaring and Quiescent Intervals}\label{sec:classify}

Our Stage~1 HMMs use continuous underlying processes to model stellar flare activity (see Section~\ref{sec:cont-HMM}). In practice, however, we also wish to identify those time intervals when the star is in its quiescent state and those when it is in its flaring state. In this section we introduce our Stage~2 analysis, which uses a finite mixture model to classify the $\hat{\bX}_1, \ldots, \hat{\bX}_T$ fitted in Stage~1 into the quiescent and flaring states. 

We consider two scenarios: semi-supervised and unsupervised classification. The semi-supervised scenario applies in cases where we are able to identify a subsample of size $m$ of the predicted states, $A_\text{q} = \{\hat{\bX}_{t_1}, \ldots, \hat{\bX}_{t_m}\}$, where $m$ is reasonably large and the subsample is assumed to arise from a period of quiescence. Identifying a quiescent subsample invariably involves a degree of subjectivity (e.g., through visual inspection). We refer to this scenario as semi-supervised because some, but not all, of the data is assumed to be classified \emph{a priori}. If there is a clearly identifiable interval of quiescence, $A_\text{q}$ can be selected using a range of time bins where the light curves appear to be in equilibrium and do not exhibit flaring behaviour. In the unsupervised scenario, we do not have such a subsample. 

In both the semi-supervised and unsupervised scenarios, we propose to model the full set of Stage~1 predicted states, $\hat{\bX}_1, \ldots, \hat{\bX}_T$, as a \emph{mixture of two distributions}, one corresponding to the quiescent state and the other corresponding to the flaring state.\footnote{In the unsupervised scenario, one could, in principle, apply a non-parametric unsupervised clustering method such as $k$-means (with $k=2$) to the $\hat{\bX}_t$ to classify observations into quiescent and flaring intervals. Such methods have the benefit of being fully automatic, and are easy to implement using built-in routines within any statistical software package. However, quantification of uncertainty for the classifications produced by these ``black-box'' algorithms are difficult to interpret (and are often not available at all), particularly when there is not a probabilistic model underlying the algorithm. Furthermore, different unsupervised clustering algorithms (e.g., $k$-means, $k$-medians, DBSCAN, etc.) use different loss/objective functions and can yield different classifications of the same data; aside from computational complexity, there are few clear reasons for choosing one clustering algorithm over another. Thus, we deploy a more statistical approach, using a likelihood-based finite mixture model.} This modelling approach is corroborated by the histogram of the \evlac state predictions shown in Section~\ref{sec:intervaldetermination}. Formally, we assign the label `1' to the quiescent distribution and `2' to the flaring distribution, and for the purpose of classification, suppose
\begin{equation}\label{eq:FMM}
    \hat{\bX}_1, \ldots, \hat{\bX}_T \stackrel{\rm iid} {\sim} \alpha \cdot F_1 + (1-\alpha) \cdot F_2,
\end{equation}
where $F_1$ and $F_2$ are cdfs and $\alpha \in (0,1)$ is a mixing parameter, all to be inferred from the data. The mixing parameter corresponds to the proportion of time that the star spent in the quiescent state. Model (\ref{eq:FMM}) can be equivalently represented by introducing a sequence of latent variables, $Z_1,\ldots,Z_T \stackrel{\rm iid}{\sim} \text{Bernoulli}(\alpha)$ and declaring 
\begin{equation}\label{eq:FMMlatent}
    \hat{\bX}_t \mid Z_t = k \sim F_k, \quad \hbox{for each } t \hbox{ and for } k \in \{1,2\}.
\end{equation} 
Note that neither of these model representations accounts for the autocorrelation (or more generally, the time series nature) of $\hat{\bX}_1, \ldots, \hat{\bX}_T$ implied by the Stage~1 HMMs (e.g., \eqref{eq:PoissonSSM5}, \eqref{eq:modelone}, \eqref{eq:model2}, and \eqref{eq:uncorVAR1}) and observed in the actual \evlac fits (see, e.g., Figure~\ref{fig:statepreds}). Instead, we assume that temporal characteristics are captured by the Stage~1 HMM fit, and here we merely aim to classify the light curve into flaring and quiescent intervals.

For simplicity, we assume henceforth that as for Models~1 and 2, the predicted states are univariate, although our theory generalizes to higher-dimensional state predictions (as in Model~3). While mixture models often involve component distributions belonging to the same parametric family -- normal distributions or other exponential family distributions are especially popular -- we consider a less rigid approach to the choices of $F_1$ and $F_2$. Ultimately, the estimated probability that the star is in a flaring state depends on the relative size of $f_1(x)$ and $f_2(x)$ at each value of $x$, where $f_1$ and $f_2$ are the probability density functions corresponding to $F_1$ and $F_2$, respectively. The choice of $f_1$ and $f_2$ is particularly influential for ranges of $x$ at the transition between states, where $f_1(x)$ and $f_2(x)$ are both moderate and are both well above zero; thus, the choice of densities is important, and poor approximations using standard parametric families can potentially yield inaccurate flaring state probabilities for such $x$.

Note that in both the semi-supervised and unsupervised procedures, we are not concerned with overfitting the relevant mixture distributions to the data, as each fitted distribution pertains specifically to the predicted states output by a particular fitted state-space model and are not intended to be used elsewhere.

\subsection{Semi-Supervised Classification}\label{subsub:semisupervised}
There is a distinct advantage in the semi-supervised scenario where $A_\text{q}$ can be used to form a robust non-parametric estimate of $f_1(x)$. Under the mixture model, we have $\hat{X}_{t_j} \sim F_1$ for $j=1,\ldots,m$ (i.e. for $\hat{X}_{t_j}\in A_\text{q}$) and we can use a \emph{kernel density estimate} (KDE), $\hat{f}_1$, to approximate $f_1(x)$. The KDE  essentially traces out a smoothed version of the histogram of the sample $A_\text{q}$. 

The flaring component density $f_2(x)$, on the other hand, does not yield as easily to a KDE because an analogous subsample of data known to be from the flaring state is usually unavailable. Instead, we approximate $f_2(x)$ by a step function $\hat{f}_2(x; \bpi)$ parameterized by the constant value $\pi_k$ that it takes within a pre-specified bounded interval $[b_{k-1}, b_k)$ for a fixed number $K$ of intervals; that is, 
\begin{equation}\label{eq:stepfunction}
    \hat{f}_2(x; \bpi) = \sum_{k=1}^K \frac{\pi_k}{b_k - b_{k-1}} \cdot \one{x \in [b_{k-1}, b_k)},  
\end{equation}
where the $\pi_k$ are unknown non-negative parameters subject to $\sum_{k=1}^K \pi_k = 1$. When the intervals $[b_{k-1}, b_k)$ are evenly spaced, $\hat{f}_2$ is essentially a histogram function. We choose $b_K = \sup A$, where $A$ is the essential domain used to approximate the domain of the $X_t$ (see Section~\ref{sub:approximation}). This is because the values of $\hat{X}_t$ produced by the local decoding algorithm (see Appendix~\ref{subsub:fwdbckwdalgo}) take values in the set of representative points $\{c_1^*, \ldots, c_m^*\} \subseteq A$; thus $\hat{X}_t \in A$ for all $t$. On the other hand, we assume that the smallest values of $X_t$ are reserved for the quiescent state and thus we choose $b_0$ as the median of $\hat{f}_1(x)$, although other choices are possible. 

The unknown parameters in the model \eqref{eq:FMM}, namely $\alpha$ and $\bpi$, can be estimated using the EM algorithm, which is a standard tool for computing maximum likelihood estimates in finite mixture models \citep[see][]{dempster1977maximum} and is easily derived for \eqref{eq:FMM} (see Appendix~\ref{app:EMsemisupervised}). We run the EM algorithm on the subset $A_\text{r} = \{\hat{X}_1, \ldots, \hat{X}_T\} \setminus A_\text{q}$ of predicted states \emph{not} used to fit $\hat{f}_1(x)$, so as not to use $A_\text{q}$ twice in the estimation process; this requires a minor adjustment to the mixing parameter $\alpha$ to account for the proportion of quiescent state data removed (see Appendix~\ref{app:EMsemisupervised}). 

Once the estimation of \eqref{eq:FMM} is complete, the estimated posterior probability that each $X_t$ is in a flaring state (i.e., state `2') can be derived using the representation in \eqref{eq:FMMlatent}, which yields
\begin{equation}\label{eq:posteriorprob}
    \mathbb{P}(Z_t = 2 \mid \hat{X}_t = \hat{x}_t) 
    = \frac{(1-\hat{\alpha}) \cdot \hat{f}_2(\hat{x}_t; \hat{\bpi}) }{\hat\alpha \cdot \hat{f}_1(\hat{x}_t) + (1-\hat{\alpha}) \cdot \hat{f}_2(\hat{x}_t; \hat{\bpi})},
\end{equation}
where $\hat\alpha$ and $\hat\bpi$ are the maximum likelihood estimates computed with the EM algorithm.

\subsection{Unsupervised Classification}\label{subsub:unsupervised}

In situations where there is no subsample of the data that can reasonably be assumed to have arisen from the quiescent state, 
inference must be fully unsupervised and there is no immediate way to use KDE to approximate $f_1(x)$. In this case, we have found that for the \evlac data  a mixture of three normal distributions provides a reasonable approximation to the distribution of the $\hat{X}_t$: that is, 
\begin{equation}\label{eq:normalFMM}
    \hat{X}_1, \ldots, \hat{X}_T \stackrel{\rm iid} {\sim} \alpha_1 \cdot \sN(\mu_1, \tau_1^2) + \alpha_2 \cdot \sN(\mu_2, \tau^2_2) + \alpha_3 \cdot \sN(\mu, \tau_3^2),
\end{equation}
where $\alpha_1, \alpha_2,$ and $\alpha_3$ are non-negative mixing parameters subject to $\sum_{k=1}^3 \alpha_k = 1$ and each $\mu_k$ and $\tau^2_k$ are mean and variance parameters (respectively), all to be estimated. This distribution is also fit using the EM algorithm (see Appendix~\ref{app:EMunsupervised}).

In this instance, we assume that one component of the model corresponds to the flaring state, while the remaining two components together correspond to the quiescent state (see Section~\ref{subsub:unsupervised10679} for further discussion in the context of \evlacnospace). We may assume without loss of generality that $\mu_1 < \mu_2 < \mu_3$, and since a lower $X_t$ corresponds to a lower Poisson intensity for the emission $Y_{t,1}$ (see \eqref{eq:PoissonSSM1}), we regard the first two normal distributions in \eqref{eq:normalFMM} as those corresponding to quiescence, with $\alpha_1 + \alpha_2$ representing the proportion of time spent in that state. By using two normal distributions, we are able to better represent the skew in the quiescent distribution. Once \eqref{eq:normalFMM} has been fitted, the posterior probability that each $X_t$ is in a flaring state is given by
\begin{equation}\label{eq:posteriorprobuns}
    \mathbb{P}(Z_t \neq 1 \mid \hat{X}_t = \hat{x}_t) 
    = \frac{\hat{\alpha}_3 \cdot f(\hat{x}; \hat{\mu}_3, \hat{\tau}_3^2) }{\sum_{k=1}^3 \hat{\alpha}_k \cdot f(\hat{x}_t; \hat{\mu}_k, \hat{\tau}^2_k)},
\end{equation}
where $f(\cdot; \mu, \tau^2)$ is the density of the $\sN(\mu, \tau^2)$ distribution.

\section{Analysis of EV Lac}
\label{sec:analysis}

In this section, we illustrate the statistical methods developed in the Sections~\ref{sec:HMMs},  \ref{sec:model}, and \ref{sec:classify} by applying them to the \evlac data described in Section~\ref{sec:motivate}. In particular, we derive a classification of the light curves in Figure~\ref{fig:EVlacTS1} into quiescent and flaring intervals.

\subsection{Stage~1: HMM Selection and Fit for EV Lac}\label{sub:SSMfit}

We analyzed the two long-duration \chandra\ observations of \evlacnospace, ObsID~01885 obtained in 2001 and ObsID~10679 obtained in 2009. For both observations, we used the dispersed data from the combined HEG and MEG arms, and from the combined positive and negative orders, which avoids pileup effects seen during the stronger flares in the zeroth~order. We split the data into soft (0.3--1.5~keV) and hard (1.5--8~keV) passbands, and binned them into time bins of $w = 50$ s (see Figure~\ref{fig:EVlacTS1}). We also tested the sensitivity of our model fits to these binning schemes by replicating the results using other passbands (i.e., 0.3--0.9~keV and 0.9--8~keV) and changing the binning phase by 25~sec, and found no qualitative differences; see Appendix~\ref{app:othertimebins} for details.

\begin{figure}
    \centering
    \includegraphics[width=\columnwidth]{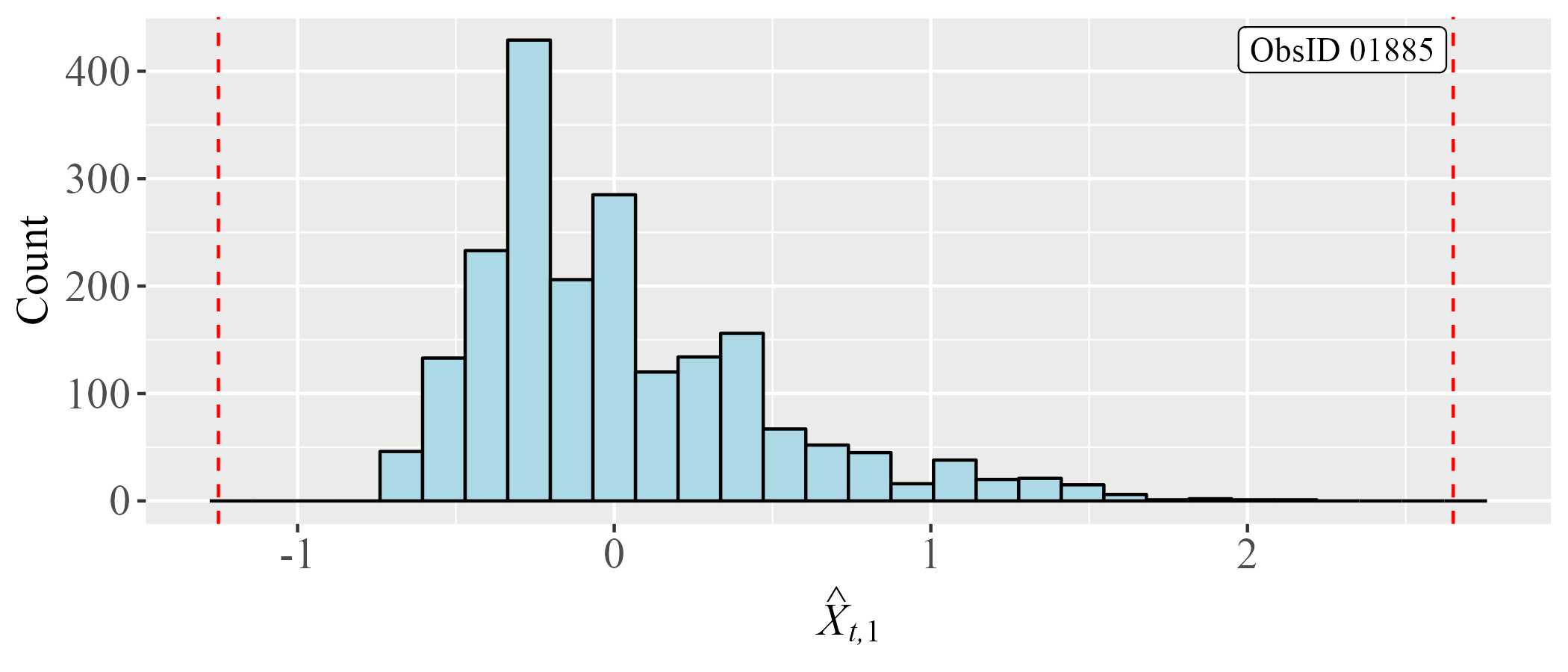}
    \includegraphics[width=\columnwidth]{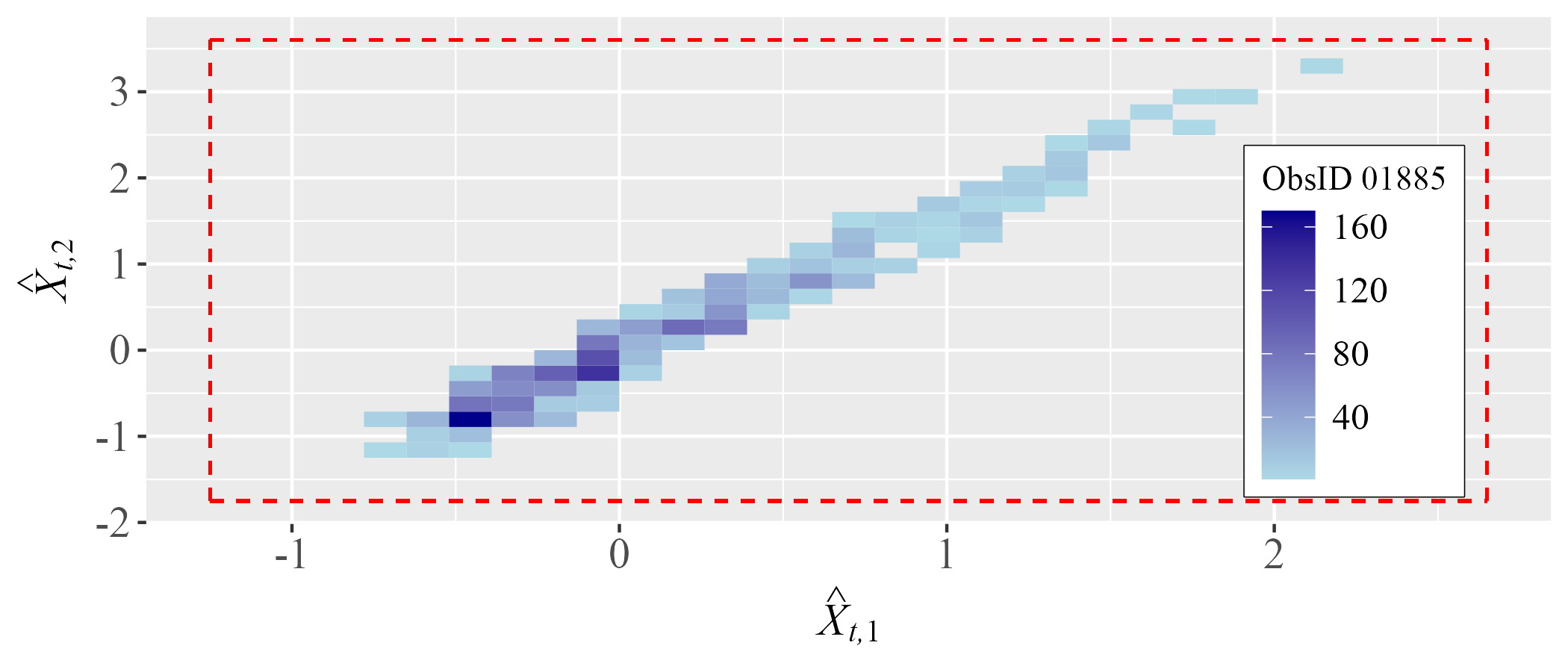}
    \caption{Univariate (top panel) and bivariate (bottom panel) histograms of predicted states $\hat{X}_t$ based on an initial fits of Models~2 and 3 to ObsID~01885; above, the dashed red lines enclose the essential domain $A = [-1.25, 2.65]$ chosen for the discrete-space approximation of Model~2, and below, they enclose the essential domain $A = [-1.25, 2.56] \times [-1.75, 3.6]$ chosen for Model~3}
    \label{fig:essential_dom}
\end{figure}

We fit the three state-space models described in Section~\ref{sub:nestedHMMs} to both observations. For brevity, we present only the fitted models for ObsID~01885; classifications into flaring and quiescent intervals are presented for both observations in Section~\ref{sec:intervaldetermination}. We employed visual diagnostics to determine the parameters of the discretizations of the continuous state spaces. For Model 2, for example, we chose the essential domain $A = [-1.25, 2.65]$ and partitioned $A$ into $m = 40$ evenly-spaced subintervals and chose the representative points $\{c_1^*, \ldots, c_{40}^*\}$ as the midpoints of these subintervals; a histogram of the states $\hat{X}_t$ predicted by the model via local decoding shows that this choice of essential domain was conservative in that it easily covers the range of the $\hat{X}_t$ (see the upper panel of Figure~\ref{fig:essential_dom}). The estimates can be sensitive to the choice of $m$ when $m$ is small and we chose $m=40$ because this is the approximate number of sub-intervals at which the parameter estimates and maximized log-likelihood stabilized. Similarly, for Model~3 we chose the essential domain $A=[-1.25, 2.56] \times [-1.75, 3.6]$ (see the lower panel of Figure~\ref{fig:essential_dom}) and $m=40^2$.

Bias-corrected parameter estimates and confidence intervals computed using the parametric bootstrap (see Section~\ref{sec:cont-HMM}) under Models~1, 2, and 3 appear in Tables~\ref{tab:M1_w50main}, \ref{tab:M2_w50main}, and \ref{tab:M3_w50main}, respectively. The estimates of the parameters common to the models are broadly consistent with each other, as are their standard errors. The estimates are also very similar to those produced with a passband split at $0.9$ keV (omitted for brevity), demonstrating robustness to that choice.

As a byproduct of the optimization procedure used to fit the models, we extracted the values of the maximized log-likelihood function \eqref{eq:CSHMMlikeapp} for each model (shown in Table~\ref{tab:LRTs}). The standard LRT decisively rejected Model~1 in favour of Model~2, with a test statistic of $918.64$ far exceeding the asymptotic $\chi^2_{(1)}$ distribution at the 95\% significance level. For a comparison between Models~2 and 3, we turned to the bias-corrected parameter estimates and their corresponding bootstrap standard errors shown in Tables~\ref{tab:M2_w50main} and \ref{tab:M3_w50main}, respectively. These tables show that the correlation parameter $\rho$ in Model 3 is estimated at $1$ -- precisely its value fixed by Model~2 -- with virtually no uncertainty in the estimate (all values have been rounded to six significant figures). Moreover, the remaining parameters shared by Model~2 and Model~3 are estimated very consistently between the two models, as are their standard errors, and the Model~3 estimates of $\phi_1 = \text{Cor}(X_{t,1}, X_{t-1,1})$ and $\phi_2 = \text{Cor}(X_{t,2}, X_{t-1,2})$ are very close. We thus have substantial evidence that the additional structure of Model~3 is unnecessary for the \evlac data, and we proceed with an analysis of Model~2.

\begin{table}
\centering
\begin{tabular}{c|r|rrr}
  \toprule
 Parameter & Estimate & Standard Error & CI (Lower) & CI (Upper) \\ 
  \midrule
  $\phi_{1}$ & 0.987235 & 0.004579 & 0.978260 & 0.996209 \\ 
  $\sigma_{1}$ & 0.128329 & 0.006212 & 0.116155 & 0.140504 \\ 
  $\beta_{1}$ & 0.184642 & 0.045141 & 0.096165 & 0.273119 \\ 
  $\beta_{2}$ & 0.075158 & 0.018178 & 0.039529 & 0.110788 \\ 
   \bottomrule
\end{tabular}
\caption{Bias-corrected parameter estimates for Model~1 fit to  ObsID~01885, with bias correction and standard errors obtained via the parametric bootstrap}
\label{tab:M1_w50main}
\end{table}

\begin{table}
\centering
\begin{tabular}{c|r|rrr}
  \toprule
 Parameter & Estimate & Standard Error & CI (Lower) & CI (Upper) \\ 
  \midrule
$\phi_{1}$ & 0.979644 & 0.006456 & 0.966991 & 0.992297 \\ 
  $\sigma_{1}$ & 0.100712 & 0.004811 & 0.091282 & 0.110142 \\ 
  $\sigma_{2}$ & 0.161689 & 0.007409 & 0.147168 & 0.176210 \\ 
  $\beta_{1}$ & 0.193817 & 0.022021 & 0.150656 & 0.236978 \\ 
  $\beta_{2}$ & 0.062417 & 0.010696 & 0.041453 & 0.083380 \\ 
   \bottomrule
\end{tabular}
\caption{Bias-corrected parameter estimates for Model~2 fit to  ObsID~01885, with bias correction and standard errors obtained via the parametric bootstrap}
\label{tab:M2_w50main}
\end{table}

\begin{table}
\centering
\begin{tabular}{c|r|rrr}
  \toprule
 Parameter & Estimate & Standard Error & CI (Lower) & CI (Upper) \\ 
  \midrule
$\phi_{1}$ & 0.981721 & 0.008663 & 0.964742 & 0.998700 \\ 
  $\phi_{2}$ & 0.976232 & 0.007997 & 0.960558 & 0.991906 \\ 
  $\sigma_{1}$ & 0.096086 & 0.006253 & 0.083829 & 0.108342 \\ 
  $\sigma_{2}$ & 0.171667 & 0.008918 & 0.154188 & 0.189147 \\ 
  $\beta_{1}$ & 0.206301 & 0.021047 & 0.165048 & 0.247554 \\ 
  $\beta_{2}$ & 0.066548 & 0.009203 & 0.048510 & 0.084585 \\ 
  $\rho$ & 1.000000 & 0.000000 & 1.000000 & 1.000000 \\
   \bottomrule
\end{tabular}
\caption{Bias-corrected parameter estimates for Model~3 fit to ObsID~01885, with bias correction and standard errors obtained via the parametric bootstrap}
\label{tab:M3_w50main}
\end{table}

\begin{table}
\centering
\begin{tabular}{l|c}
  \toprule
 Model & Maximized Log-Likelihood\\ 
  \midrule
Model 1: AR(1) Process & $-9914.53$ \\
Model 2: VAR(1) Process on a Line & $-9455.21$\\
Model 3: Uncorrelated VAR(1) Process & $-9424.47$\\
   \bottomrule
\end{tabular}
\caption{Maximized log-likelihoods for all three models based on ObsID 01885}
\label{tab:LRTs}
\end{table}

\subsection{Stage~2: Flaring/Quiescent Interval Estimates for EV Lac} \label{sec:intervaldetermination}

In this section, we demonstrate our Stage~2 methods for classifying the light curve, $\bY\oneT$, into flaring and quiescent intervals by fitting finite mixture distributions to the predicted states $\hat{X}_1, \ldots, \hat{X}_T$. All calculations in this section are under the preferred Model~2.

\subsubsection{Semi-Supervised Classification for ObsID 01885}\label{subsub:semisupervised01885}

\begin{figure}
    \centering
    \includegraphics[width=\columnwidth]{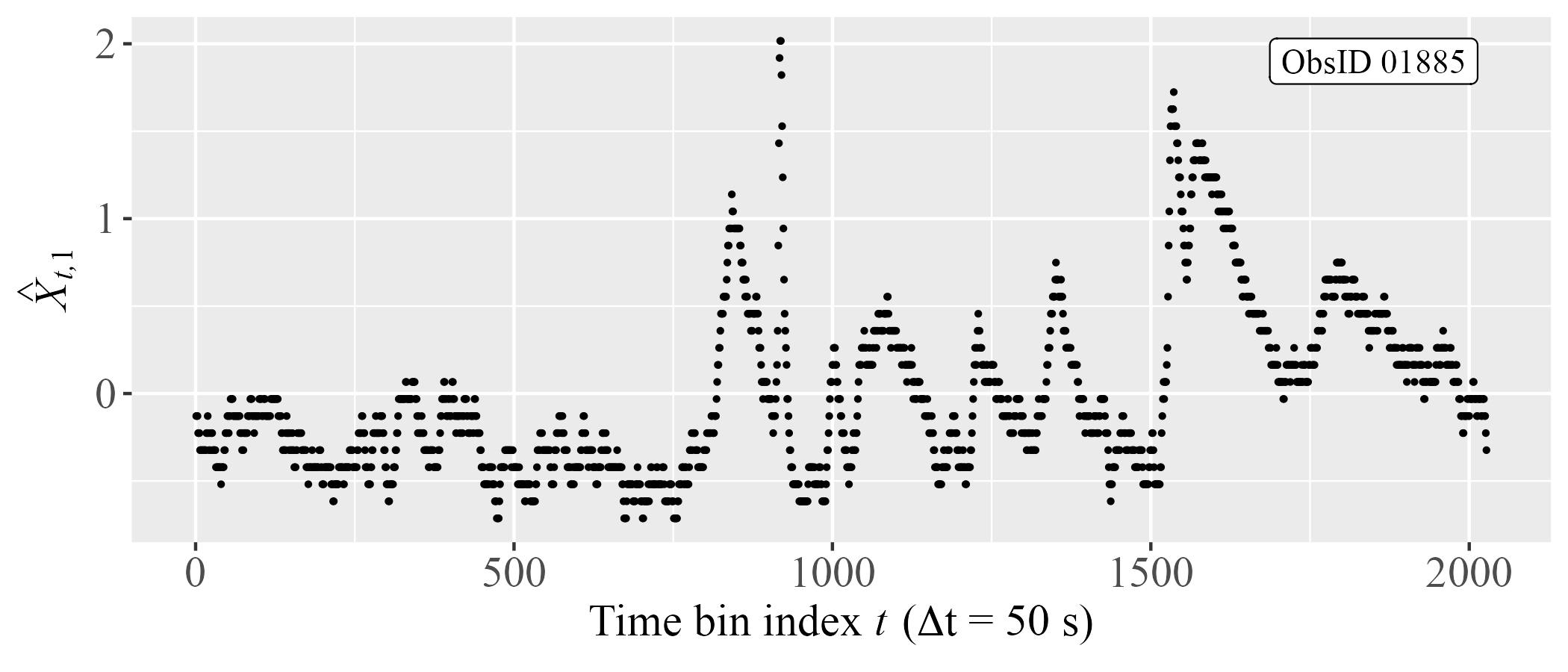}
    \caption{Predicted soft band states $\hat{X}_1, \ldots, \hat{X}_{2027}$ for ObsID 01885}
    \label{fig:statepreds}
\end{figure}

The predicted state variables, given by
\begin{equation}\label{eq:predictedstates}
\hat{X}_t = \argmax{x \in \sX} \mathbb{P}_{\hat{\bveta}} \mleft( X_t = x \mid \bY\oneT = \by\oneT\mright), \quad t=1,\ldots,T
\end{equation}
with $T = 2027$ are computed using the local decoding procedure described in Appendix~\ref{app:HMMalgos} and plotted for ObsID~01185 in Figure~\ref{fig:statepreds}.

\begin{figure}
    \centering
    \includegraphics[width=\columnwidth]{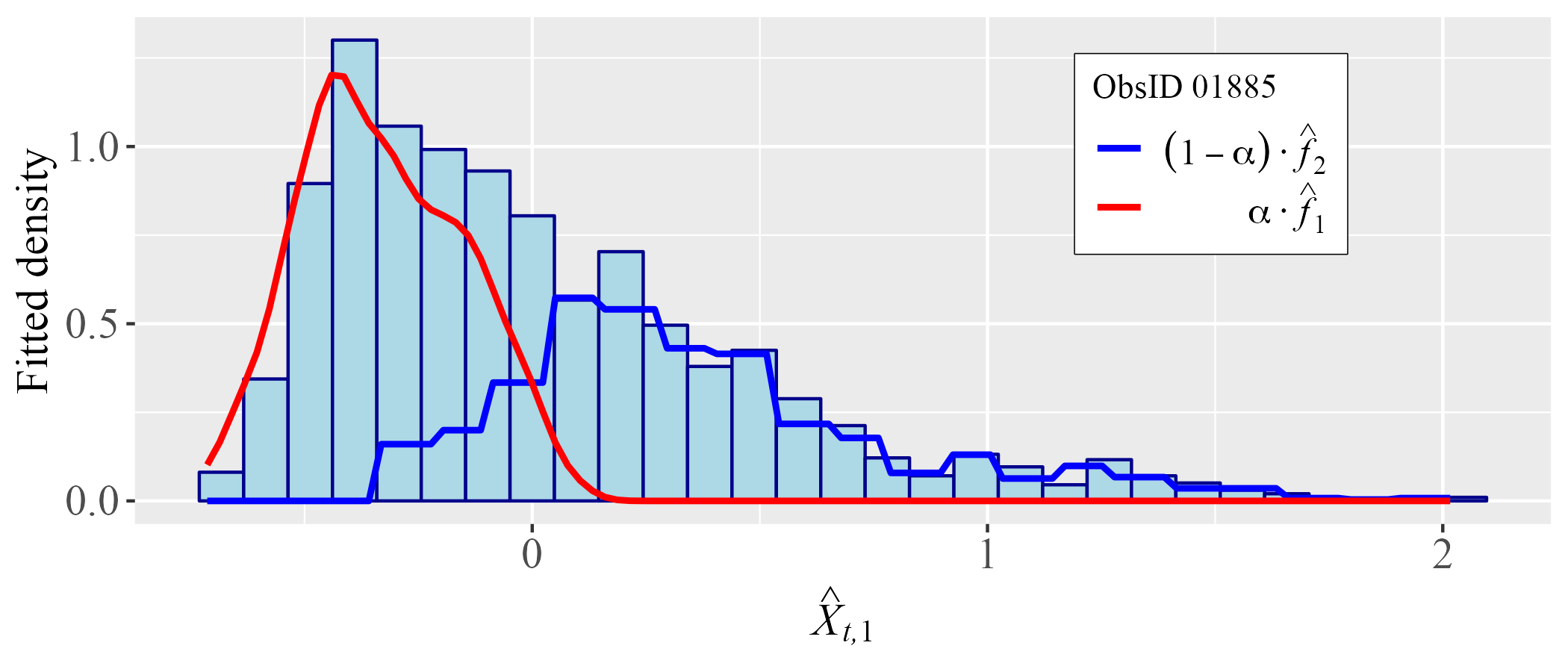}
    \includegraphics[width=\columnwidth]{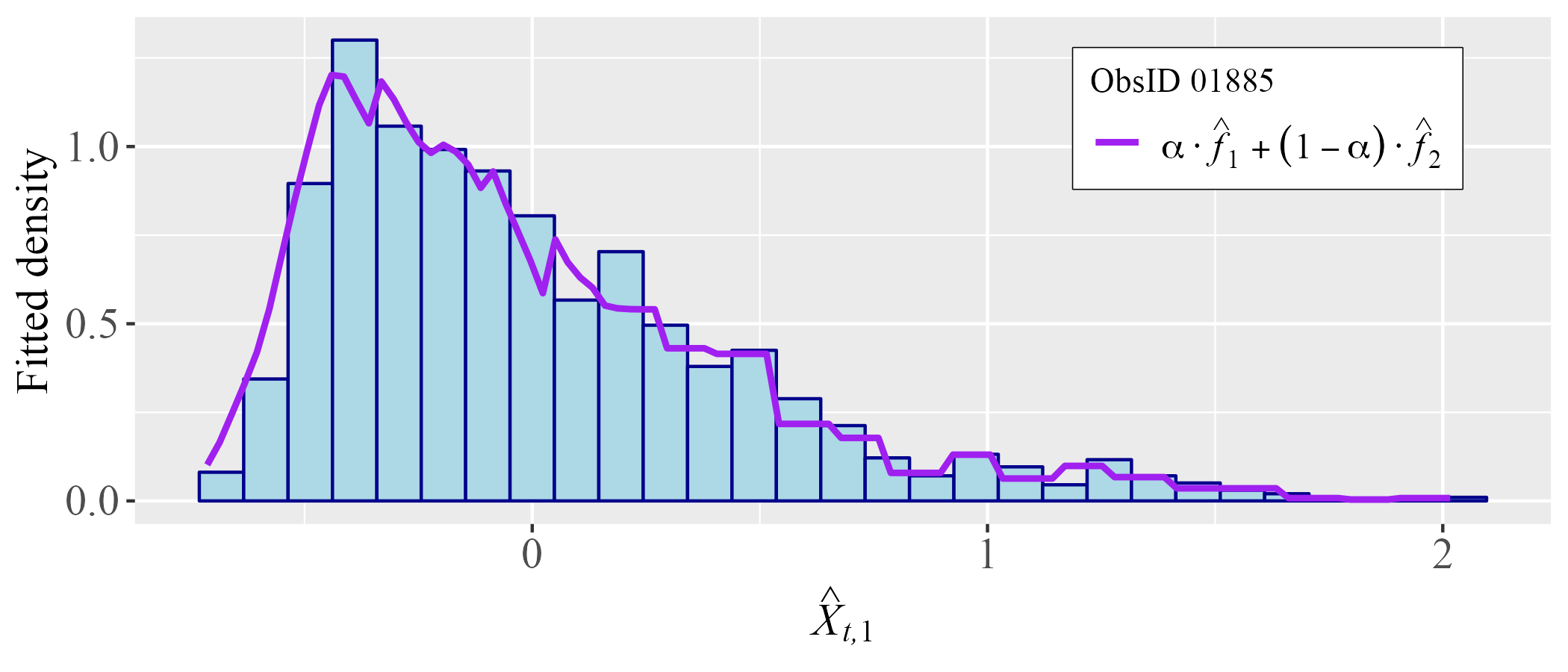}
    \caption{Fitted component densities (top panel) and mixture density (bottom panel) for ObsID~01885; the densities are overlaid on a histogram of $\{\hat{X}_1, \ldots \hat{X}_{2027}\}$}
    \label{fig:FMM_semisupervised}
\end{figure}

A visual inspection of the ObsID~01185 light curve in Figure~\ref{fig:EVlacTS1} and its predicted states in Figure~\ref{fig:statepreds} reveals a clear period of quiescent equilibrium over the first 750 time bins. Thus, we could apply the semi-supervised approach of Section~\ref{subsub:semisupervised} to model the distribution of the $\hat{X}_t$. After fitting the KDE $\hat{f}_1$ to $\{\hat{X}_{1}, \ldots, \hat{X}_{750}\}$,  we chose $K=25$ ``steps'' for the step function in \eqref{eq:stepfunction}, setting the intervals $[b_{k-1}, b_k)$ to be $25$ evenly spaced subintervals in $[b_0, b_K]$, where $b_0 \, \approx -0.35$ is the median of $\hat{f}_1$ and $b_K = \sup A = 2.65$. (We assume that the lowest levels of activity correspond to quiescence.) We fit the mixture in \eqref{eq:FMM} to $\{\hat{X}_{751}, \ldots, \hat{X}_{2027}\}$ using the EM algorithm described in Appendix~\ref{app:EMsemisupervised}, which yielded a mixing parameter estimate $\hat{\alpha} = 0.5528$, indicating that the estimated proportion of time that \evlac spends in a flaring state based on the ObsID~01885 time bin is $100\% \cdot (1 - \hat{\alpha}) \approx 45\%$. The resulting component densities and mixture density are illustrated in Figure~\ref{fig:FMM_semisupervised}.

\begin{figure}
    \centering
    \includegraphics[width=\columnwidth]{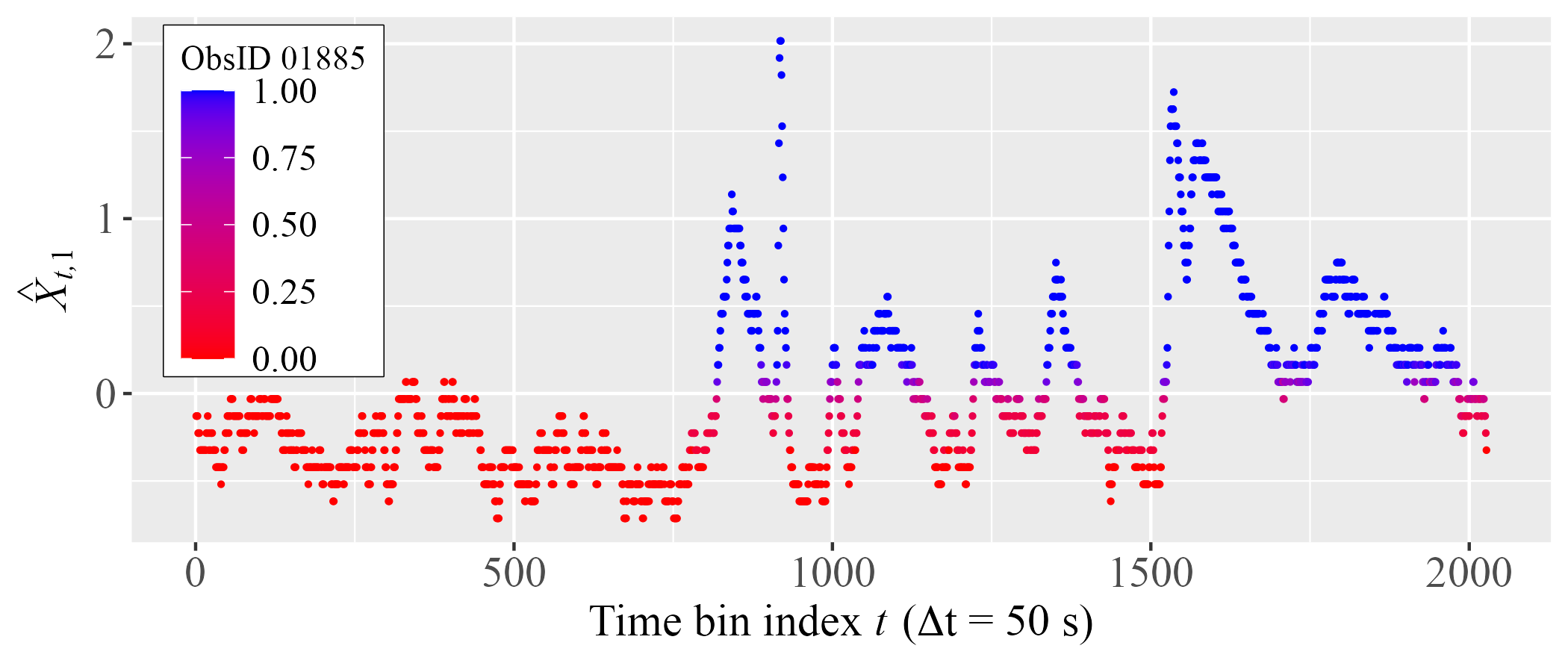}
    \includegraphics[width=\columnwidth]{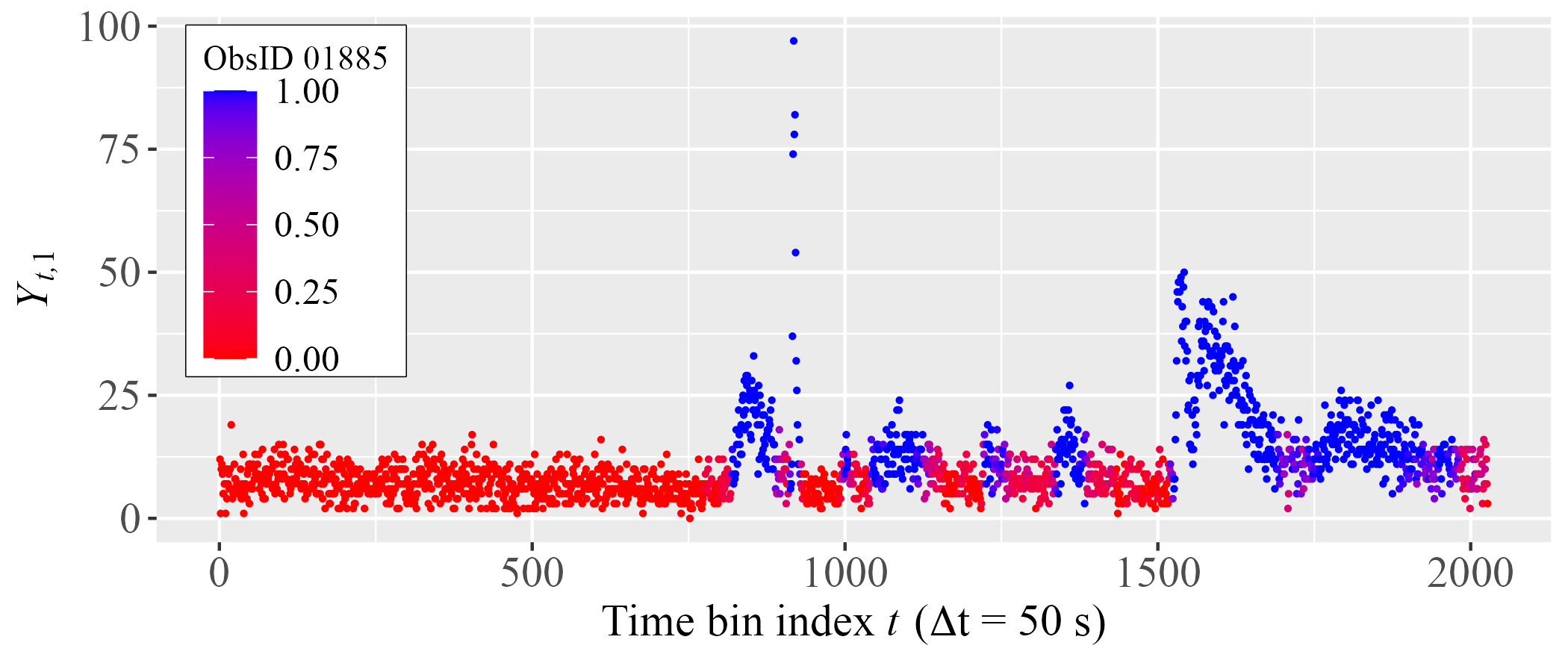}
    \caption{Posterior flaring state probabilities used to colour the predicted states $\hat{X}_1, \ldots, \hat{X}_t$ (top panel) and the observed soft-band counts $Y_{1,1},\ldots,Y_{T,1}$ (bottom panel) for ObsID~01885}
    \label{fig:posteriors}
\end{figure}

Using \eqref{eq:posteriorprob}, we computed the posterior flaring state probability for each $\hat{X}_t$; these are shown on a colour gradient in Figure~\ref{fig:posteriors}, both for the predicted states and the original soft-band counts, $Y_{1,1:T}$ (Figure~\ref{fig:response} in Appendix~\ref{app:othertimebins} shows the soft and hard band counts coloured by the same probabilities.) From the posterior flaring state probabilities, we created binary quiescent/flaring classifications $\hat{z}_1, \ldots, \hat{z}_T \in \{1,2\}$ using a simple classification rule, which is the basis for the results given in Section~\ref{subsub:EVLacstates} below: letting $\hat{p}_t = \mathbb{P}(Z_t = 2 \mid \hat{X}_t = \hat{x}_t)$ as in \eqref{eq:posteriorprob}, \evlac was classified as being in a flaring state at time index $t$ if and only if $\hat{p}_t > 0.5$; equivalently \begin{equation}\label{eq:classrule1}
    \hat{z}_t = 1 \cdot \one{\hat{p}_t \leq 0.5} + 2 \cdot \one{\hat{p}_t > 0.5}.
\end{equation}

\begin{table}
    \centering
    \caption{Flaring time intervals for ObsID~1885, in spacecraft clock time. The times are offset from the observation start time of 117315383.3~s, corresponding to a calendar time of 2001-09-19 19:36:23.}
    \begin{tabular}{crcc}
    \toprule
    Interval & Duration [s] & Start time [s] & Stop time [s] \\
    \midrule
    1 &    4000 &    41624.1 &     45624.1 \\
    2 &     950 &    46224.1 &     47174.1 \\
    3 &     700 &    50474.1 &     51174.1 \\
    4 &    4900 &    52724.1 &     57624.1 \\
    5 &     100 &    58124.1 &     58224.1 \\
    6 &    2100 &    61774.1 &     63874.1 \\
    7 &     100 &    64474.1 &     64574.1 \\
    8 &     150 &    65324.1 &     65474.1 \\
    9 &     100 &    66874.1 &     66974.1 \\
   10 &     100 &    67174.1 &     67274.1 \\
   11 &    3000 &    67474.1 &     70474.1 \\
   12 &     300 &    71724.1 &     72024.1 \\
   13 &   23250 &    76674.1 &     99924.1 \\
   14 &     600 &   100724.1 &    101324.1 \\
   15 &     100 &   101524.1 &    101624.1 \\
    \bottomrule
    \end{tabular}
    \label{tab:flr1885}
\end{table}

\subsubsection{Unsupervised Classification for ObsID~10679}\label{subsub:unsupervised10679}

\begin{figure}
    \centering
    \includegraphics[width=\columnwidth]{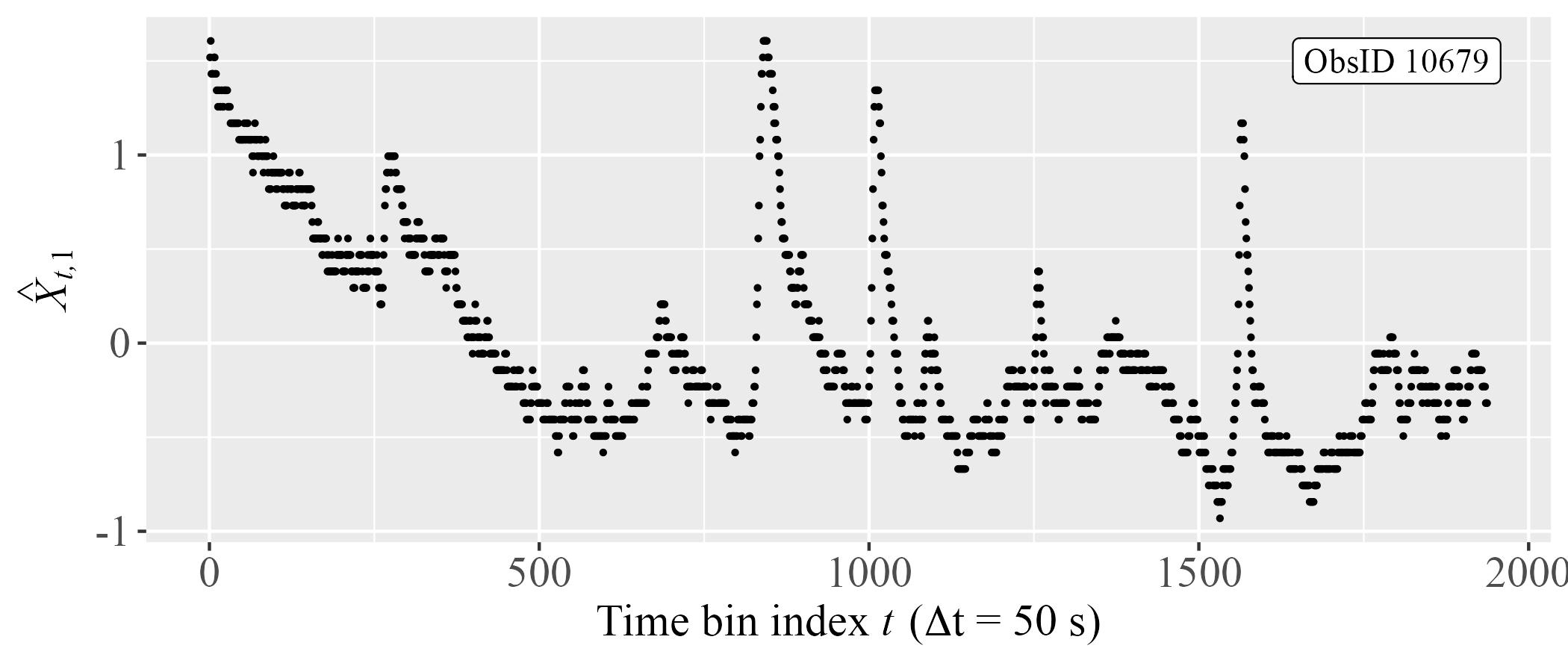}
    \caption{Predicted soft band states $\hat{X}_1, \ldots \hat{X}_{2027}$ for ObsID~10679}
    \label{fig:statepreds79}
\end{figure}

For the light curves in ObsID~10679, there is no clear period of quiescence (see the lower panel of Figure~\ref{fig:EVlacTS1}). The soft-band predicted state variables, $\hat{X}_1, \ldots, \hat{X}_{1937}$, under Model~2 are plotted in Figure~\ref{fig:statepreds79}, again illustrating the lack of a clearly sustained period of quiescence. Thus, we used this data to demonstrate the unsupervised classification method described in Section~\ref{subsub:unsupervised}, fitting a mixture of three normal distributions to the complete set of predicted state variables.

\begin{figure}
    \centering
    \includegraphics[width=\columnwidth]{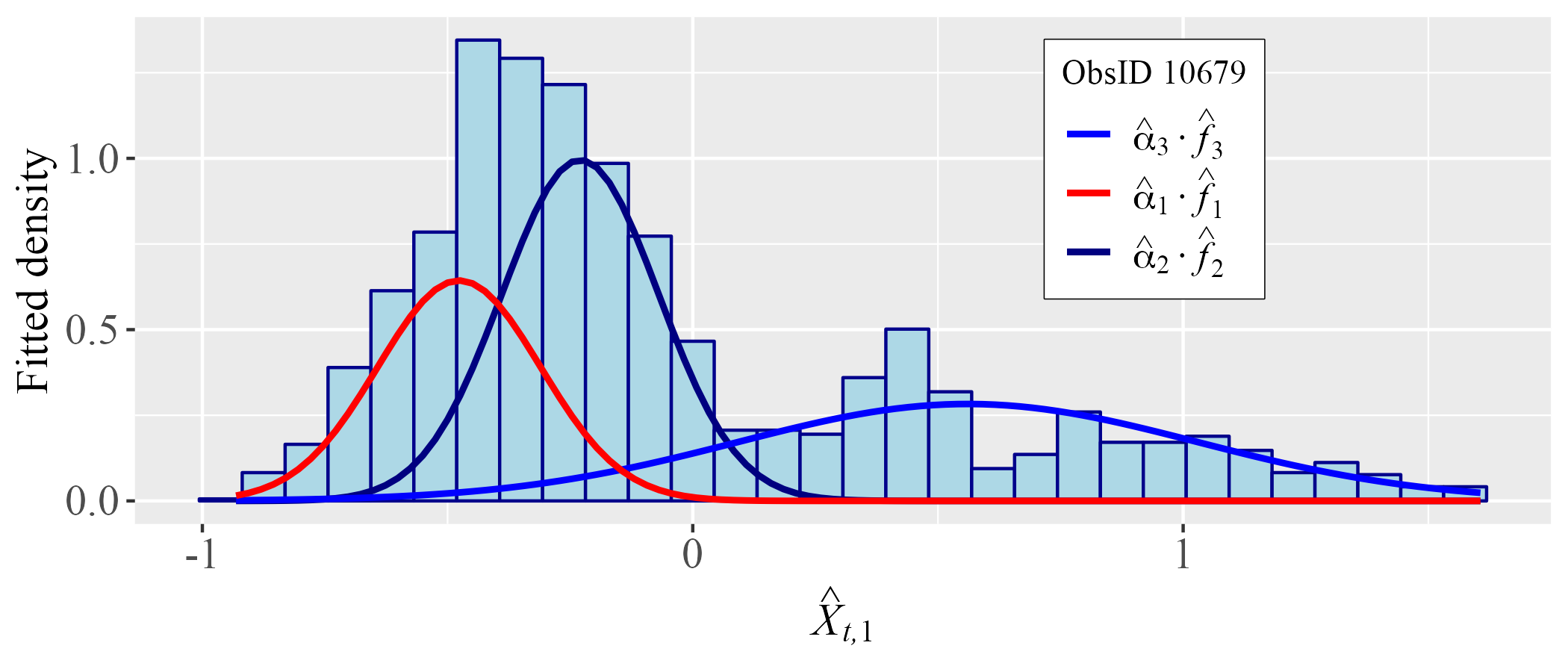}
    \includegraphics[width=\columnwidth]{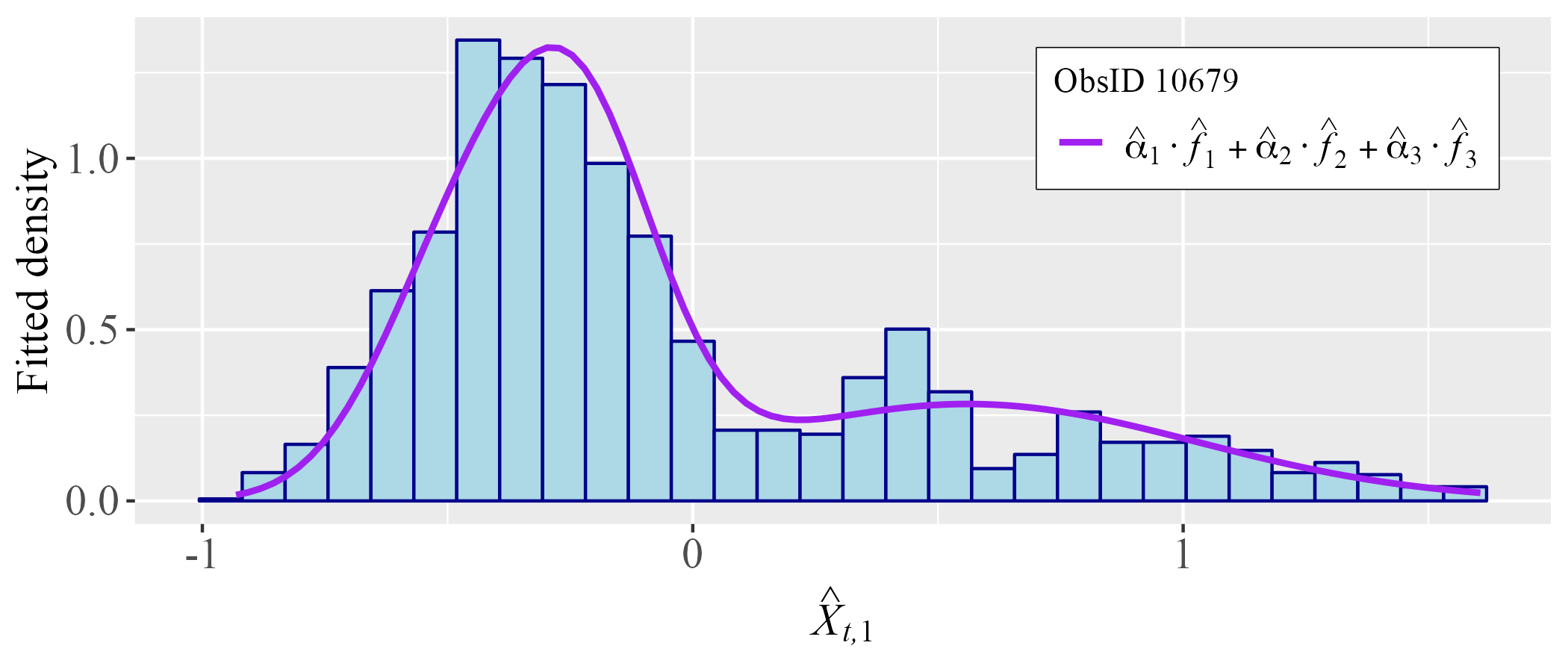}
    \caption{Fitted component densities (above) and mixture density (below) for ObsID~10697; the densities are overlaid on a histogram of $\{\hat{X}_1, \ldots \hat{X}_{1937}\}$}
    \label{fig:FMM_unsupervised}
\end{figure}

\begin{figure}
    \centering
    \includegraphics[width=\columnwidth]{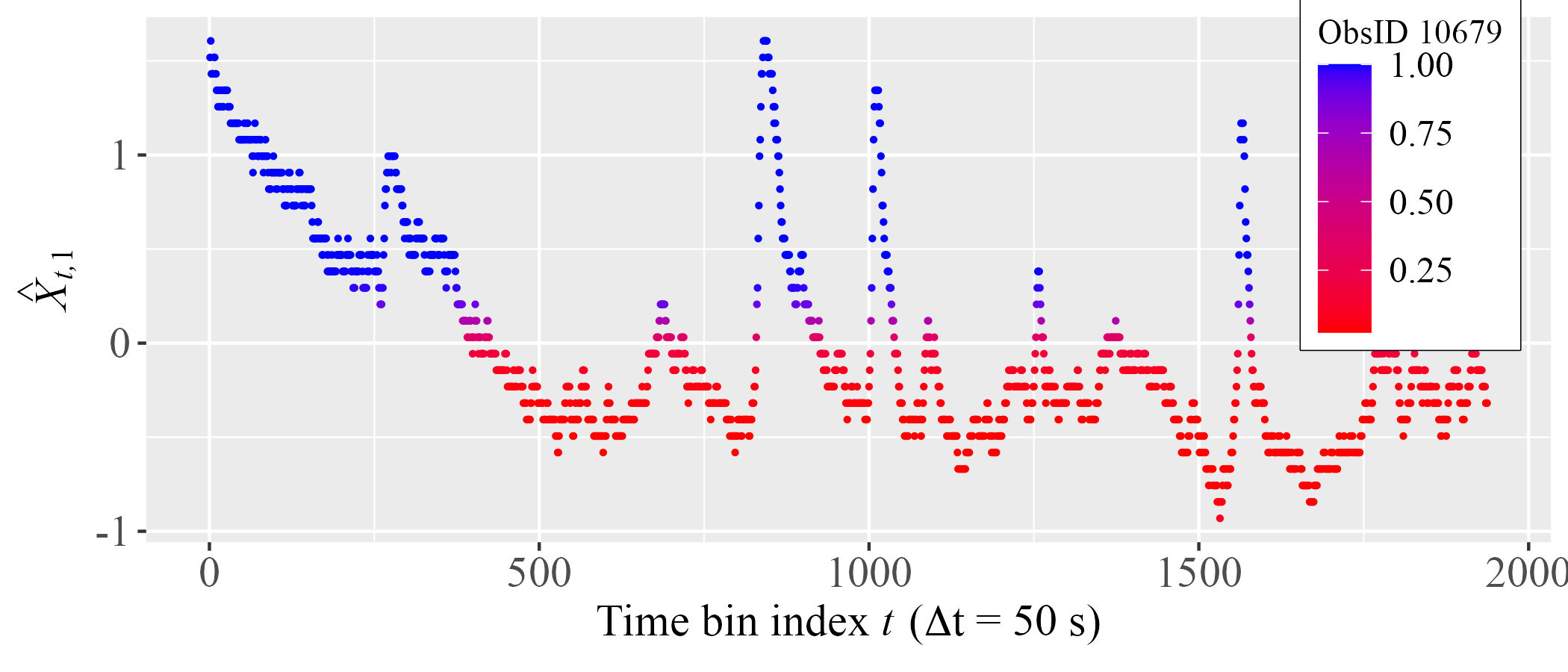}
    \includegraphics[width=\columnwidth]{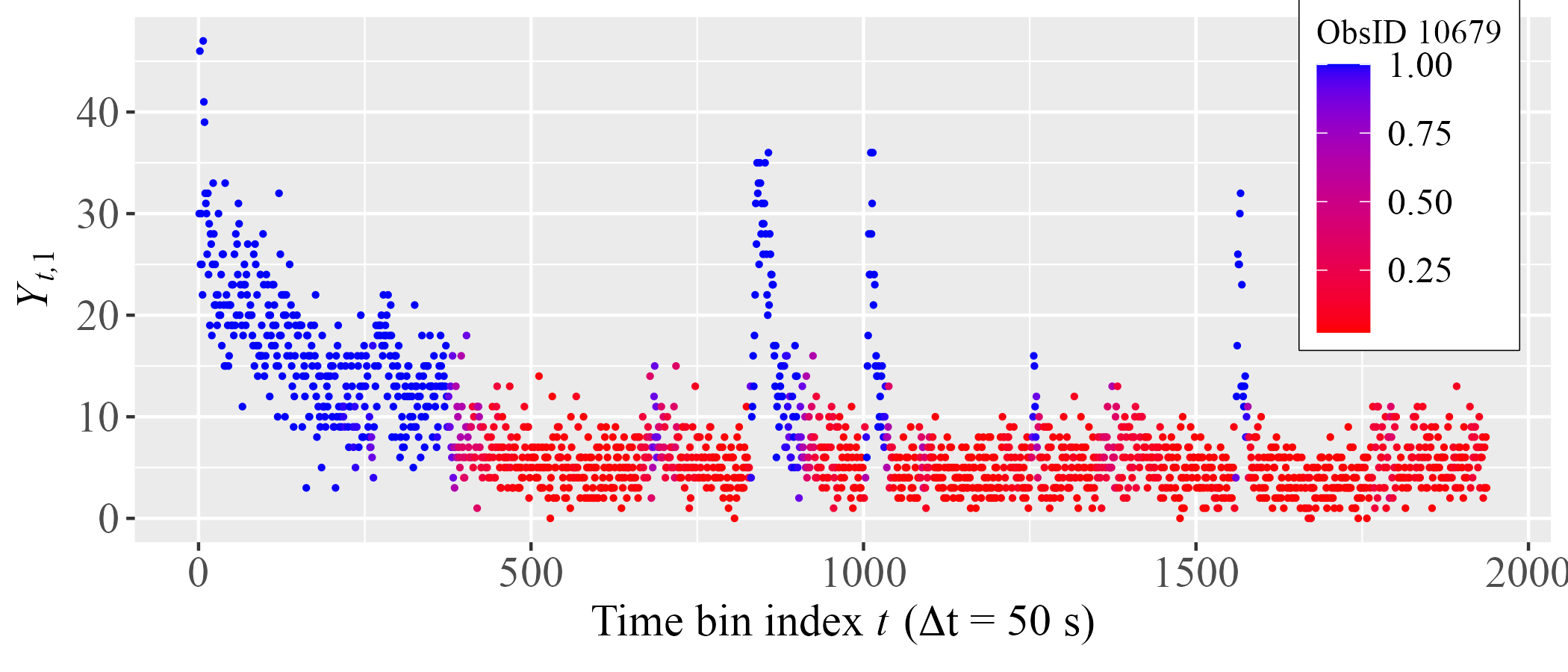}
    \caption{Posterior flaring state probabilities used to colour the predicted states $\hat{X}_1, \ldots, \hat{X}_t$ (above) and the observed soft band data $Y_{1,1},\ldots,Y_{T,1}$ (below) for ObsID~10697}
    \label{fig:posteriors79}
\end{figure}

The estimated parameters of the mixture components are given in Table~\ref{tab:3FMMests} and the estimated component densities and mixture density are shown in Figure~\ref{fig:FMM_unsupervised}. Under the assumption that the third component corresponds strictly to the flaring state, the estimated proportion of time that \evlac spends in a flaring state based on the ObsID~10679 data is $100\% \cdot \hat{\alpha}_3 \approx 27\%$.\footnote{We associate the first two components of the mixture distribution with quiescence because the fitted density shown in Figure~\ref{fig:FMM_unsupervised} indicates considerable overlap between these two components. Alternatively, one could postulate that the second component is composed of both flaring and quiescent states and/or corresponds to the transition between the two states. Thus, $\hat{\alpha}_3$ may slightly underestimate the proportion of time that \evlac spends in its flaring state. Of course, the mixture model is completely agnostic to our own astrophysical interpretations of its components.  Possibly, both the second \emph{and} the third component together correspond to a flaring state; under this assumption, the estimated proportion of time spent in this state is $100\% \cdot (\hat{\alpha}_2 + \hat{\alpha}_3) \approx 60\%$.  In general, the interpretation of the distinction between quiescent and flaring states must be done on a case by case basis, as it can depend on the source, the epoch of observation, and the instrument being used.  The spectral variability analysis presented in Section~\ref{subsub:EVLacstates} strongly supports our interpretation in this case.} Corresponding posterior flare probabilities, which associate the third component of the mixture model \eqref{eq:normalFMM} to the flaring state, are shown in Figure~\ref{fig:posteriors79} 
(and in Figure~\ref{fig:response} for counts in both bands). A binary classification rule nearly identical to that described in Section~\ref{subsub:semisupervised01885} was created, the only difference being that now $\hat{p}_t = \P(Z_t = 3 \mid \hat{X}_t = \hat{x}_t)$; binary quiescent/flaring classifications were again constructed according to \eqref{eq:classrule1}.

\begin{table}
\centering
\begin{tabular}{c|ccc}
  \toprule
Component $k$ & $\hat{\alpha}_k$ & $\hat{\mu}_k$ & $\hat{\tau}_k^2$ \\ 
  \midrule
1 & 0.3988 & -0.2294 & 0.0255 \\ 
  2 & 0.3328 & 0.5608 & 0.2202 \\ 
  3 & 0.2683 & -0.4764 & 0.0277 \\ 
   \bottomrule
\end{tabular}
\caption{Parameter estimates for the 3-component mixture of normal distributions}
\label{tab:3FMMests}
\end{table}

\begin{table}
    \centering
    \caption{Flaring time intervals for ObsID 10679, in spacecraft clock time.  The times are offset from the observation start time of 353314077.3~s, corresponding to a calendar time of 2009-03-13 06:47:57.}
    \begin{tabular}{crcc}
    \toprule
    Interval & Duration [s] & Start time [s] & Stop time [s] \\
    \midrule
    1 &   19850 &     1742.7 &     21592.7 \\
    2 &     250 &    21792.7 &     22042.7 \\
    3 &     150 &    22742.7 &     22892.7 \\
    4 &     600 &    35792.7 &     36392.7 \\
    5 &    4300 &    43192.7 &     47492.7 \\
    6 &     100 &    47892.7 &     47992.7 \\
    7 &    1800 &    51842.7 &     53642.7 \\
    8 &     150 &    56142.7 &     56292.7 \\
    9 &     450 &    64392.7 &     64842.7 \\
   10 &     100 &    70392.7 &     70492.7 \\
   11 &    1000 &    79692.7 &     80692.7 \\
    \bottomrule
    \end{tabular}
    \label{tab:flr10679}
\end{table}

\subsubsection{The Quiescent and Flaring States of EV Lac}\label{subsub:EVLacstates}

We also carried out sensitivity checks on the flaring intervals determined as above in Sections~\ref{subsub:semisupervised01885} and~\ref{subsub:unsupervised10679} by jittering the phase of the binning by $\pm 25$~s, changing the passband intervals (using $0.3$--$0.9$~keV for the softer and $0.9$--$8.0$~keV for the harder bands), and checking different time bin widths (see Appendix~\ref{app:othertimebins}). We found that the flaring intervals thus determined remain stable and repeatable to within $2$--$3$ time bin widths in all cases. We thus adopted a $3\times$ time bin width as a nominal systematic uncertainty on the intervals, and merged all gaps smaller than that. We further inflated the intervals by adding $25$~s (half the width $w$ of the adopted time bins) both before and after the ends of each interval. This resulted in $15$ distinct intervals for ObsID~01885 and $11$ intervals for ObsID~10679 (see Tables~\ref{tab:flr1885} and \ref{tab:flr10679}, respectively). The durations of the interval correspond to approximately $30$\% and $40$\% of the total observation interval for the first and second epochs respectively. This is consistent with the expected flare rates seen on \evlac before: flares occurring at rates of $0.2$--$0.4$~hr$^{-1}$ \citep{huenemoerder2010x} lasting approximately 5~ks cover a fraction of $0.28$--$0.55$ of the exposure durations, assuming no overlaps. Note that our method does not distinguish the number of flares within a flare state (e.g., the first interval in ObsID~10679 covers a duration that clearly includes a smaller flare that overlaps another with a longer decay time scale).

This separation between flaring and quiescent states allows us to explore changes in the energy spectrum of the star. The overall spectrum is well fit with a 2-temperature component {\tt xsapec} model in CIAO/Sherpa v4.16 \citep{2009pysc.conf...51R} with similar temperature, abundance, and normalisations for both epochs (see Table~\ref{tab:fitparams}).

\begin{table*}
    \centering
    \caption{Sherpa 2-temperature {\tt apec} model fits to the full spectra}
    \begin{tabular}{lcccccc}
    \toprule
    ObsID & $T_\textrm{low}$ & $T_\textrm{high}$ & Metallicity & Norm$_\textrm{low}$ & Norm$_\textrm{high}$ & {\tt cstat}/dof \\
    \hfil & \multicolumn{2}{c}{[keV]} & Z$_\odot$ & \multicolumn{2}{c}{[${\times}10^{14}$~cm$^{-5}$]} & \hfil \\
    \midrule
    01885 & $0.35{\pm}0.0024$ & $1.26{\pm}0.007$ & $0.17{\pm}0.004$ & $0.016{\pm}0.0004$ & $0.0099{\pm}0.0001$ & 24850.1/24980 \\
    10679 & $0.35{\pm}0.003$ & $1.35{\pm}0.009$ & $0.17{\pm}0.005$ & $0.015{\pm}0.0005$ & $0.0095{\pm}0.0001$ & 21668.7.1/24980 \\
    \bottomrule
    \end{tabular}
    \label{tab:fitparams}
\end{table*}

\begin{figure*}
    \centering
    \includegraphics[width=0.48\linewidth]{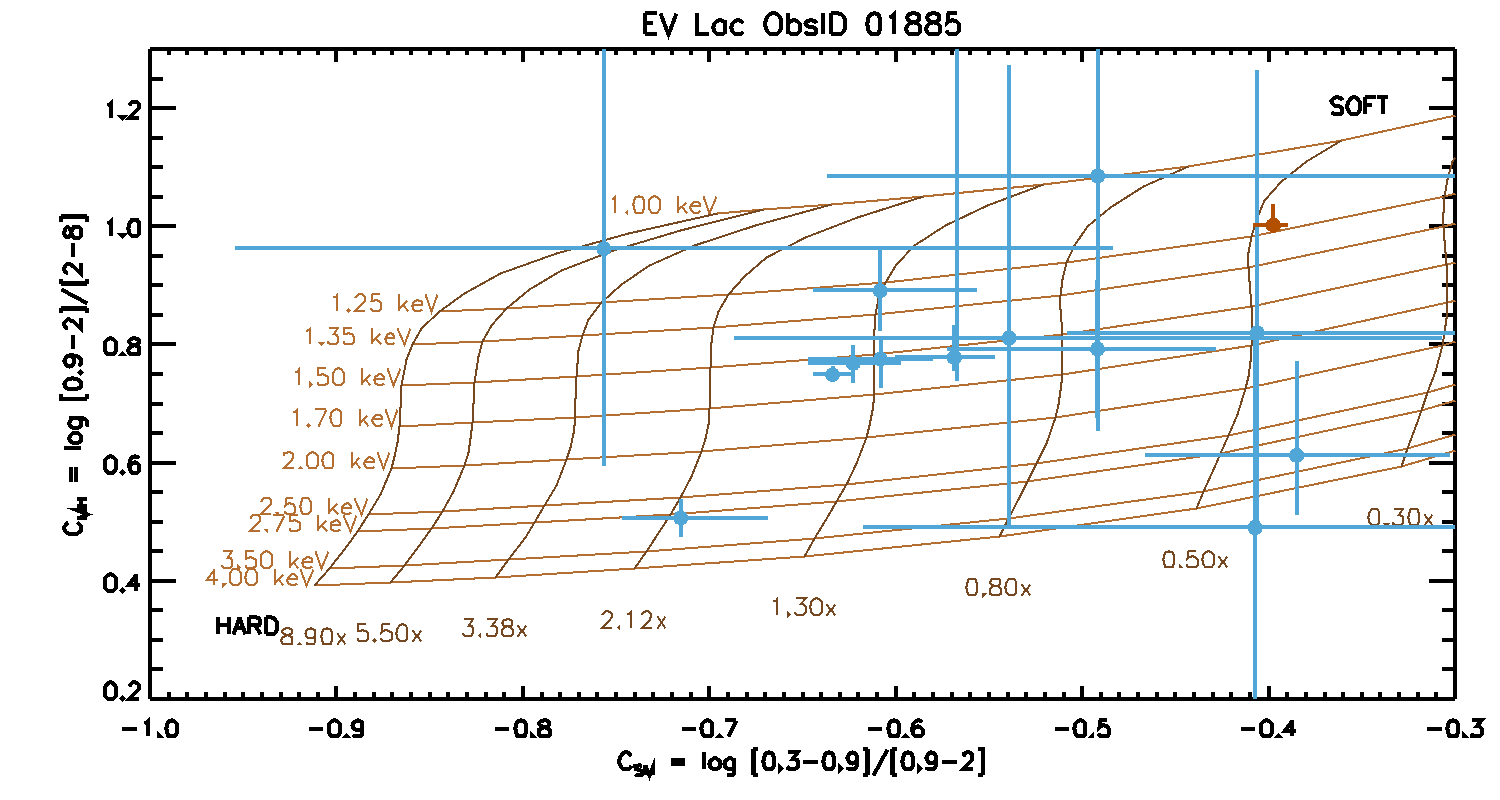}
    \includegraphics[width=0.48\linewidth]{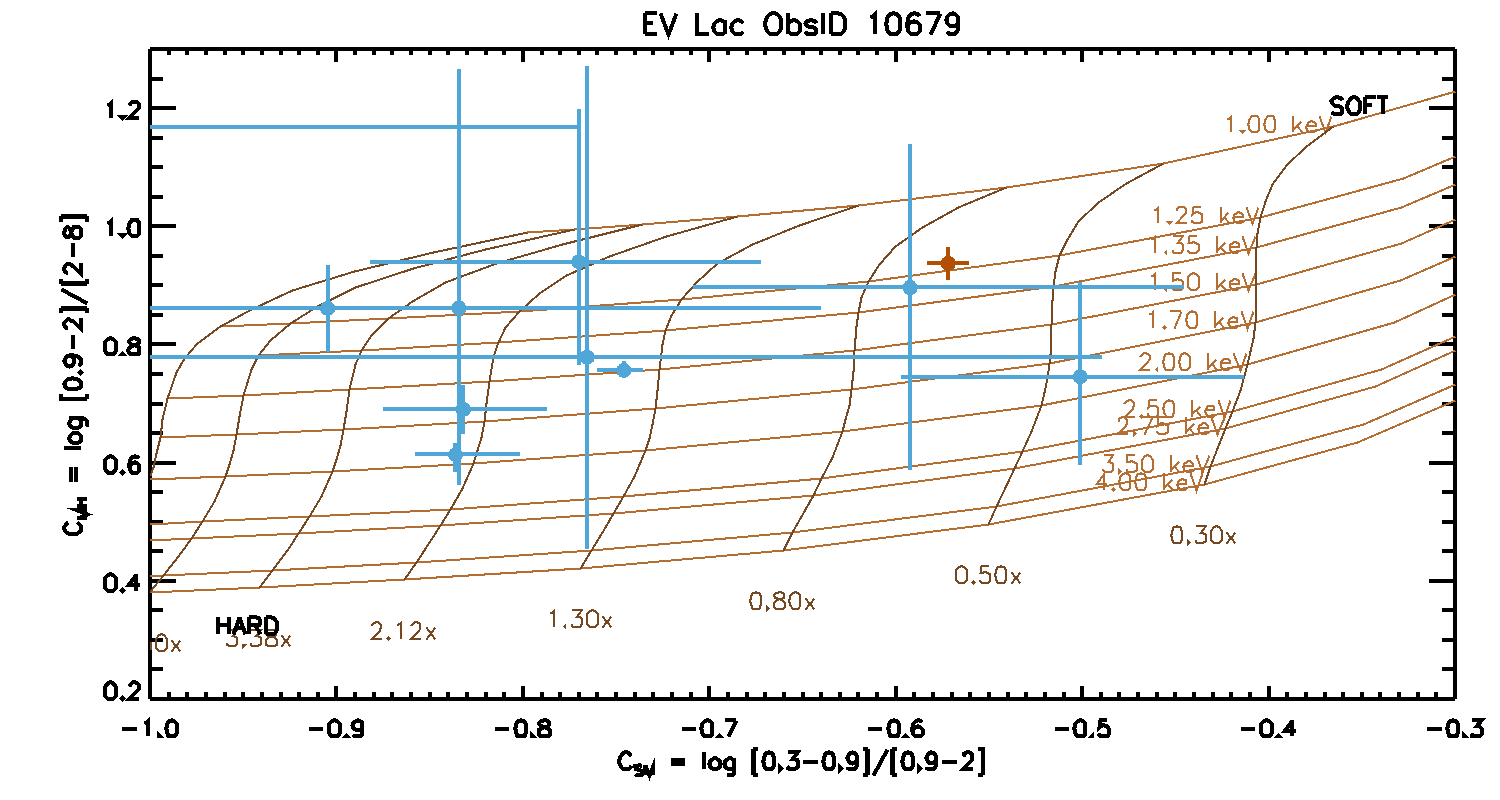}
    \caption{Spectro-temporal treatment of flaring. For both \evlac datasets, the panels show the hardness ratio colours $C_{SM}=\log(S/M)$, $C_{MH}=\log(M/H)$, where $S$, $M$, and $H$ are counts in passbands $0.3$--$0.9$~keV, $0.9$--$1.2$~keV, and $1.2$--$8.0$~keV respectively. The colours during each distinct flaring interval (blue points with error bars) are compared to the quiescent interval (sole red point with error bars). The grid in the background (brown curves) shows the predicted colours for spectra with two temperature components: a low temperature plasma at $T_\textrm{low}=0.35$~keV ($\approx$4~MK), and a high temperature component with a variety of temperatures $T_\textrm{high}$ ranging from $1$~keV ($\approx$12~MK) to $4$~keV ($\approx$46~MK), with the relative emission measure of the high temperature component ranging between $0.1$ to $8.9$ times that of the low temperature component. We adopt a metallicity of $0.16$, commensurate with a 2-temperature {\tt apec} fit to the spectra. Note that in both epochs, the quiescent interval has a softer spectrum than any of the flaring intervals. The shift in the grid is due to changes in ACIS effective area between the epochs.}
    \label{fig:EVLac_lc_color}
\end{figure*}

Figure~\ref{fig:EVLac_lc_color} shows the changes in spectral colour for each of the flare intervals (marked in blue) compared to the combined quiescent interval (marked in red); all error bars were computed using BEHR \citep{2006ApJ...652..610P}). The colours were computed as log-ratios of counts in the soft ($S$: $0.3$--$0.9$~keV) to medium ($M$: $0.9$--$2.0$~keV) and medium to hard ($H$: $2.0$--$8.0$~keV) bands. It is clear that all of the flaring intervals have harder spectra than the quiescent spectrum. The underlying grid, constructed for a 2-temperature {\tt apec} model as for the full spectra (see Table~\ref{tab:fitparams}) but with varying normalization and temperature for the high-temperature component, also demonstrates this quantitatively. The flaring intervals include the low-temperature component because the flares are likely confined to small regions in the corona, so that the quiescent corona continues to contribute to the emission, even as the emission is dominated by the flare. Note that the grids shift leftwards from the earlier epoch to the later, which is a consequence of the increased contamination buildup on the ACIS detector which reduces the soft effective area.

\begin{figure*}
    \centering
    \includegraphics[width=\linewidth]{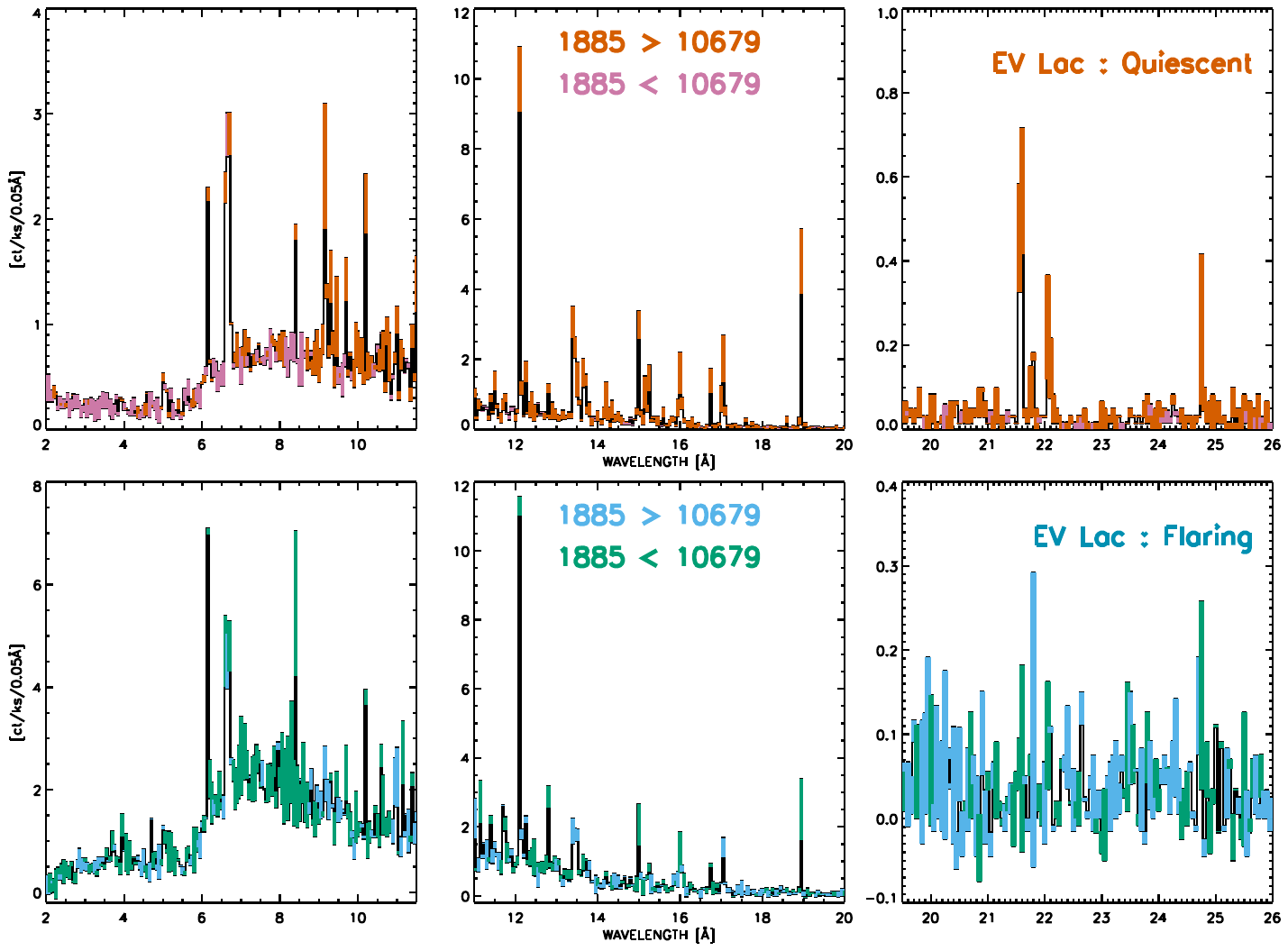}
    \caption{Comparing high-resolution spectra of quiescent (top) and flaring (bottom) states of active dMe star \evlacnospace.  The quiescent spectra are subtracted out from the corresponding spectra obtained during the flaring state. Each state is split into three panels in order to better show weak lines. Spectra from the two epochs are shown superposed for both cases; the difference between the epochs is marked in red/pink (quiescent) and blue/green (flaring) shades depending on which epoch had more counts within a given bin (see label in middle panels). Prominent lines from several species are visible in both spectra, with some resonance line ratios changing from quiescence to a flaring state favouring the higher ionization state (see especially Si\,XIV/Si\,XIII at $6.2$ and $6.7$~\AA, Mg\,XII/Mg\,XI at $8.4$ and $9.2$~\AA). The overall brightness is higher, and the continuum is stronger and more prominent during the flaring state, signifying a different thermal signature. The differences between the quiescent and flaring states are greater than the differences between epochs for the same state, which suggests that there are distinct quiescent and flaring states present on the star.}
    \label{fig:QvFspectra}
\end{figure*}

Finally, we show in Figure~\ref{fig:QvFspectra} the full resolution combined HEG+MEG first-order spectra separately for the quiescent (upper panels) and flaring states (lower panels). Spectra from both epochs are overplotted, and deviations where the counts from one epoch exceed the other are marked in different shades. As is expected from the evolution in the soft effective area, the earlier epochs have systematically higher counts at longer wavelengths. The spectra are dominated by several prominent lines, such as those from Ne\,X ($12.15$~\AA), Fe\,XVII ($15.01$, $17.05$~\AA), and O\,VIII ($18.96$~\AA) (see middle panels of Figure~\ref{fig:QvFspectra}). The density- and temperature-sensitive He-like O\,VII triplet ($21.6$, $21.8$, $22.1$~\AA\ of the resonance, intercombination, and forbidden lines) is visible in the right panels; higher density plasma is present in the flaring state, as shown by the higher ratio of the intercombination to forbidden lines. In the left panels, several high-temperature lines appear during the flaring state at short wavelengths (Ar\,XVIII $2.92$~\AA, Ar\,XVII $3.95$~\AA, S\,XVI $4.73$~\AA, S\,XV $5.0$~\AA). The ratios of the temperature sensitive resonance lines of Si\,XIV ($6.2$~\AA) and Si\,XIII ($6.74$~\AA), and Mg\,XII ($8.4$~\AA) and Mg\,XI ($9.2$~\AA) change to favour the higher temperature species and the continuum becomes more prominent, all indicating the presence of higher temperature plasma, and thus supporting the conclusions of \cite{huenemoerder2010x}.

In addition, the Ne\,X/O\,VIII counts ratio increases from $2.1$ during quiescence to $3.5 \pm 0.2$ in the first epoch, and from $2.6$ to $3.4 \pm 0.3$ during the second epoch. The Ne\,X/Fe\,XVII counts ratio also increases, from approximately $2.8$ during quiescence to approximately $3.5$--$4.0$ during flaring in both epochs, indicating that there could be an increase in Ne abundance during flaring. In contrast, the O\,VIII/Fe\,XVII counts ratio decreases by approximately $10$\% during flaring in both epochs; detailed modelling is necessary to establish whether this decrease is simply a temperature effect or whether oxygen abundance variations are also required to explain it.

Crucially, the differences between epochs for each state are minuscule compared to the changes seen between the quiescent and flaring states. This is a strong indication that our method can clearly identify and separate these states. Furthermore, the similarity in the apparent thermal characteristics in both states, as evidenced by the similar shapes of the continuum, shows that the two states are strongly differentiated: that is, the star has a very well defined quiescent state, suggesting that there may be a distinct heating mechanism that operates during quiescence.

\section{Discussion and Future Work}\label{sec:discussion}

This paper combines state-space models and finite mixture models as a means of classifying periods of quiescence and flaring in multi-band astronomical light curves. Specifically, we apply our models to high-energy X-ray data of the active binary \evlacnospace, grouping the photons into two passbands and classifying the light curves into flaring and quiescent states. In Stage~1 of our analysis, our state-space models (HMMs) assume that the underlying physical process driving the flaring activity can be represented by a Markov chain defined on a continuous multidimensional state space. When the component of the Markov chain corresponding to a particular energy band migrates to higher or lower values, the rate of photon emissions within that band tends to increase or decrease in kind. We propose a series of nested HMMs to capture this underlying process with increasing levels of generality. We tabulate emissions in the soft and hard energy bands separately in order to capture the more complete information contained in the bivariate data. The state-space models allow us to predict the individual states of the underlying chain that are most likely to have generated the observed data. Using finite mixture models in Stage~2, we devise two situation-specific schemes to classify the predictions and ultimately dichotomize the observed time periods into flaring and quiescent intervals. 

\subsection{Quiescence}\label{sec:quiescence}

We demonstrate our method on two sets of observations of the dMe star \evlacnospace, leading to a clear separation of flaring activity and quiescence, as well as to the discovery of a well-defined and persistent quiescent state. The presence of such persistent quiescent emission in counterpoint to flaring has been recognized and analyzed ubiquitously in astronomical literature. Exemplar treatments include that of the Sun by \cite{2008A&A...488.1069A} and of an active M dwarf YZ\,CMi by \cite{2007MNRAS.379.1075R}; see also a review by \cite{2004A&ARv..12...71G}. The possibility of a persistent quiescent state has also been suggested for the active binary AR\,Lac \citep{2014ApJ...783....2D}, and for the young stellar binary XZ\,Tau \citep{2023AJ....166..148S}.  Our analysis of spectral variability supports the idea that steady and persistent non-flaring emission is present even on active stars. 

The continuous-space HMMs that we propose in our Stage~1 analysis (Section~\ref{sec:HMMs}) do not alone clearly differentiate between the quiescent and active states of the source, instead allowing for variability within the states and a smooth transition between them. The time intervals during which flaring emission dominates are identified from the distribution of the fitted HMM states in our Stage~2 analysis (Section~\ref{sec:classify}).  Alternatively, we could posit a model where the quiescent emission is present at all times, with the intermittent and variable flaring emission (presumably arising in localized active regions on the star) superposed over it. For example, the observed counts could be modelled as the sum of two Poisson processes, the first an iid process representing quiescence and the second representing the flare state alone. Such a model would be more complex than the HMMs we consider here in that its second Poisson process (for the flaring state) would be as complex as the HMMs we propose in Section~\ref{sec:model}. As we expect that the flexibility of the continuous-space HMM may render the more complex model unidentifiable or only weakly identifiable, we leave its study for future research.

\subsection{Future Directions} \label{sec:discussion-improve}

We propose several avenues for future modifications and generalizations of our HMMs. The discrete-space HMM approximation to the state-space likelihood developed by \cite{langrock2011some} is, in the end, only an approximation, which can potentially be made more accurate using adaptive binning \citep{borowska2023semi} or other procedures that further refine the discretization of the continuum  (i.e., the choice of essential domain and the partition thereof). From a computational perspective, it would also be desirable to eliminate the need for manual verification that the essential domain adequately covers the distribution of the underlying Markov chain.

The state-space models themselves can be modified or augmented with additional features. In Section~\ref{sec:cont-HMM}, for example, we discuss the use of state-dependent bivariate distributions for the observed data. In the general case, this avoids a conditional independence assumption for the hard and soft energy bands, and allows for more involved bivariate distributions capable of capturing potential dependence between the bands at the observed data level. Even more generally, one could split the counts into any number $d$ of bands (the hard and soft bands we used for \evlac correspond to $d=2$). The $d$-band generalization of Model~2 is straightforward: for each additional band $h$, we introduce one new parameter $\beta_h$ controlling the Poisson rate for $Y_{t,h}$, as well as a rescaling parameter $\sigma_h$ so that $X_{t,h} = \sigma_h X_{t,1}/\sigma_1$. The generalization of Model~3 is more challenging due to the increased complexity of the (non-diagonal) covariance matrix $\bSigma$ in the error terms: in addition to new parameters $\beta_h$, $\phi_h$, and $\sigma_h$, each band $h$ requires pairwise correlation terms with every other dimension, resulting in a $d\times d$ covariance matrix $\bSigma$. Depending on the covariance structure selected for the model, $\bSigma$ can include as few as two free parameters (for a first-order autoregressive covariance) or as many as $(d^2 + d)/2$ (for a completely unstructured covariance). For large $d$, this would effectively model the evolution of the spectrum over time. This model is in contrast to Automark, which looks for breakpoints in the spectrum but assumes the spectrum is unchanging between breakpoints \citep{wong2016detecting}.

It is also possible to generalize the distribution of the state process $\bX\oneT$ by replacing the multivariate normal distributions with other multivariate distributions. For instance, one could account for potentially heavier tails in the distribution of $\bX_t \mid \bX_{t-1}$ by assuming a multivariate $t$-distribution; alternatively, one could assume that $\bX_t \mid \bX_{t-1}$ follows a mixture of conditional multivariate normal distributions with common mean $\bX_{t-1}$ but differing variances, which could potentially model a discrete latent process taking place in some physical process within the star itself. Both of these generalize the multivariate normal distribution, and their associated stationary distributions are available \citep[e.g.,][]{meitz2023mixture}; however, stationary distributions corresponding to other choices of the distribution of $\bX_t \mid \bX_{t-1}$ may not be known, and therefore a different distribution would be needed for the initial state $\bX_0$. The effect of this choice is likely small with large datasets.

Even when adhering to multivariate normal conditional distributions for the state process, our models can be generalized in several other ways. For example, the $\text{VAR}(1)$ model \eqref{eq:uncorVAR1} can be generalized to allow $\Phi$ to be a generic asymmetric non-diagonal matrix; in this case, stationarity is characterized by a rather complex set of nonlinear constraints on the entries of $\Phi$. This generalization would allow for dependence of $X_{t,1}$ on $X_{t-1,2}$, and vice versa. Such dependencies can be used to capture physical processes where hot coronal plasma in a magnetic flux tube cools sequentially from higher to lower temperatures \citep[e.g.,][]{2012ApJ...753...35V}. One can also consider more general $\text{VAR}(p)$ processes (i.e., where the distribution of $\bX_t$ depends linearly on $\bX_{t-1}, \ldots, \bX_{t-p}$). Any discrete-time stochastic process $(X_t)_t$ on a state-space $\sX$ for which the distribution of $X_t$ depends on the history of the chain through $X_{t-1}, \ldots, X_{t-p}$ (a so-called \emph{higher-order} Markov chain) induces a standard vector-valued Markov chain $(\bX'_t)_t$ on $\sX^{p}$, and so, in principle, a $\text{VAR}(p)$ process on $\sX^d$ can be recast as a first-order matrix (or ``tensor'') autoregressive process on $\sX^{d \times p}$ for which the discrete-space HMM approximation can be applied. However, the calculations required for the initial distribution and transition probabilities would involve the so-called matrix normal distribution, which can be quite computationally involved.

Additionally, rather than binning the photon counts into discrete intervals, one could model the series of photon counts directly in continuous time. This would involve modelling the exponentially-distributed waiting time between the Poisson process of photon arrivals. The underlying state process would evolve in continuous time and could be modelled as an Ornstein-Uhlenbeck (OU) process, the continuous-time analogue of the $\text{AR}(1)$ process. The OU process has been applied in astrophysical settings by \cite{kelly2009variations,kelly2014flexible} and \cite{meyer2023td}; such processes generalize fairly naturally to the multivariate case \citep{gardiner2004handbook}. Perhaps the most natural continuous-time analogue of our state-space model is a bivariate time-heterogeneous Poisson process \citep{cox1972multivariate} whose parameters are driven by components of the aforementioned Ornstein-Uhlenbeck process.

Finally, our methods can also be generalized to apply to sources other than stars that also exhibit intermittent or episodic flaring (e.g., Sgr\,A*, the jet of M87, or dipping sources such as LMXBs). Such generalizations would require our underlying HMMs to be extended in order to model additional passbands and their possible correlations.

\section*{Acknowledgements}

This work was conducted under the auspices of the CHASC International Astrostatistics Center. CHASC is supported by DMS-21-13615, DMS-21-13397, and DMS-21-13605; by the UK Engineering and Physical Sciences Research Council [EP/W015080/1]; and by NASA grant 80-NSSC21-K0285. RZ was supported by a Mitacs Globalink Research Award. DvD was supported in part by a Marie-Sk\l odowska-Curie RISE (H2020-MSCA-RISE-2019-873089) Grant provided by the European Commission. VLK and AS acknowledge support from NASA Contract NAS8-03060 to the {\sl Chandra X-ray Center}. We thank Yanbo Tang for helpful discussions.

\section*{Data Availability}

This paper employs a list of \chandra datasets, obtained by the Chandra X-ray Observatory, contained in the Chandra Data Collection (CDC) 235~\href{https://doi.org/10.25574/cdc.235}{doi:10.25574/cdc.235}.

Data were obtained and reduced using CIAO \citep{2006SPIE.6270E..1VF}.  Spectral fitting and parameter grid calculations were carried out using Sherpa \citep{2001SPIE.4477...76F,2009pysc.conf...51R}.  Hardness ratios were computed using BEHR \citep{2006ApJ...652..610P}. Flux estimates and line identifications were carried out in PINTofALE \citep{2000BASI...28..475K}. All state-space modelling was conducted in \texttt{R} \citep{Rsoftware2023} with the aid of several packages; all \texttt{R} code is publicly available at \href{https://github.com/rob-zimmerman/SSM-flare}{github.com/rob-zimmerman/SSM-flare}.


\bibliographystyle{mnras}
\bibliography{References}

%


\appendix
\onecolumn

\section{Algorithms for Discrete-Space HMMs}\label{app:HMMalgos}

Likelihood computation and state decoding for discrete-space HMMs generally rely on several related algorithms, including the \emph{forward algorithm}, the \emph{backward algorithm}, and the \emph{forward-backward algorithm}. Here we briefly derive these three algorithms, each of which can be succinctly described by an iterated sequence of matrix multiplications; each step of the algorithms thus involves multiplying the matrix obtained in the previous step by a new matrix, yielding in the end a product of matrices which we show to be equivalent to a quantity of interest, such as the value of the likelihood function. The algorithms use dynamic programming to efficiently compute certain quantities; for example, a na\"ive computation of \eqref{eq:DSHMMlike} via direct summation would require a number of operations exponential in $T$, while the forward algorithm reduces the computation to being only polynomial in $T$. In the HMM literature \citep[e.g.,][]{zucchini2017hidden} these algorithms are typically expressed in terms of quantities known as forward and backward variables, but these are unnecessary for our purposes as the relevant equations can be succinctly expressed in terms of matrices alone. Nevertheless, the forward and backward variables play key roles in the theory of HMMs; interested readers may consult \cite{rabiner1989tutorial}, \cite{CappeHMM}, and \cite{zucchini2017hidden}.

To simplify notation, we assume that the $\by_t$ are discrete in our derivations, as is the case in the Poisson models developed in Section~\ref{sec:model}. Nonetheless, our calculations carry through verbatim for continuous observations, with probability mass functions replaced by their analogous density functions.

\subsection{Likelihood Computation via the Forward Algorithm}\label{subsub:fwdalgo}

The forward algorithm for discrete-space HMMs evaluates the HMM likelihood $L(\bveta \mid \by\oneT)$ given by \eqref{eq:DSHMMlike} via an efficient computation of the right-hand side of the identity 
\begin{equation}
    L(\bveta \mid \by\oneT) = {\bb{P}}_{\bveta}(\bY\oneT = \by\oneT).
\end{equation}
For the remainder of this section, we drop the subscript $\bveta$ from $\mathbb{P}_{\bveta}(\cdot)$ for notational simplicity; however, all probabilities should be understood as being taken with respect to the model with parameter $\bveta$. 

Define the matrix-valued function $\bP: \sY \to [0,\infty)^{K \times K}$ by
\begin{equation}
    \bP(\by) = \begin{bmatrix}
        h_1(\by \mid \blambda_1) & 0 & \cdots & 0\\
        0 & h_2(\by \mid \blambda_2) & \cdots & 0\\
        \vdots & \vdots & \ddots & \vdots \\
        0 & 0 & \cdots & h_K(\by \mid \blambda_K) \\
    \end{bmatrix};
\end{equation}
that is, $\bP(\by)$ is a diagonal matrix with the state-dependent mass/density functions $h_1, \ldots, h_K$ evaluated at $\by$ along its diagonal. Recalling that $h_k(\by \mid \blambda_k) = {\bb{P}}(\bY_t = \by \mid X_t = k)$ for any $t$ we have
\begin{equation}\bdelta^\top \bP(\by_1) 
= 
\begin{bmatrix}
    {\bb{P}}(X_1 = 1) \cdot {\bb{P}}(\bY_1 = \by_1 \mid X_1 = 1) \\
    \vdots \\
    {\bb{P}}(X_1 = K) \cdot {\bb{P}}(\bY_1 = \by_1 \mid X_1 = K) \\
\end{bmatrix}^\top
= 
\begin{bmatrix}
    {\bb{P}}(\bY_1 = \by_1, X_1 = 1) \\
    \vdots \\
    {\bb{P}}(\bY_1 = \by_1, X_1 = K) \\
\end{bmatrix}^\top
\end{equation}
and
\begin{align}
\begin{split}
\bdelta^\top \bP(\by_1) \bGamma \bP(\by_2)
&= 
\begin{bmatrix}
    \sum_{k=1}^K {\bb{P}}(\bY_1 = \by_1, X_1 = k) \cdot {\bb{P}}(X_2 = k \mid X_1 = 1) \cdot {\bb{P}}(\bY_2 = \by_2 \mid X_2 = k) \\
    \vdots \\
    \sum_{k=1}^K {\bb{P}}(\bY_1 = \by_1, X_1 = k) \cdot {\bb{P}}(X_2 = k \mid X_1 = K) \cdot {\bb{P}}(\bY_2 = \by_2 \mid X_2 = k) \\
\end{bmatrix}^\top \\
&= 
\begin{bmatrix}
    {\bb{P}}(\bY_1 = \by_1, \bY_2 = \by_2, X_2 = 1) \\
    \vdots \\
    {\bb{P}}(\bY_1 = \by_1, \bY_2 = \by_2, X_2 = K)\\
\end{bmatrix}^\top.
\end{split}
\end{align}
The forward algorithm iterates this matrix multiplication, and by induction at iteration $t\geq 2$, the algorithm returns
\begin{equation}
    \bdelta^\top \bP(\by_1) \prod_{s=2}^t \bGamma \bP(\by_s)
    =
    \begin{bmatrix}
    {\bb{P}}(\bY_1 = \by_1, \bY_2 = \by_2, \cdots, \bY_t = \by_t, X_t = 1) \\
    \vdots \\
    {\bb{P}}(\bY_1 = \by_1, \bY_2 = \by_2, \cdots, \bY_t = \by_t, X_t = K) \\
\end{bmatrix}^\top
= 
\begin{bmatrix}
    {\bb{P}}(\bY\onet = \by\onet, X_t = 1) \\
    \vdots \\
    {\bb{P}}(\bY\onet = \by\onet, X_t = K) \\
\end{bmatrix}^\top.
\end{equation}
Finally, the likelihood in (\ref{eq:DSHMMlike}) can be computed as
\begin{equation}\label{eq:fwdalgofinal}
    \mleft(\bdelta^\top \bP(\by_1) \prod_{s=2}^T \bGamma \bP(\by_s)\mright)\bone = \sum_{k=1}^K {\bb{P}}(\bY\oneT = \by\oneT, X_T = k) = {\bb{P}}(\bY\oneT = \by\oneT)
    = L(\bveta \mid \by\oneT).
\end{equation}
Thus, the forward algorithm computes the likelihood via the left-hand side of \eqref{eq:fwdalgofinal}. (In practice, one must usually rescale the probabilities with each additional matrix product to prevent numerical underflow.) This computation has a time complexity of $O(TK^2)$. Note that as a byproduct of the likelihood computations, the forward algorithm also yields the joint probabilities $\bb{P}(\bY\onet = \by\onet, X_t = k)$ for any $t \in \{1,\ldots,T\}$, which are used in the local decoding algorithm (see Appendix~\ref{subsub:fwdbckwdalgo}).

\subsection{Local Decoding via the Forward-Backward Algorithm}\label{subsub:fwdbckwdalgo}

The \emph{forward-backward algorithm} for discrete-space HMMs evaluates the conditional state-membership probabilities given the full data set (i.e., ${\bb{P}}(X_t = k \mid \bY\oneT = \by\oneT)$, for each $k \in \sX$ and each $t \geq 1$); these are then used to predict the state variables via \eqref{eq:predictedstates}, as we describe below. The forward-backward algorithm itself comprises of two sub-algorithms -- the forward algorithm, which computes the joint probabilities $\bb{P}(\bY\onet = \by\onet, X_t = k)$, and the \emph{backward algorithm}, which computes the conditional probabilities ${\bb{P}}(\bY_{t:T} = \by_{t:T} \mid X_{t-1} = k)$ for each $t \geq 2$. A final combination of the forward and backward algorithms yields the desired conditional state-membership probabilities (i.e., ${\bb{P}}(X_t = k \mid \bY\oneT = \by\oneT)$). The forward algorithm, which also outputs the HMM likelihood $L(\bveta \mid \by\oneT) = \mathbb{P}(\bY\oneT = \by\oneT)$, is detailed in Appendix~\ref{subsub:fwdalgo}; we present the backward algorithm and the final combination step here.

Using the same notation as in Appendix~\ref{subsub:fwdalgo}, we first note that 
\begin{equation}
    \mleft(\bGamma \bP(\by_T) \mright) \bone
    =
    \begin{bmatrix}
        \sum_{k=1}^K {\bb{P}}(X_T = k \mid X_{T-1} = 1) \cdot {\bb{P}}(\bY_T = \by_T \mid X_T = k) \\
        \vdots \\
        \sum_{k=1}^K {\bb{P}}(X_T = k \mid X_{T-1} = K) \cdot {\bb{P}}(\bY_T = \by_T \mid X_T = k)
    \end{bmatrix}
    = 
    \begin{bmatrix}
        {\bb{P}}(\bY_T = \by_T \mid X_{T-1} = 1) \\
        \vdots \\
        {\bb{P}}(\bY_T = \by_T \mid X_{T-1} = K) \\
    \end{bmatrix}
\end{equation}
and 
\begin{align}
\begin{split}
    &\phantom{{}={}}\mleft(\bGamma \bP(\by_{T-1})\bGamma \bP(\by_T) \mright) \bone\\
    &=
    \begin{bmatrix}
        \sum_{k=1}^K {\bb{P}}(X_{T-1} = k \mid X_{T-2} = 1) \cdot {\bb{P}}(\bY_{T-1} = \by_{T-1} \mid X_{T-1} = k) \cdot {\bb{P}}(\bY_T = \by_T \mid X_{T-1} = k) \\
        \vdots \\
        \sum_{k=1}^K {\bb{P}}(X_{T-1} = k \mid X_{T-2} = K) \cdot {\bb{P}}(\bY_{T-1} = \by_{T-1} \mid X_{T-1} = k) \cdot {\bb{P}}(\bY_T = \by_T \mid X_{T-1} = k)\\
    \end{bmatrix}\\
    &= 
    \begin{bmatrix}
        {\bb{P}}(\bY_{T-1} = \by_T, \bY_{T} = \by_{T-1} \mid X_{T-2} = 1) \\
        \vdots \\
        {\bb{P}}(\bY_{T-1} = \by_T, \bY_{T} = \by_{T-1} \mid X_{T-2} = K) \\
    \end{bmatrix}.
\end{split}
\end{align}
It then follows by induction that for any $t \in \{2, \ldots, T\}$,
\begin{equation}\label{eq:bkwdalgofinal}
    \mleft( \prod_{s=t}^T \bGamma \bP(\by_{s}) \mright) \bone
    = 
    \begin{bmatrix}
    {\bb{P}}(\bY_t = \by_t, \ldots, \bY_{T} = \by_{T} \mid X_{t-1} = 1) \\
        \vdots \\
        {\bb{P}}(\bY_t = \by_t, \ldots, \bY_{T} = \by_{T} \mid X_{t-1} = K) \\
    \end{bmatrix}
    = 
    \begin{bmatrix}
    {\bb{P}}(\bY_{t:T} = \by_{t:T} \mid X_{t-1} = 1) \\
        \vdots \\
    {\bb{P}}(\bY_{t:T} = \by_{t:T} \mid X_{t-1} = K) \\
    \end{bmatrix}
\end{equation}

The backward algorithm computes the conditional probabilities ${\bb{P}}(\bY_{t:T} = \by_{t:T} \mid X_{t-1} = k)$ for each $t \geq 2$, via the left-hand side of \eqref{eq:bkwdalgofinal}. The time complexity of this algorithm is also $O(TK^2)$.

With the quantities $\bb{P}(\bY\onet = \by\onet, X_t = k)$ and ${\bb{P}}(\bY_{t:T} = \by_{t:T} \mid X_{t-1} = k)$ in hand for each $t \in \{2,\ldots,T\}$, we have that for any $t \in \{2,\ldots,T-1\}$,
\begin{align}\label{eq:fwdbckwd}
\begin{split}
    &\phantom{{}={}}
    \frac{1}{\sum_{k=1}^K {\bb{P}}(\bY\oneT = \by\oneT, X_T = k)} \begin{bmatrix}
        {\bb{P}}(\bY\onet = \by\onet, X_t = 1) \cdot {\bb{P}}(\bY_{(t+1):T} = \by_{(t+1):T} \mid X_t = 1) \\
        \vdots \\
        {\bb{P}}(\bY\onet = \by\onet, X_t = 1) \cdot {\bb{P}}(\bY_{(t+1):T} = \by_{(t+1):T} \mid X_t = K)
    \end{bmatrix} \\
    &=
    \frac{1}{{\bb{P}}(\bY\oneT = \by\oneT)} \begin{bmatrix}
        {\bb{P}}(X_t = 1) \cdot {\bb{P}}(\bY\oneT = \by\oneT \mid X_t = 1) \\
        \vdots \\
        {\bb{P}}(X_t = K) \cdot {\bb{P}}(\bY\oneT = \by\oneT \mid X_t = K) \\
    \end{bmatrix} \\
    &=
    \begin{bmatrix}
        {\bb{P}}(X_t = 1 \mid \bY\oneT = \by\oneT) \\
        \vdots \\
        {\bb{P}}(X_t = K \mid \bY\oneT = \by\oneT) \\
    \end{bmatrix},
\end{split}
\end{align}
which is now a vector consisting of the desired conditional state-membership probabilities. Replacing the terms in the first expression of \eqref{eq:fwdbckwd} by equivalent quantities computed efficiently using the forward and backward algorithms, the overall forward-backward algorithm can itself be summarized concisely by the equivalent identity
\begin{equation}\label{eq:fwdbwdalgo}
    \frac{1}{\mleft(\bdelta^\top \bP(\by_1) \prod_{s=2}^T \bGamma \bP(\by_s)\mright)\bone} \mleft( \bdelta^\top \bP(\by_1) \prod_{s=2}^t \bGamma \bP(\by_s) \mright)^\top \odot \mleft( \mleft( \prod_{s=t+1}^T \bGamma \bP(\by_{s}) \mright) \bone \mright) = \begin{bmatrix}
        {\bb{P}}(X_t = 1 \mid \bY\oneT = \by\oneT) \\
        \vdots \\
        {\bb{P}}(X_t = K \mid \bY\oneT = \by\oneT) \\
    \end{bmatrix}, 
\end{equation}
where $\odot$ refers to the element-wise (i.e., Hadamard) product of two matrices of equal dimension. The \emph{forward-backward algorithm} refers to the computation of the conditional state-membership probabilities via one pass each of the forward and backward algorithms in order to compute \eqref{eq:fwdbckwd} for each $t \geq 2$. The time complexity of the forward-backward algorithm remains $O(TK^2)$. 

After running the forward-backward algorithm, the \emph{local decoding} procedure computes the most likely state of the Markov chain at each time index $t$ given the observed data $\bY\oneT$ by simply selecting the coordinate corresponding to the largest entry in \eqref{eq:fwdbwdalgo}. That is, we select
\begin{equation}
    \hat{X}_t = \argmax{k \in \sX} {\bb{P}}(X_t = k \mid \bY\oneT = \by\oneT)
\end{equation}
for each $t=1,\ldots,T$, as required in \eqref{eq:predictedstates}.

\section{Likelihood Approximation by Discrete-Space HMM\MakeLowercase{s}}\label{app:approximation}

In this appendix, we show how the continuous-space HMM likelihood \eqref{CSHMMlike}, which involves $T$ iterated integrals over the state space $\sX$, can be approximated by a quantity which is essentially of the form \eqref{eq:DSHMMlike} and can be computed efficiently via the forward algorithm (see Appendix~\ref{subsub:fwdalgo}). Our presentation is based closely on the derivation of the univariate case in \cite{langrock2012some}, but applies to all three state-space models presented in Section~\ref{sub:nestedHMMs}, including the bivariate process described in Section~\ref{subsub:model3}. For generality, we present the approximation for an arbitrary continuous state space $\sX$; for Models~1 and 2, $\sX = \R$, and for Model~3, $\sX = \R^2$. In the former case, each $\sX$-valued vector below (e.g., $\bx_t$, $\bc^*_i$, etc.) is a univariate quantity.

The first step of the approximation is to identify an ``essential domain'' $A \subset \sX$ \citep{kitagawa1987non} for the $\bX_t$, such that $A$ is bounded and ${\bb{P}}(\bX_t \not \in A) = {\bb{P}}(\bX_t \in A^c)$ is small for each $t$. We then partition $A$ into a large number of subregions, $A_1, \ldots, A_m$; when $\sX = \R$ it is convenient to use intervals and when $\sX = \R^2$ we can use rectangles,  possibly of different lengths and widths. We  choose within each $A_i$ a representative point $\bc_i^*$, such as its center. If the area of each $A_i$ comprises a sufficiently small proportion of the total area of $A = \cup_{i=1}^m A_i$, then 
\makeatletter
\newcommand{\pushright}[1]{\ifmeasuring@#1\else\omit\hfill$\displaystyle#1$\fi\ignorespaces}
\makeatother
\begin{align}\label{eq:LLapprox1}
\begin{split}
    \iint_{\sX} \gamma(\bx_{T-1}, \bx_T) \cdot h_{\bx_T}(\by_T \mid \blambda_{\bx_T}) \dif \bx_T
    &= \iint_{A} \gamma(\bx_{T-1}, \bx_T) \cdot h_{\bx_T}(\by_T \mid \blambda_{\bx_T}) \dif \bx_T + \iint_{A^c} \gamma(\bx_{T-1}, \bx_T) \cdot h_{\bx_T}(\by_T \mid \blambda_{\bx_T}) \dif \bx_T\\
    &\approx \iint_{A} \gamma(\bx_{T-1}, \bx_T) \cdot h_{\bx_T}(\by_T \mid \blambda_{\bx_T}) \dif \bx_T \quad \mbox{since the integral over $A^c$ is assumed small}\\
    &= \sum_{i_T=1}^m \iint_{A_{i_T}} \gamma(\bx_{T-1}, \bx_T) \cdot h_{\bx_T}(\by_T \mid \blambda_{\bx_T}) \dif \bx_T\\
    &\approx \sum_{i_T=1}^m \iint_{A_{i_T}} \gamma(\bx_{T-1}, \bx_T) \dif \bx_T \cdot h_{\bc_{i_T}^*}(\by_T \mid \blambda_{\bc_{i_T}^*}) \quad \mbox{since $\bx_T \approx \bc_{i_T}^*$ when $\bx_T \in A_i$}\\
    &= \sum_{i_T=1}^m {{\bb P}}\left(\bX_T \in A_{i_T} \mid \bX_{T-1} = \bx_{T-1}\right) \cdot h_{\bc_{i_T}^*}(\by_T \mid \blambda_{\bc_{i_T}^*}).
\end{split}
\end{align}
Thus, as $A \to \sX$ and each $A_i \to \{\bc_i^*\}$ (i.e., as the essential domain becomes larger and its partition becomes finer), the approximations above become more exact. Applying the same reasoning,
\begin{align}
\begin{split}
&\phantom{{}={}}\iint_{\sX} \iint_{\sX} \gamma(\bx_{T-2}, \bx_{T-1}) \cdot h_{\bx_{T-1}}(\by_{T-1} \mid \blambda_{\bx_{T-1}}) \cdot \gamma(\bx_{T-1}, \bx_T) \cdot h_{\bx_T}(\by_T \mid \blambda_{\bx_{T}}) \dif \bx_{T} \dif \bx_{T-1}\\
&= \iint_{\sX} \gamma(\bx_{T-2}, \bx_{T-1}) \cdot h_{\bx_{T-1}}(\by_{T-1} \mid \blambda_{\bx_{T-1}}) \cdot \left(\iint_{\sX} \gamma(\bx_{T-1}, \bx_T) \cdot h_{\bx_T}(\by_T \mid \blambda_{\bx_{T}}) \dif \bx_{T}\right) \dif \bx_{T-1}\\
&\approx \iint_{\sX} \gamma(\bx_{T-2}, \bx_{T-1}) \cdot h_{\bx_{T-1}}(\by_{T-1} \mid \blambda_{\bx_{T-1}}) \cdot \left(\sum_{i_T=1}^m {{\bb P}}\left(\bX_T \in A_{i_T} \mid \bX_{T-1} = \bx_{T-1}\right) \cdot h_{\bc_{i_T}^*}(\by_T \mid \blambda_{\bc_{i_T}^*})\right) \dif \bx_{T-1}\\
&\hspace{4.75in}{\mbox{approximating the inner integral by \eqref{eq:LLapprox1}}}\\
&\approx \iint_{A} \gamma(\bx_{T-2}, \bx_{T-1}) \cdot h_{\bx_{T-1}}(\by_{T-1} \mid \blambda_{\bx_{T-1}}) \cdot \left(\sum_{i_T=1}^m {{\bb P}}\left(\bX_T \in A_{i_T} \mid \bX_{T-1} = \bx_{T-1}\right) \cdot h_{\bc_{i_T}^*}(\by_T \mid \blambda_{\bc_{i_T}^*})\right) \dif \bx_{T-1}\\
&\hspace{4.75in}\mbox{since the integral over $A^c$ is assumed small}\\
&= \sum_{i_{T-1} = 1}^m \iint_{A_{i_{T-1}}} \gamma(\bx_{T-2}, \bx_{T-1}) \cdot h_{\bx_{T-1}}(\by_{T-1} \mid \blambda_{\bx_{T-1}}) \cdot \left(\sum_{i_T=1}^m {{\bb P}}\left(\bX_T \in A_{i_T} \mid \bX_{T-1} = \bx_{T-1}\right) \cdot h_{\bc_{i_T}^*}(\by_T \mid \blambda_{\bc_{i_T}^*})\right) \dif \bx_{T-1}\\
&\approx \sum_{i_{T-1} = 1}^m \iint_{A_{i_{T-1}}} \gamma(\bx_{T-2}, \bx_{T-1}) \cdot h_{\bc^*_{i_{T-1}}}(\by_{T-1} \mid \blambda_{\bc^*_{i_{T-1}}}) \cdot \left(\sum_{i_T=1}^m {{\bb P}}\left(\bX_T \in A_{i_T} \mid \bX_{T-1} = \bc^*_{i_{T-1}}\right) \cdot h_{\bc_{i_T}^*}(\by_T \mid \blambda_{\bc_{i_T}^*})\right) \dif \bx_{T-1}\\
&\hspace{4.75in}\mbox{since $\bx_{T-1} \approx \bc_{i_{T-1}}^*$ when $\bx_{T-1} \in A_i$}\\
&= \sum_{i_{T-1} = 1}^m  \left[ \iint_{A_{i_{T-1}}} \gamma(\bx_{T-2}, \bx_{T-1}) \dif \bx_{T-1} \cdot h_{\bc^*_{i_{T-1}}}(\by_{T-1} \mid \blambda_{\bc^*_{i_{T-1}}}) \cdot \left(\sum_{i_T=1}^m {{\bb P}}\left(\bX_T \in A_{i_T} \mid \bX_{T-1} = \bc^*_{i_{T-1}}\right) \cdot h_{\bc_{i_T}^*}(\by_T \mid \blambda_{\bc_{i_T}^*})\right)\right]\\
&= \sum_{i_{T-1} = 1}^m \sum_{i_T = 1}^m \left( {\bb{P}}(\bX_{T-1} \in A_{i_{T-1}} \mid \bX_{T-2} = \bx_{T-2}) \cdot h_{\bc^*_{i_{T-1}}}(\by_{T-1} \mid \blambda_{\bc^*_{i_{T-1}}}) \cdot {\bb{P}}(\bX_T \in A_{i_T} \mid \bX_{T-1} = \bc^*_{i_{T-1}}) \cdot h_{\bc^*_{i_T}}(\by_T \mid \blambda_{\bc_{i_T}^*}) \right).
\end{split}
\end{align}

Proceeding inductively and handling the edge case of $\bX_1$ similarly, we obtain the approximation
\begin{equation}\label{eq:CSHMMlikeapp1appendix}
    L(\bveta \mid \by\oneT) \approx \sum_{i_1=1}^m \cdots \sum_{i_T=1}^m \left( {\bb{P}}(\bX_1 \in A_{i_1}) \cdot h_{\bc^*_{i_1}}(\by_1 \mid \blambda_{\bc^*_{i_{1}}}) \prod_{t=2}^T \left( {\bb{P}}(\bX_t \in A_{i_t} \mid \bX_{t-1} = \bc^*_{i_{t-1}}) \cdot h_{\bc^*_{i_t}}(\by_t \mid \blambda_{\bc^*_{i_{t}}})\right)\right),
\end{equation} 
which is exactly \eqref{eq:CSHMMlikeapp1}.

\section{EM Algorithms}\label{app:EMalgos}

The EM algorithm \citep{dempster1977maximum} is a popular tool used to fit statistical models in the presence of latent (or unobserved) data. Latent data may have a natural interpretation within the context of the problem (e.g., the $\bX\oneT$ in \eqref{eq:PoissonSSM2} representing the underlying physical process driving flaring activity is unobserved), or it may arise purely as a mathematical convenience to aid in inference (e.g., the $Z_t$ in \eqref{eq:FMMlatent} representing component membership when a finite mixture distribution is used for non-parametric density estimation). The essential idea is to augment the observed data, $\bx$, with ``missing'' data, $\bZ$, to form a \emph{complete data} set, $(\bx, \bZ)$, which induces a \emph{complete-data log-likelihood function} $\ell_\text{com}(\bveta \mid \bx, \bZ)$. Similar to the ordinary log-likelihood function, $\ell_\text{com}(\bveta \mid \bx, \bZ)$ is simply the logarithm of the joint density of $(\bX, \bZ)$, but viewed as a function of the underlying model parameter $\bveta$. The missing data, $\bZ$, is user-selected and chosen to make $\ell_\text{com}(\bveta \mid \bx, \bZ)$ more analytically tractable than the ordinary observed-data log-likelihood $\ell(\bveta \mid \bx)$. The EM algorithm is designed to compute the maximum likelihood estimate -- that is, the value of $\bveta$ that maximizes $\ell(\bveta \mid \bx)$ -- by iteratively maximizing the conditional expectation of $\ell_\text{com}(\bveta \mid \bx, \bZ)$, conditioned on the observed data $\bx$. More formally, given a starting value of the parameter $\bveta^{(0)}$, the algorithm iterates between the following two steps,
\begin{description}
    \item[E-step:] Compute $Q(\bveta \mid \bveta^{(r)}) = \mathbb{E}_{\bveta^{(r)}}\mleft[ \ell_\text{com}(\bveta \mid \bX, \bZ) \mid \bX = \bx \mright]$
    \item[M-step:] Set $\bveta^{(r+1)} = \argmax{\bveta} Q(\bveta \mid \bveta^{(r)})$
\end{description}
for $r=1,2,\ldots$ until convergence is achieved. 
In practice, if the equations in the E and M-step admit closed-form solutions, the resulting algorithm can be specified as a set of recursive updates for the components of $\bveta^{(r+1)}$ in terms of $\bveta^{(r)}$.

The convergence properties of the EM algorithm have been well studied. One primary benefit of the EM algorithm is that at each step of the algorithm, the maximizing value produced by the M-step can never decrease the observed-data log-likelihood $\ell(\cdot \mid \bx)$ from its value at the previous iteration. Thus, the EM algorithm can only converge to a stationary point of the likelihood function (assuming such a point exists), and under broad regularity conditions this guarantees convergence to the MLE when the likelihood is unimodal. There is a rich literature on the EM algorithm and the numerous algorithms related to it; for more information, we refer the reader to the seminal paper by \cite{dempster1977maximum}, the monograph by \cite{mclachlan2007algorithm}, and the review paper by \cite{vand:meng:10}.

In the following subsections, we briefly derive the EM algorithms used to fit the finite mixture models described in Section~\ref{sec:intervaldetermination}. 

\subsection{For Semi-Supervised Classification}\label{app:EMsemisupervised}

Here, the observed data $X\oneT = (X_1, \ldots, X_T)$ is assumed to be an independent and identically distributed sample from the mixture distribution
\begin{equation}\label{eq:FMM1app}
    F = \alpha \cdot F_1 + (1-\alpha) \cdot F_2(\cdot; \bpi),
\end{equation}
where $F_1$ is a \emph{known} distribution and $F_2(\cdot; \bpi)$ is a distribution with density
\begin{equation}
    f_2(x; \bpi) = \sum_{k=1}^K \frac{\pi_k}{b_k - b_{k-1}} \cdot \one{x \in [b_{k-1}, b_k)} = \prod_{k=1}^K \mleft( \frac{\pi_k}{b_k - b_{k-1}} \mright)^{\one{x \in [b_{k-1}, b_k)}}.
\end{equation}
Here $b_0, \cdots, b_K$ are known with $b_0 < b_1 < \cdots < b_K$, while $\alpha \in (0,1)$ and $\bpi = (\pi_1, \ldots, \pi_K)$ are parameters to be estimated,  with  $\sum_{k=1}^K \pi_k = 1$ and $\pi_k \in (0,1)$ for each $k=1, \ldots, K$. We define the independent and identically distributed latent variables, $Z\oneT = (Z_1, \ldots, Z_T)$, such that $Z_t \sim \text{Bernoulli}(\alpha)$ and $X_t \mid (Z_t = k-1) \sim F_k$, for $k=1,2$.  It is easy to verify that \eqref{eq:FMM1app} gives the marginal distribution of the $X_t$. The complete-data log-likelihood is 
\begin{align}
\begin{split}
    \ell_\text{com}(\bpi, \alpha \mid x\oneT, Z\oneT) &= \log\mleft( \prod_{t=1}^T (\alpha \cdot f_1(x_t))^{Z_t} \cdot ((1-\alpha) \cdot f_2(x_t; \bpi))^{1-Z_i} \mright) \\
    &= \sum_{t=1}^T \mleft\{ Z_t \cdot \mleft[\log\mleft( \frac{\alpha}{1-\alpha}\mright) + \log(f_1(x_t))  - \log(f_2(x_t;\bpi))\mright] + \log(1-\alpha) + \log(f_2(x_t; \bpi)) \mright\},
\end{split}
\end{align}
where 
\begin{equation}
\log(f_2(x_t;\bpi)) = \sum_{k=1}^K \mleft(\log(\pi_k) - \log(b_k - b_{k-1}) \mright) \cdot \one{x_t \in [b_{k-1},b_k)}
\end{equation}

The E-step requires the computation of $\mathbb{E}_{\bveta^{(r)}}[\ell_\text{com}(\bpi, \alpha; X\oneT, Z\oneT) \mid X\oneT = x\oneT]$, which by linearity requires only $\mathbb{E}_{\bveta^{(r)}}[Z_t \mid X\oneT = x\oneT] = \mathbb{P}_{\bveta^{(r)}}(Z_t = 1 \mid X_t = x_t)$. Using Bayes' rule and the law of total probability, we find that
\begin{equation}
    \mathbb{P}_{\bveta^{(r)}}(Z_t = 1 \mid X_t = x_t) = \frac{\alpha^{(r)} \cdot f_1(x_t)}{\alpha^{(r)} \cdot f_1(x_t) + (1-\alpha^{(r)}) \cdot f_2(x_t; \bpi^{(r)})} =: \gamma_1(x_t; \bpi^{(r)}, \alpha^{(r)}).
\end{equation}
The M-step requires that we maximize
\begin{align}
    &\phantom{{}={}}\mathbb{E}_{\bveta^{(r)}}[\ell_\text{com}(\bpi, \alpha; X\oneT, Z\oneT) \mid X\oneT = x\oneT] \nonumber \\
    &= \sum_{t=1}^T \mleft\{ \gamma_1(x_t; \bpi^{(r)}, \alpha^{(r)}) \cdot \mleft[\log\mleft( \frac{\alpha}{1-\alpha}\mright) + \log(f_1(x_t))  - \log(f_2(x_t;\bpi))\mright] + \log(1-\alpha) + \log(f_2(x_t; \bpi)) \mright\} \label{eq:Qfunction1}
\end{align}
with respect to both $\alpha$ and $\bpi$; these optimizations can be carried out separately because these parameters are functionally independent in \eqref{eq:Qfunction1}. Basic calculus shows that the maximizing value of $\alpha$ is 
\begin{equation}
    \hat{\alpha} = \frac{1}{T} \sum_{t=1}^T \gamma_1(x_t; \bpi^{(r)}, \alpha^{(r)}).
\end{equation}
Optimizing $\bpi$ is only slightly more complicated due to the constraint $\sum_{k=1}^K \pi_k = 1$, for which the method of Lagrange multipliers is particularly suitable. Applying this technique shows that $\pi_k$ is maximized by
\begin{equation}
    \hat{\pi}_k = \frac{\sum_{t=1}^T \gamma_2(x_t; \bpi^{(r)}, \alpha^{(r)}) \cdot \one{x_t \in [b_{k-1}, b_k]}}{\sum_{l=1}^K \sum_{t=1}^T \gamma_2(x_t; \bpi^{(r)}, \alpha^{(r)}) \cdot \one{x_t \in [b_{l-1}, b_l]}},
\end{equation}
where $\gamma_2(x_t; \bpi^{(r)}, \alpha^{(r)})  = 1 - \gamma_1(x_t; \bpi^{(r)}, \alpha^{(r)})$. The EM algorithm to estimate \eqref{eq:FMM1app} then simply amounts to repeating the following two steps for $r=1, 2, \ldots$ until convergence is reached, starting with initial values $\alpha^{(0)}$ and $\bpi^{(0)}$:
\begin{enumerate}
    \item Set \begin{equation}\alpha^{(r+1)} = \frac{1}{T} \sum_{t=1}^T \gamma_1(x_t; \bpi^{(r)}, \alpha^{(r)})\end{equation}
    \item Set \begin{equation} \bpi^{(r+1)} = \mleft(\frac{\sum_{t=1}^T \gamma_2(x_t; \bpi^{(r)}, \alpha^{(r)}) \cdot \one{X_t \in [b_{k-1}, b_k]}}{\sum_{l=1}^K \sum_{t=1}^T \gamma_2(x_t; \bpi^{(r)}, \alpha^{(r)}) \cdot \one{X_t \in [b_{l-1}, b_l]}}, \ldots, \frac{\sum_{t=1}^T \gamma_2(x_t; \bpi^{(r)}, \alpha^{(r)}) \cdot \one{X_t \in [b_{k-1}, b_k]}}{\sum_{l=1}^K \sum_{t=1}^T \gamma_2(x_t; \bpi^{(r)}, \alpha^{(r)}) \cdot \one{X_t \in [b_{l-1}, b_l]}} \mright). \end{equation}
\end{enumerate}

\subsection{For Unsupervised Classification}\label{app:EMunsupervised}

For full generality, we assume the data $\bX\oneT = (\bX_1, \ldots, \bX_T)$ is an independent and identically distributed sample from a mixture of $K$ multivariate normal distributions
\begin{equation}\label{eq:mixtureofMVN}
    F = \sum_{k=1}^K \alpha_k \cdot \sN_d(\cdot; \bmu_k, \bSigma_k)
\end{equation}
where $\alpha_1, \ldots, \alpha_K \in (0,1)$ with $\sum_{k=1}^K \alpha_k = 1$. We define the independent and identically distributed latent variables $Z\oneT = (Z_1, \ldots, Z_T)$ such that $Z_t \sim \text{Categorical}(K;\balpha)$ (i.e., each $Z_t$ is a discrete $\{1,\ldots,K\}$-valued random variable with $\mathbb{P}(X_t = k) = \pi_k$) and $\bX_t \mid (Z_t = k) \sim \sN_d(\bmu_k, \bSigma_k)$. Writing $\bveta = (\bmu_1, \ldots, \bmu_K, \bSigma_1, \ldots, \bSigma_K, \balpha)$, the complete-data log-likelihood is then
\begin{align}
    \ell_\text{com}(\bveta \mid \bx, Z\oneT) 
    &= \log \mleft[ \prod_{t=1}^T \prod_{k=1}^K \mleft(  \frac{\alpha_k}{\sqrt{(2\pi)^d |\bSigma_k|}} \cdot \exp\mleft(-\frac{1}{2}(\bx_t - \bmu_k)^\top \bSigma_k^{-1} (\bx_t - \bmu_k)\mright)\mright)^{\one{Z_t = k}} \mright] \\
    &= \sum_{t=1}^T \sum_{k=1}^K \one{Z_t = k} \cdot \mleft[ \log(\alpha_k) - \frac{1}{2} \mleft(d \log{2\pi} + \log{|\bSigma_k| + (\bx_t - \bmu_k)^\top \bSigma_k^{-1} (\bx_t - \bmu_k)}\mright) \mright].
\end{align}

The E-step requires the computation of $\mathbb{E}_{\bveta^{(r)}}[\ell_\text{com}(\bveta \mid \bX\oneT, Z\oneT) \mid \bX\oneT = \bx\oneT]$, which this time requires $\mathbb{E}_{\bveta^{(r)}}[\one{Z_t = k} \mid \bX\oneT = \bx\oneT] = \mathbb{P}_{\bveta^{(r)}}(Z_t = k \mid \bX_t = \bx_t)$ to be computed. Again, Bayes' rule and the law of total probability yield
\begin{equation}
    \mathbb{P}_{\bveta^{(r)}}(Z_t = k \mid \bX_t = \bx_t) = \frac{\alpha_k^{(r)} \cdot \phi_d(\bx_t; \bmu_k^{(r)}, \bSigma_k^{(r)})}{\sum_{l=1}^K \alpha^{(r)}_l \cdot \phi_d(\bx_t; \bmu_l^{(r)}, \bSigma_l^{(r)})} =: \gamma_k(\bx_t; \bveta^{(r)}),
\end{equation}
where $\phi_d(\cdot; \bmu, \bSigma)$ is the $\sN_d(\bmu, \bSigma)$ density function. The M-step thus requires the maximization of 
\begin{equation}
    \mathbb{E}_{\bveta^{(r)}}[\ell_\text{com}(\bveta \mid \bX\oneT, Z\oneT) \mid \bX\oneT = \bx\oneT] 
    = \sum_{t=1}^T \sum_{k=1}^K \gamma_k(\bx_t; \bveta^{(r)}) \cdot \mleft[ \log(\alpha_k) - \frac{1}{2} \mleft(d \log{2\pi} + \log{|\bSigma_k| + (\bx_t - \bmu_k)^\top \bSigma_k^{-1} (\bx_t - \bmu_k)}\mright) \mright]
\end{equation}
with respect to each $\bmu_k$, $\bSigma_k$, and $\balpha$. It is straightforward to show that the maximizing value of $\alpha_k$ is 
\begin{equation}
    \hat{\alpha}_k = \frac{1}{T} \sum_{t=1}^T \gamma_k(\bX_t; \bveta^{(r)}).
\end{equation}
The remaining parameters are most easily optimized using matrix calculus (we omit details but see, e.g., \cite{muirhead2009aspects}), which yield the optima
\begin{equation}
    \hat{\bmu}_k = \frac{1}{\sum_{t=1}^T \gamma_k(\bx_t; \bveta^{(r)})} \sum_{t=1}^T \gamma_k(\bX_t; \bveta^{(r)}) \bx_t
\end{equation}
and
\begin{equation}
    \hat{\bSigma}_k = \frac{1}{\sum_{t=1}^T \gamma_k(\bx_t; \bveta^{(r)})} \sum_{t=1}^T \gamma_k(\bx_t; \bveta^{(r)}) (\bx_t - \hat{\bmu}_k) (\bx_t - \hat{\bmu}_k)^\top.
\end{equation}
The EM algorithm to estimate \eqref{eq:FMM1app} then simply amounts to repeating the following two steps for $r=1, 2, \ldots$ until convergence is reached, starting with initial values $\balpha^{(0)}$ and $\bmu_1^{(0)}, \ldots, \bmu_K^{(0)}, \bSigma_1^{(0)}, \ldots, \bSigma_K^{(0)}$:
\begin{enumerate}
    \item Set \begin{equation}\balpha^{(r+1)} = \mleft(\frac{1}{T} \sum_{t=1}^T \gamma_1(\bx_t; \bveta^{(r)}), \ldots, \frac{1}{T} \sum_{t=1}^T \gamma_K(\bx_t; \bveta^{(r)}) \mright)\end{equation}
    \item Set\begin{equation}\bmu_k^{(r+1)} = \frac{1}{\sum_{t=1}^T \gamma_k(\bx_t; \bveta^{(r)})} \sum_{t=1}^T \gamma_k(\bx_t; \bveta^{(r)}) \bx_t, \quad k=1,\ldots, K \end{equation}
    \item Set \begin{equation}
    \bSigma^{(r+1)}_k = \frac{1}{\sum_{t=1}^T \gamma_k(\bx_t; \bveta^{(r)})} \sum_{t=1}^T \gamma_k(\bx_t; \bveta^{(r)}) (\bx_t - \bmu_k^{(j+1)}) (\bx_t - \bmu_k^{(r+1)})^\top, \quad k=1,\ldots, K 
    \end{equation}
\end{enumerate}

The initial values for mixture models such as \eqref{eq:mixtureofMVN} are typically obtained using the $k$-means algorithm. In the present case, one can run this algorithm (available in any statistical software package) on $\bX\oneT$ with the number of centers specified as $K$, which partitions the data into $K$ distinct subsets; for each mixture component $k \in \{1,\ldots,K\}$, $\bmu_k^{(0)}$ and $\bSigma_k^{(0)}$ are respectively set to the sample mean and covariance matrix from subset $k$, while $\alpha_k^{(0)}$ is set to the proportion of $\bX\oneT$ that comprises subset $k$.

Note that when a mixture model involves all component distributions within the same parametric family (such as \eqref{eq:mixtureofMVN} but not \eqref{eq:FMM1app}), any permutation of the component ``labels'' $1,\ldots,K$ produces the same value of the ordinary log-likelihood function, in the sense that
\begin{equation}
    \ell(\blambda_1, \ldots, \blambda_K \mid \bX\oneT) = \ell(\blambda_{\sigma(1)}, \ldots, \blambda_{\sigma(K)} \mid \bX\oneT),
\end{equation}
where $\blambda_k$ is the set of parameters associated with the $k$'th component distribution (including the mixing parameter $\alpha_k$) and $\sigma$ is any permutation of $(1,\ldots,K)$. This is an example of a phenomenon known as \emph{unidentifiability}, which results in $K$ modes in the log-likelihood surface; the value of the log-likelihood at each such mode is the same, and so the EM algorithm can converge to any one of them. Thus, from a computational perspective, one cannot \emph{a priori} associate any particular physical state (such as quiescence) to a specific component distribution of \eqref{eq:mixtureofMVN}. 

When the $\blambda_1,\ldots,\blambda_K$ can be ordered in some way, any particular ordering of the ``labels'' can be imposed on the likelihood function; for example, if \eqref{eq:mixtureofMVN} comprises of $K$ \emph{univariate} normal distributions and one desires the component distributions to be ordered increasingly with respect to their means, then one can set the log-likelihood to $-\infty$ whenever $\mu_1 < \mu_2 < \cdots < \mu_K$ fails to hold. Alternatively, the EM algorithm can sometimes be coaxed towards a particular labelling by judiciously choosing initial values. When $K$ is small, however, one may assign meaning to the components settled on by the algorithm following estimation. If one desires a specific ordering of the components, the labels of the estimated parameters can simply be permuted.

\newpage

\section{Additional Results for EV~Lac}\label{app:othertimebins}

\subsection{Model Estimates for Varying Time Bins}

In this section, we provide maximum likelihood estimates from the first stage of the model estimation procedure (prior to the bootstrap-based de-biasing procedure) for each of the three models described in Section~\ref{sub:nestedHMMs} fit to ObsID~01885, as the time bin $w$ (in seconds) varies among $\{25, 50, 75, 100\}$. The estimates are given in Tables~\ref{tab:mod1diffW}, \ref{tab:mod2diffW}, and \ref{tab:mod3diffW}, respectively and do not vary materially with $w$. Our experiments show that the de-biased estimates are similarly stable.

We do observe that in general, the $\sigma_1$ and $\sigma_2$ increase steadily as $w$ increases. This behaviour is expected, because these parameters control the step size (per unit time) of the underlying Markov chain. Heuristically, suppose that $X\oneT$ is the underlying soft-band process associated with the original time bin $w$, and $X'_{1:T'}$ is that associated with a larger time bin $w'$, where $T' < T$. Then the resolution of $X'_{1:T'}$ is lower than that of $X\oneT$, and so within each time bin of length $w'$, $X\oneT$ takes several independent steps (say $s$ of them, where $s > 1$) while $X'_{1:T'}$ takes a single step; that is, $X_{t_1:t_s}$ occur at the same time as $X'_t$, and the error of the latter is approximately the sum of the errors of each component of $X_{t_1:t_s}$. If $\sigma_1$ and $\sigma_1'$ are the parameters associated with $X\oneT$ and $X'_{1:T'}$, respectively, then $\sigma_1' \approx \sqrt{s}\sigma_1 > \sigma_1$. (This argument makes several simplifying assumptions, but can be made rigorous.)

\begin{table}
\centering
\begin{tabular}{r|cccc}
  \toprule
 & $w = 25$ & $w = 50$ & $w = 75$ & $w = 100$ \\ 
  \midrule
$\phi_{1}$ & 0.9874 & 0.9755 & 0.9636 & 0.9563 \\ 
  $\sigma_{1}$ & 0.0794 & 0.1161 & 0.1383 & 0.1576 \\ 
  $\beta_{1}$ & 0.1769 & 0.1787 & 0.1861 & 0.1823 \\ 
  $\beta_{2}$ & 0.0726 & 0.0733 & 0.07636 & 0.0748 \\ 
   \bottomrule
\end{tabular}
\caption{Maximum likelihood estimates for Model~1 fit using ObsID~01885 with $w \in \{25, 50, 75, 100\}$}
\label{tab:mod1diffW}
\end{table}

\begin{table}
\centering
\begin{tabular}{r|cccc}
  \toprule
 & $w = 25$ & $w = 50$ & $w = 75$ & $w = 100$ \\ 
  \midrule
$\phi_{1}$ & 0.9883 & 0.9773 & 0.9672 & 0.9591 \\ 
  $\sigma_{1}$ & 0.0667 & 0.0961 & 0.1147 & 0.1309 \\ 
  $\sigma_{2}$ & 0.1069 & 0.1539 & 0.1843 & 0.2099 \\ 
  $\beta_{1}$ & 0.1872 & 0.1864 & 0.1921 & 0.1869 \\ 
  $\beta_{2}$ & 0.0597 & 0.0593 & 0.0622 & 0.0596 \\ 
   \bottomrule
\end{tabular}
\caption{Maximum likelihood estimates for Model~2 fit using ObsID~01885 with $w \in \{25, 50, 75, 100\}$}
\label{tab:mod2diffW}
\end{table}

\begin{table}
\centering
\begin{tabular}{r|cccc}
  \toprule
 & $w = 25$ & $w = 50$ & $w = 75$ & $w = 100$ \\ 
  \midrule
$\phi_{1}$ & 0.9884 & 0.9768 & 0.9693 & 0.9612 \\ 
  $\phi_{2}$ & 0.9878 & 0.9744 & 0.9647 & 0.9549 \\ 
  $\sigma_{1}$ & 0.0711 & 0.0987 & 0.1143 & 0.1305 \\ 
  $\sigma_{2}$ & 0.1064 & 0.1568 & 0.1905 & 0.2152 \\ 
  $\beta_{1}$ & 0.1865 & 0.1860 & 0.1877 & 0.1836 \\ 
  $\beta_{2}$ & 0.0596 & 0.0591 & 0.0594 & 0.0580 \\ 
  $\rho$ & 1.0000 & 1.0000 & 1.0000 & 1.0000 \\ 
   \bottomrule
\end{tabular}
\caption{Maximum likelihood estimates for Model~3 fit using ObsID~01885 with $w \in \{25, 50, 75, 100\}$}
\label{tab:mod3diffW}
\end{table}

\subsection{Scatterplots of Hard and Soft Counts}

\begin{figure}
\includegraphics[width=0.5\linewidth]{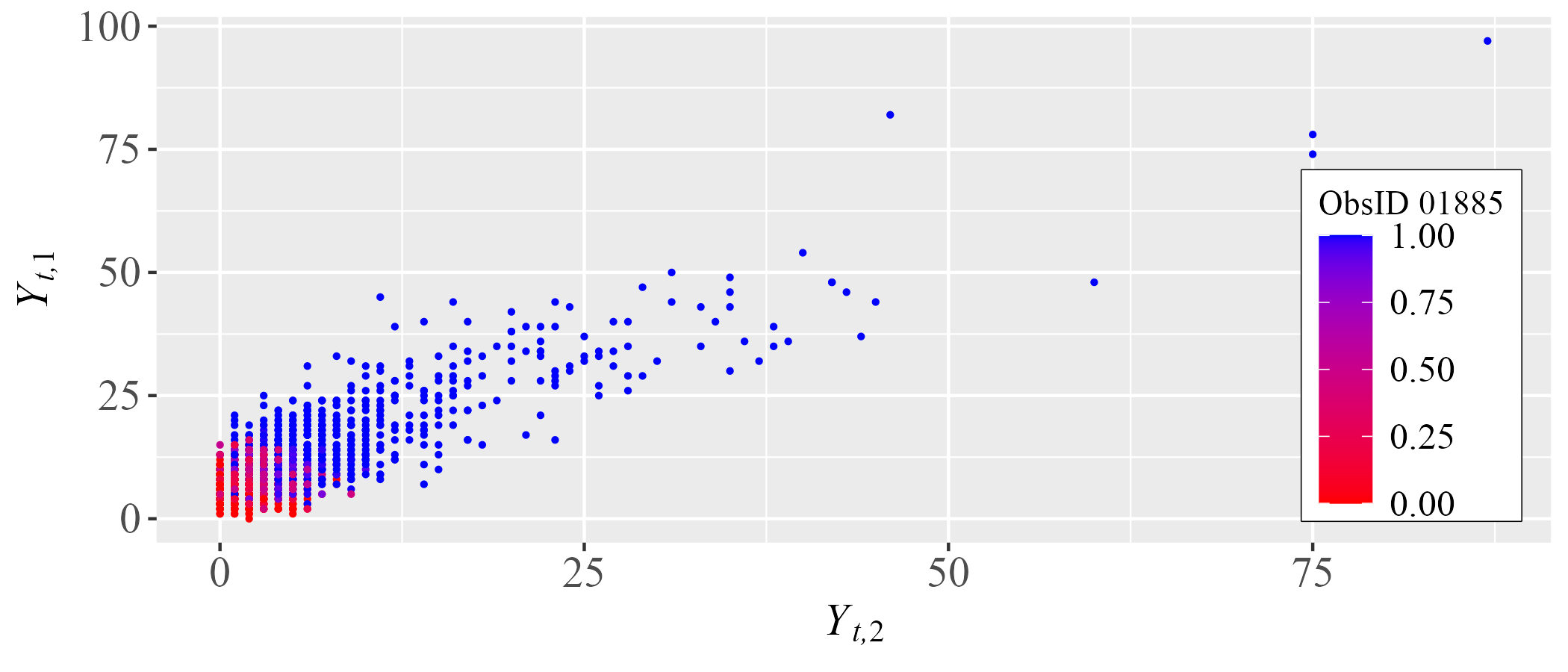}
\includegraphics[width=0.5\linewidth]{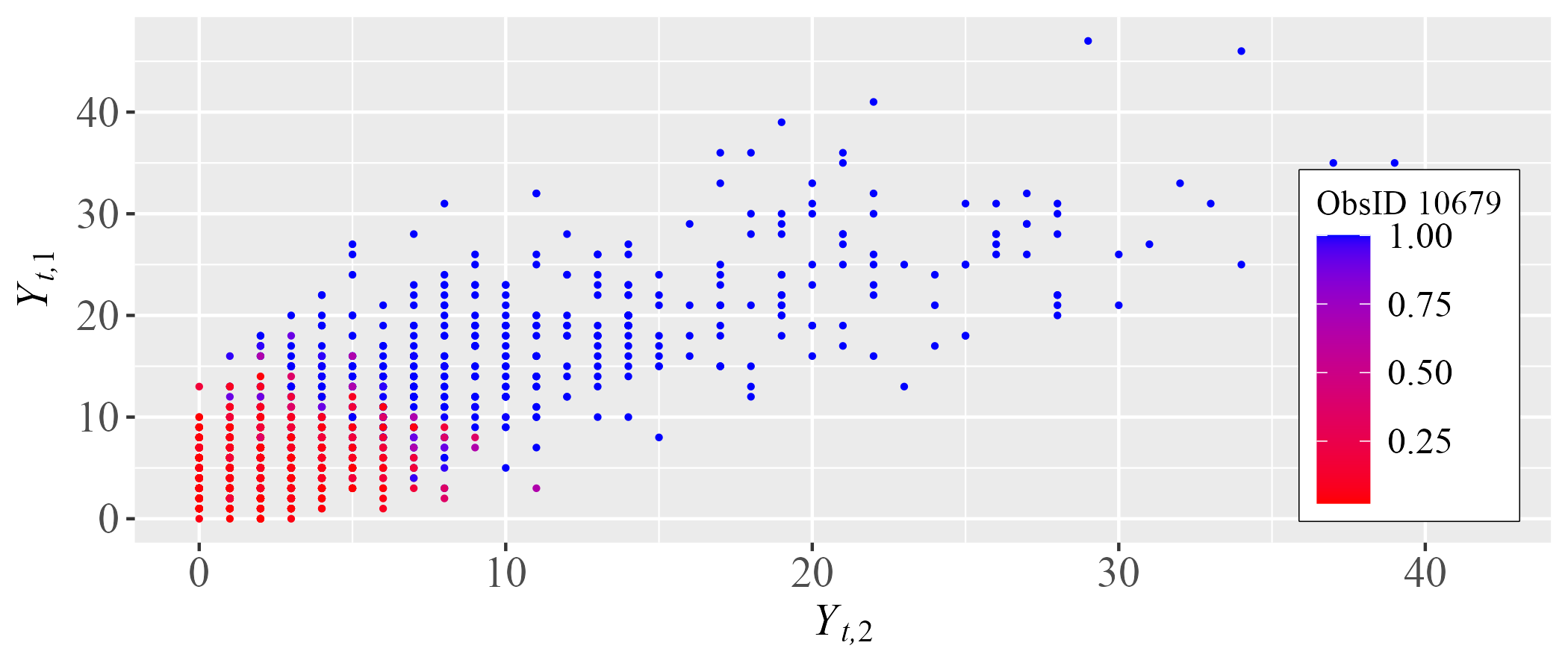}
\caption{Posterior flaring state probabilities computed via \eqref{eq:posteriorprob} for ObsID~01885 and via \eqref{eq:posteriorprobuns} for ObsID~10679. The probabilities are plotted as a function of the observed soft-band counts $Y_{1,1},\ldots,Y_{T,1}$ and the observed hard-band counts $Y_{1,2}, \ldots, Y_{T,2}$ for ObsID~01885 (left) and ObsID~10679 (right). Colour represents the posterior probability of the flaring state. It would not be possible for a flux threshold to reproduce these probabilities.}
 \label{fig:response}
\end{figure}

Scatterplots of the hard and soft counts coloured according to their posterior probabilities of being associated with the flaring state of \evlac (ObsID~01885 and ObsID~10679) appear in Figure~\ref{fig:response}. The scatterplots confirm that these probabilities could not be obtained with a threshhold on the observed counts.

\bsp	
\label{lastpage}
\end{document}